# Phase-field simulations of the morphology formation in evaporating crystalline multicomponent films

*Olivier J.J. Ronsin* [a,b]* *and Jens Harting* [a,c]*

[a] Helmholtz Institute Erlangen-Nürnberg for Renewable Energy, Forschungszentrum Jülich, Fürther Straße 248, 90429 Nürnberg, Germany

[b] Department of Chemical and Biological Engineering, Friedrich-Alexander-Universität Erlangen-Nürnberg, Fürther Straße 248, 90429 Nürnberg, Germany

[c] Department of Chemical and Biological Engineering and Department of Physics, Friedrich-Alexander-Universität Erlangen-Nürnberg, Fürther Straße 248, 90429 Nürnberg, Germany

E-mail: o.ronsin@fz-juelich.de, j.harting@fz-juelich.de

## Abstract

In numerous solution-processed thin films, a complex morphology resulting from liquid-liquid phase separation (LLPS) or from polycrystallization arises during the drying or subsequent processing steps. The morphology has a strong influence on the performance of the final device but unfortunately the process-structure relationship is often poorly and only qualitatively understood. This is because many different physical mechanisms (miscibility, evaporation, crystallization, diffusion, advection) are active at potentially different time scales, and because the kinetics plays a crucial role: the morphology develops until it is kinetically quenched far from equilibrium. In order to unravel the various possible structure formation pathways, we propose a unified theoretical framework that takes into account all these physical phenomena. This phase-field simulation tool is based on the Cahn-Hilliard equations for diffusion and the Allen-Cahn equation for crystallization and evaporation, which are coupled to the equations for the dynamics of the fluid. We discuss and verify the behavior of the coupled model based on simple test cases. Furthermore, we illustrate how this framework allows to investigate the morphology formation in a drying film undergoing evaporation-induced LLPS and crystallization, which is typically a situation encountered, e.g., in organic photovoltaics applications.

# 1. Introduction

In the field of renewable energies, organic electronics or membrane technologies, devices containing thin films or a stack of thin layers are very common. Solution processing is a method of choice for the fabrication of many of these devices. This processing route is often simple, low-energy demanding, low-cost and scalable, which makes it very attractive especially in an industrial context. Therefore, successfully producing well-performing devices with solution processing can be a fundamental milestone on the way to the market for future technologies. Typically, a thin film consists of one or several materials dissolved in a solvent or a solvent blend. The mixture is deposited on a substrate by various methods such as spin coating, doctor blading, slot-die coating or inkjet printing. [1] Then, the film is dried until the solvents fully evaporate. Finally, the dry film might undergo additional processing steps, for instance thermal annealing or solvent vapor annealing. The performance of the fabricated device depends not only on the properties of the selected materials, but also on the morphology of the dry thin films. This morphology develops during the fabrication process, especially during the drying phase. Therefore, it is highly desirable to understand the physical processes driving the morphology formation in order to gain control over the process-structure relationship and to propose improved processing routes for better device performance.

Solution-processed organic photovoltaics (OPV) is a very good example of such a system where the process parameters are of highest importance for device performance. The organic photoactive layers are typically 100-300nm thick and made of two materials, one electron donor (frequently a polymer material) and one electron acceptor. The current understanding of the structure-property relationship can be summarized as follows: [2] [3] [4] [5] the desired structure is a so-called 'bulk heterojunction', a co-continuous nanostructure of separated, relatively pure donor and acceptor regions with a significant crystallinity (typical crystal sizes of 10nm), and a mixed phase in between. This allows for high exciton separation efficiency, low recombination rates, high charge carrier mobilities and pathways to the electrodes for electrons and holes. The bulk heterojunction concept has led to very successful results over the past two decades, the best solar cells efficiencies now reaching 16-18%. [6] [7] [8] [9]. By contrast, the process-structure relationship is poorly and only qualitatively understood since the direct experimental assessment of the arising morphology is difficult. The general picture is the following: [3] [10] [11] [12] [13] starting from a very dilute, mixed wet film after deposition, the concentration increase upon drying leads to the onset of crystallization of one or both materials and/or liquid-liquid phase separation (LLPS). Whether these phase transitions occur and in which order of appearance depends on the one hand on the thermodynamic properties of the chosen material system. On the other hand, the kinetic properties of the drying mixture strongly influence the final morphology. This is because they vary over orders of magnitude upon drying, especially when polymer materials are involved, so that the system is kinetically quenched at a point far from its thermodynamic equilibrium. The available time before this quench typically determines the domain sizes, topology and crystallinity of the bulk heterojunction. Therefore, the final morphology is strongly influenced by the fabrication process. It can be optimized by changing parameters such as the temperature, the choice of the solvent, addition of an antisolvent, or post-processing steps allowing further evolution of the film (thermal annealing, solvent annealing). [3] [5] [14] [15] [16] Moreover, since the film morphology is not at equilibrium, it might in principle evolve during operation of the device and contribute to lifetime limitations or loss of efficiency. [17]

The question to be solved to understand the morphology formation of such systems is a non-equilibrium thermodynamics problem. The evolution of a mixture with variable composition (due to solvent removal) towards its thermodynamic equilibrium should be described depending on time, until the evolution becomes too slow and no noticeable changes can be observed, even if this 'final' state is still far from equilibrium. The objective of this paper is to present a simulation framework which is able to describe such a situation. Considering the situation for solution-processed OPV, the following features and physical processes are considered:

- The framework should be able to handle multicomponent mixtures with very different materials. For OPV, a mixture of three materials (polymer donor, small molecule acceptor and solvent) is the minimum requested, and a mixture of four materials (use of an antisolvent or a third photoactive material) is after all quite common.
- The thermodynamics should take into account the liquid-vapor phase transition for solvents (for evaporation in the case of drying or absorption by the film for solvent vapor annealing), the liquid-solid phase transition for all other materials (crystallization) including the handling of polycrystalline structures, and the miscibility of the mixture (LLPS by spinodal decomposition or nucleation and growth).
- The mass transport is expected to occur either by diffusion, by advection or by both, so that both processes should be considered.
- The kinetic properties (diffusion coefficients, crystallization rate, viscosities) are crucial for the proper description of the morphology formation. They have to be strongly composition-dependent.
- Even if it is not a critical problem for OPV, drying polymer mixtures or crystallizing systems (like solution-processed photoactive perovskite layers) often lead to rough dry structures featuring even sometimes uncovered substrate regions, which might dramatically hamper the device quality. The flexibility of the film surface and occurrence of dewetting processes should be handled consequently.

- The various rate processes (diffusion, fluid flows, evaporation, nucleation and growth, phase coarsening) occur with time scales that not only vary with composition, but also often differ by orders of magnitude (consider for instance the diffusion time over 100nm in a solvent, which is roughly 10 microseconds, with evaporation times usually exceeding one second). This has to be possible in the framework to perform simulations with realistic parameters and to obtain satisfactory agreement with experimental measurements.

These requirements are inspired by the typical example of OPV, but obviously they are generic and applicable to many other similar material systems. Thus, the framework presented in this paper is not restricted to OPV, and it can be used for other applications where part or all of the features described above have to be taken into account. Similarly, it is not restricted to the simulation of drying films but can also be used for instance for solvent vapor annealing, thermal annealing or morphology evolution during device lifetime.

It has been highlighted that the kinetic evolution is of highest importance for the structure determination. Together with the considered time scales and length scales of the structures (typically from a few nanometers to more than a micrometer wet film thickness for OPV), this makes small-scale simulation methods not well suited for this problem, even if very relevant results have been obtained by molecular dynamics, dissipative particle dynamics or self-consistent field theory. [18] [19] [20] [21] Mesoscale or continuum mechanics methods are more appropriate in terms of reachable time and lengths scales. Monte Carlo based lattice models [22] or the Lattice Boltzmann method [23] [24] [25] [26] [27] have been successfully applied to the simulation of phase separating mixtures, crystallization and evaporation. In addition, the phase-field (PF) method is a very attractive alternative and the most widely used method to deal with these topics. It is a well-established, versatile technique to handle interfacial problems with diffuse interfaces, starting with the thermodynamic description of the mixture through a free energy functional, so that phase transitions are taken into account in a very natural way. It is not the objective of this paper to propose an exhaustive review of the numerous scientific questions PF methods have been applied to, and the reader is referred to the existing literature for this purpose. [28] [29] [30] [31] [32] [33] [34] [35] [36] [37] [38] Different parts of the requirements described above have already been investigated with PF simulations for a long time and we will give in the following some striking examples picked from the available literature. The description of pure LLPS or pure crystallization goes back to the early work of Allen, Cahn and Hilliard. [39] [40] [41] More recently, PF simulations have been used extensively to investigate spinodal decomposition in multicomponent systems, [42] [43] [44] [45] boiling and evaporation, [46] [47] [48] [49] crystallization in liquid mixtures or solid blends, [50] [51] [52] [53] [54] [55] [56] [57] interplay between crystallization and LLPS, [58] [59] [60] evaporation induced LLPS [61] [62] [63] [64] [65] [66] [67] [68] [69] or evaporation induced crystallization [70]. The most relevant work to the problem handled in the current paper has probably been published by Saylor and Kim, who investigated LLPS and crystallization in evaporating polymer films for drug delivery applications. [71] [72] [73] [74] The coupling to fluid dynamics has been investigated and used by many authors. [75] [76] [77] [78] [79] [80] [81] [82] [83] [84] [85] [86] [87]

However, to the best of our knowledge, no PF framework that meets all of the specifications listed above has been proposed so far. Recently, we developed on the one hand a PF framework for the investigation of miscible or immiscible multicomponent crystallizing mixtures. [88] On the other hand, we proposed a new general PF framework for the description of evaporating liquid mixtures taking into account surface deformation, [89] which we modified and improved in order to better match theoretical and experimental results. [90] In the current paper, we now couple and extend these models, in particular taking into account thermal fluctuations and coupling them to the dynamics of the fluids, finally building a general framework that fulfills all the requirement detailed above. This enables us to investigate the morphology formation of multicomponent mixtures upon drying, even if the materials are immiscible and/or crystalline. The model equations are given in the following section (section 2). Then, we discuss the numerical implementation (section 3) and present benchmarks and simple test cases (section 4). We present simulations of structure evolution upon drying using the full coupled model (section 5) and finally discuss our results and possible perspectives (section 6).

## 2. Model Equations

### 2.1. Free energy functional

The phase-field equations described in this paper result from the coupling of the models reported in our previous work, [88] [90] and the reader is referred to these papers for more details. The simulated mixture is composed of $n$ fluids which can have a liquid and a vapor phase. Among them, $n_{cryst}$ materials are able to crystallize. The composition of the system is described at any point in time and space by the respective volume fractions of these materials $\varphi_i$. The phase state of each crystalline material is characterized by $n_{cryst}$ order parameters $\phi_k$ which vary from 0 in the liquid/amorphous phase to 1 in the solid/crystal phase. Additionally, for each crystalline material, one marker field $\theta_k$ allows for identification of the distinct crystallites. The value of $\theta_k$ is defined only where the crystals are present, and remains undefined in the liquid/amorphous or in the gas phase (see more details below). In the case where evaporation plays a role, the simulation domain includes not only the condensed phase, but also part of the vapor phase. $n_{solv}$ materials are solvents that can evaporate from the mixture and go into the vapor phase, which is nevertheless mainly composed of a further material (which will be called the 'air'). The solvents progressively escape from the simulation box and are replaced by the air. [90] The transition between the drying mixture and the gas phase is tracked with a single order parameter $\phi_{vap}$ which varies from 0 in the condensed phase to 1 in the vapor phase. We start writing the total free energy of the system as

$$G_{tot} = \int_V \left(\Delta G_V^{loc} + \Delta G_V^{nonloc}\right) dV \tag{1}$$

where $V$ denotes volume of the system. $\Delta G_V^{loc}$ is the local free energy density and $\Delta G_V^{nonloc}$ the non-local contribution due to the field gradients. The local part of the free energy is defined as

$$\Delta G_V^{loc}(\{\varphi_i\},\{\phi_k\},\{\theta_k\},\phi_{vap}) = \begin{aligned} & \left(1-p(\phi_{vap},1)\right)\Delta G_V^{cond}(\{\varphi_i\},\{\phi_k\}) \\ + & \; p(\phi_{vap},1)\Delta G_V^{vap}(\{\varphi_i\}) \\ + & \; \Delta G_V^{crystvap}(\{\phi_k\},\phi_{vap}) \\ + & \; \Delta G_V^{num}(\{\varphi_i\}) \end{aligned} \tag{2}$$

The first term on the right-hand side of Equation 2 stands for the change of the free energy density in the condensed phase. It describes the mixing and crystallization properties of the mixtures:

$$\Delta G_V^{cond}(\{\varphi_i\},\{\phi_k\}) = \begin{aligned} & \sum_{k=1}^{n_{cryst}} \rho_k {\varphi_k}^{\gamma_m}\left(g(\phi_k,\xi_{0,k})W_k + p(\phi_k,\xi_{0,k})\Delta G_{V,k}^{cryst}\right) \\ & + \frac{RT}{v_0}\left( \begin{aligned} & + \sum_{i=1}^{n}\sum_{j>i}^{n} \varphi_i\varphi_j\chi_{ij,ll} \\ \sum_{i=1}^{n}\frac{\varphi_i ln\varphi_i}{N_i} & + \sum_{k=1}^{n_{cryst}}\sum_{j\neq k}^{n} {\phi_k}^2\varphi_k\varphi_j\chi_{kj,sl} \\ & + \sum_{k=1}^{n_{cryst}}\sum_{j\neq k}^{n_{cryst}} \phi_j\phi_k\varphi_k\varphi_j\chi_{kj,ss} \end{aligned} \right) \end{aligned} \tag{3}$$

The first term on the RHS of Equation 3 represents the free energy density variation upon crystallization, where $g(\phi,\xi) = \phi^2(\phi-\xi)^2$ and $p(\phi,\xi) = \phi^2(3\xi - 2\phi)/\xi^2$ are interpolation functions classically used in phase-field simulations of crystallization processes. [30] [33] $\rho_k$ is the density of the material $k$ and $\Delta G_{V,k}^{cryst} = L_k\left(\frac{T}{T_{m,k}} - 1\right)$ its free energy density of crystallization, calculated from its enthalpy of fusion $L_k$ and its melting temperature $T_{m,k}$, respectively. $\xi_{0,k}$ is the value of the order parameter for which, in a pure material, the free energy density of crystallization is minimized and represents the maximum crystallinity of the material. The energy barrier to be overcome during the liquid-solid (or amorphous/crystalline) phase transition is taken into account with the help of the double-well function $g$, and its height is determined by the parameter $W_k$. In a mixture, the free energy density variation upon crystallization is proportional to the volume fraction to the power $\gamma_m$. $\gamma_m$ is classically assumed to be equal to 1 but we expect other dependencies to be possible. For instance, assuming that the energy gain upon crystallization corresponds to a decrease of the pairwise interaction energy between nearest neighbors would lead to $\gamma_m = 2$.

The term in the brackets of Equation 3 refers to the free energy of mixing, basing on the concepts of the Flory-Huggins theory. [91] There, $R$ is the gas constant, $T$ the temperature, $v_0$ the molar volume of the lattice site as defined in the Flory-Huggins theory. $N_i$ is the molar size of the material $i$ in terms of units of the lattice site volume, so that its molar volume is $v_i = N_i v_0$. The $\varphi_i ln\varphi_i$ part is the ideal mixing term, while the double-sum terms represent the enthalpic interactions between the respective materials. The first double-sum, corresponding to the liquid-liquid (or amorphous/amorphous) interactions, is the usual contribution initially proposed by Flory and Huggins, $\chi_{ij,ll}$ being the interaction parameter between the amorphous phases of materials $i$ and $j$. The second and third double-sums are a generalization to multicomponent mixtures [88] of the extension of the Flory-Huggins theory to crystalline materials initially proposed by Matkar and Kyu for binary systems [92] [93]. They stand for the interactions between the liquid/amorphous phase of material $j$ and the solid/crystalline phase of material $k$ (with interaction parameter $\chi_{kj,sl}$), and for the solid-solid (crystal/crystal) interactions (with interaction parameter $\chi_{kj,ss}$), respectively.

The second term on the right-hand side of Equation 2 stands for the free energy of the gas phase. [90] Here, for simplicity, the mixture is assumed to be ideal with gases of the same molecular size, so that the local free energy contribution reads

$$\Delta G_V^{vap}(\{\varphi_i\}) = \frac{RT}{v_0}\sum_{i=1}^{n}\varphi_i ln\left(\frac{\varphi_i}{\varphi_{sat,i}}\right) \tag{4}$$

In the equation above, $\varphi_{sat,i} = P_{sat,i}/P_0$ and $\varphi_i = P_i/P_0$, where $P_{sat,i}$ is the vapor pressure of the fluid *i*, $P_i$ its partial pressure in the gas phase and $P_0$ a reference pressure. The local free energy is interpolated at the condensed-gas phase interface between $\Delta G_V^{cond}$ and $\Delta G_V^{vap}$ using again the smooth function $p$.

The third term in on the right-hand side of Equation 2 is an interaction term between the gas phase and the crystals and prevents the overlapping of the order parameters of the crystals with the one of the gas phase:

$$\Delta G_V^{crystvap}(\{\phi_k\}, \phi_{vap}) = \sum_{k=1}^{n_{cryst}} E_k(\varphi_k, \phi_k)\left(\frac{\phi_k}{\xi_{0,k}}\right)^{\gamma_c} {\phi_{vap}}^{\gamma_v} \tag{5}$$

Here, $E_k(\varphi_k, \phi_k)$ is the interaction energy which will be discussed in more details below, and $\gamma_c$ and $\gamma_v$ are exponents that are classically equal to 1 or 2 in multiple-field phase-field modelling. [33] [31]

The last local term of the free energy functional is a purely numerical contribution introduced for stability purposes. [71] It prevents the volume fractions values from leaving the desired, physical ]0,1[ interval even when the thermodynamic properties of the mixture lead to the formation of very pure phases. This numerical contribution reads as:

$$\Delta G_V^{num}(\{\varphi_i\}) = \sum_{i=1}^{n}\frac{\beta}{\varphi_i^{\gamma_b}} \tag{6}$$

$\beta$ and $\gamma_b$ are numerical coefficients. $\beta$ is chosen as small as possible in order to grant numerical stability, without significantly modifying the physical behavior of the simulation.

The non-local contribution of the free energy represents the contribution of surface tension, which originates from volume fraction gradients and from liquid-solid and liquid-gas phase changes:

$$\Delta G_V^{nonloc}(\{\varphi_i\}, \{\Phi_k\}, \{\theta_k\}, \phi_{vap}) = \begin{aligned} &\sum_{i=1}^{n}\frac{\kappa_i}{2}(\boldsymbol{\nabla}\varphi_i)^2 \\ &+ \sum_{i=1}^{n_{cryst}}\left(\frac{{\varepsilon_k}^2}{2}(\boldsymbol{\nabla}\Phi_k)^2 + p(\Phi_k, \xi_{0,k})\frac{\pi\varepsilon_{g,k}}{2}|\boldsymbol{\nabla}|\delta_D(\boldsymbol{\nabla}\theta_k)\right) \\ &+ \frac{{\varepsilon_{vap}}^2}{2}(\boldsymbol{\nabla}\phi_{vap})^2 \end{aligned} \tag{7}$$

$\kappa_i$ is the surface tension parameter for the concentration gradient of material *i*. Although $\kappa_i$ is expected to depend on the composition in the case of polymer materials, [94] it is assumed to be constant in this work for simplicity. $\varepsilon_k$ are the surface tension parameters for the gradient of the order parameter of material *k*, and represent the contribution to the surface tension of the liquid-solid phase state variation. $\varepsilon_{g,k}$ are the surface tension parameters for the marker value gradients of material *k*. Following the ideas proposed during the development of the orientation-field phase field (OFPF) model, [51] [95] [54], the corresponding term in Equation 7 stands for the orientation mismatch energy between different single crystals of a given material and is responsible for impingement of the crystallites. As stated by the delta function $\delta_D$, it is defined only where there is a defined marker value jump, namely at the boundaries between two different crystallites. Note that this contribution is the same for all grain boundaries of a given material *k*. This is a simplification as compared to OFPF models, which leads to the fact that separate crystallites never merge in our framework. Finally, $\varepsilon_{vap}$ is the surface tension parameters for the gradient of the order parameter representing the condensed-gas phase transition.

## 2.2. Kinetic equations: volume fractions

Using the free energy functional detailed above, we can define the exchange chemical potential density, for all fluids from *1* to $n - 1$ as

$$\mu_{V,j}^{gen} - \mu_{V,n}^{gen} = \frac{\delta \Delta G_V}{\delta \varphi_j} - \frac{\delta \Delta G_V}{\delta \varphi_n} = \frac{\partial \Delta G_V}{\partial \varphi_j} - \frac{\partial \Delta G_V}{\partial \varphi_n} - \left( \nabla \left( \frac{\partial \Delta G_V}{\partial (\boldsymbol{\nabla} \varphi_j)} \right) - \nabla \left( \frac{\partial \Delta G_V}{\partial (\boldsymbol{\nabla} \varphi_n)} \right) \right) \tag{8}$$

The evolution of the volume fraction fields considering purely diffusive motion is the so-called Cahn-Hilliard equation, proposed by Cahn and Hilliard for binary mixtures [39] [40] and generalized later for multicomponent mixtures. [44] [42] When thermal fluctuations are taken into account, this equation is known as the Cahn-Hilliard-Cook equation. [96] In this work, the phase-field equations are coupled to the dynamics of the fluids (see below), so that we use the advective Cahn-Hilliard-Cook equation for the $i = 1 \ldots n-1$ materials:

$$\frac{\partial \varphi_i}{\partial t} + \boldsymbol{v} \boldsymbol{\nabla} \varphi_i = \frac{v_0}{RT} \nabla \left[ \sum_{j=1}^{n-1} \Lambda_{ij} \boldsymbol{\nabla} \left( \mu_{V,j}^{gen} - \mu_{V,n}^{gen} \right) \right] + \sigma_{CH} {\zeta_{CH}}^{i} \tag{9}$$

This equation is the general version of the stochastic advection-diffusion equation for a multicomponent mixture. We use a single velocity field $\boldsymbol{v}$ for all fluids. The Onsager mobility coefficients $\Lambda_{ij}$ are symmetric, $\Lambda_{ij} = \Lambda_{ji}$. The evolution of the volume fraction for the last material is deduced from the volume conservation, $\sum_{i=1}^{n} \varphi_i = 1$. ${\zeta_{CH}}^{i}$ is a coupled Gaussian space-time white noise preserving the fluctuation-dissipation theorem, meaning that for all materials $i$ and $j$, $\langle {\zeta_{CH}}^{i}(\boldsymbol{r}, t) \rangle = 0$ and $\langle {\zeta_{CH}}^{i}(\boldsymbol{r}, t) {\zeta_{CH}}^{j}(\boldsymbol{r}', t') \rangle = -\frac{2 v_0}{N_a} \nabla \left[ \Lambda_{ij} \delta_D(t - t') \boldsymbol{\nabla} \left( \delta_D(\boldsymbol{r} - \boldsymbol{r}') \right) \right]$, where $N_a$ is the Avogadro number and $\sigma_{CH}$ is a prefactor used to adjust the intensity of the noise. The mobility is interpolated between the mobility in the condensed phase $\Lambda_{ij}^{cond}$, and the ones in the gas phase $\Lambda_{ij}^{vap}$:

$$\Lambda_{ij} = \left( \Lambda_{ij}^{cond} \right)^{(1-\phi_{vap})} \left( \Lambda_{ij}^{vap} \right)^{\phi_{vap}} \tag{10}$$

In the gas phase, we assume the composition dependence of the mutual diffusion coefficients and the coupling between fluxes to be weak so that the mobility coefficients are written as $\Lambda_{ii}^{vap} = \varphi_i D_i^{vap}$ and $\Lambda_{ij}^{vap} = 0$, where $D_i^{vap}$ is the Fickian diffusion coefficient of the gas in the air. In the condensed phase, the mobilities depend on the diffusion coefficients, but they are also composition-dependent. The "slow-mode theory" and the "fast-mode theory", proposed by De Gennes [94] and Kramer [97], respectively, are implemented in the model. The expressions of the mobility coefficients in the liquid phase read for the slow mode model as

$$\begin{cases} \Lambda_{ii}^{cond} = \omega_i \left( 1 - \frac{\omega_i}{\sum_{k=1}^{n} \omega_k} \right) \\ \Lambda_{ij}^{cond} = -\frac{\omega_i \omega_j}{\sum_{k=1}^{n} \omega_k} \end{cases} \tag{11}$$

and for the fast mode model as

$$\begin{cases} \Lambda_{ii}^{cond} = (1 - \varphi_i)^2 \omega_i + {\varphi_i}^2 \sum_{k=1, k \neq i}^{n} \omega_k \\ \Lambda_{ij}^{cond} = -(1 - \varphi_i) \varphi_j \omega_i - (1 - \varphi_j) \varphi_i \omega_j + \varphi_i \varphi_j \sum_{k=1, k \neq i \neq j}^{n} \omega_k \end{cases} \tag{12}$$

Here, the coefficients $\omega_i$ are defined as $\omega_i = N_i \varphi_i D_{s,i}^{cond}(\{\varphi_i\}, \{\phi_i\})$, whereby $D_{s,i}^{cond}$ is the self-diffusion coefficient of the material *i*, which depends on the mixture composition and of the phase state (amorphous or crystal). The self-diffusion coefficient in the liquid/amorphous phase $D_{s,i}^{liq}$ is strongly dependent on the mixture composition in a non-trivial way. We use in this work a simple power law known as the Vignes law, [98] $D_{s,i}^{liq}(\varphi) = \prod_{k=1}^{n} \left( D_{s,i}^{\varphi_k \to 1} \right)^{\varphi_k}$, where $D_{s,i}^{\varphi_k \to 1}$ is the self-diffusion coefficient of the (liquid) $i^{th}$-material in the $k^{th}$ pure (liquid) material. Nevertheless, other simple assumptions can be made (weighted arithmetic or harmonic mean for instance). Additionally, the diffusion coefficient is expected to drop over orders of magnitude upon liquid-solid transition. To take this effect into account, we introduce the interpolation function defined by

$$log\left( f(x, d, c, w) \right) = \frac{1}{2} log(d) \left( 1 + tanh\left( w(x - c) \right) \right) \tag{13}$$

The penalty for the variable *x* is defined by 3 parameters *d*, *c* and *w* determining its amplitude *d*, its center position *c* and its width *w*, respectively. Taking the product of the self-diffusion coefficient in the liquid state and of this penalty function, $D_{s,i}^{cond}$ is calculated as

$$D_{s,i}^{cond}(\{\varphi_i\}, \{\phi_k\}) = f(\phi_{tot}, d_{sl}, c_{sl}, w_{sl}) \prod_{j=1}^{n} \left( D_{s,i}^{\varphi_j \to 1} \right)^{\varphi_j} \tag{14}$$

In the equation above, $\phi_{tot} = 1 - \prod_{k=1}^{n_{cryst}}(1 - \phi_k)$ is an estimate of the overall crystallinity at a given position, and $d_{sl}$, $c_{sl}$, $w_{sl}$ are the amplitude, centering and width of the diffusion coefficient variation upon liquid-solid transition.

## 2.3. Kinetic equations: evaporation method

The evaporation model and its behavior have been already presented elsewhere [90] and we here only briefly recall the approach and the equations. It mimics the Hertz-Knudsen representation of evaporation: the solvents undergo a very fast liquid-vapor phase transition, so that the vapor on top of the drying liquid is in quasi-static equilibrium with the condensed phase. The evaporation kinetics is governed by the comparatively slow diffusion process of solvent molecules from this (high partial pressure) equilibrium layer to the (low partial pressure) environment. The evolution of the order parameter of the gas phase is given by the advective Allen-Cahn equation:

$$\frac{\partial \phi_{vap}}{\partial t} + \boldsymbol{v}\boldsymbol{\nabla}\phi_{vap} = -\frac{v_0}{RT} M_{vap} \left( \frac{\partial \Delta G_V}{\partial \phi_{vap}} - \nabla \left( \frac{\partial \Delta G_V}{\partial (\boldsymbol{\nabla}\phi_{vap})} \right) \right) \tag{15}$$

Here, $M_{vap}$ is the mobility for the condensed-gas phase interface. It is chosen to be very high, so that the quasi-static equilibrium between the condensed and the gas phase is ensured at any time. The vapor phase is always already present at the beginning of the simulation (initial condition) and thus there is no need for fluctuations in Equation (15) to trigger evaporation. What is expected from the evaporation procedure is mainly to obtain the proper evaporation kinetics and time-dependent concentrations within the film. It has been verified in a previous paper[90] that this is the case, even without noise term in the Allen-Cahn equation. All materials considered in the simulation are present in the condensed phase as well as in the gas phase, because all volume fractions have to be strictly larger than 0 and smaller than 1. However, we distinguish 3 types of materials: first, the solutes are assumed to have a very low vapor pressure, so that the vapor state is very unfavorable for them (see Equation 4), and their volume fraction in the gas phase is very small. Thus, they stay in the condensed phase. Second, the solvents have a high vapor pressure and can escape from the condensed phase to move into the vapor phase. Due to the very fast Allen-Cahn kinetics, this results in the whole vapor phase being in equilibrium with the condensed phase. In order to model the diffusion process which is responsible for the evaporation kinetics, the solvents are driven out of the simulation domain by an outflux boundary condition at the top of the box. Third, the air is assumed to have a very high vapor pressure and therefore is almost exclusively present in the vapor phase. As solvents leave the simulation box, they are replaced by air to ensure the conservation of volume, $\sum_{i=1}^{n} \varphi_i = 1$. Note that we apply the tremendous simplification that the solvent densities do not vary upon liquid-vapor phase transition, which has been shown to not affect the proper simulation of the drying kinetics. [90]

Because the condensed and the gas phase are in quasi-static equilibrium, the evaporation kinetics is fully independent of the mobility $M_{vap}$ of the surface tension parameter $\varepsilon_{vap}$ and more generally, of the interface profile. It is fully determined by the expression of the outflux. If the solvent volume fractions in the gas phase were equal to zero or negligible, the outflux would be written as

$$j_{i,HK} = \alpha \sqrt{\frac{v_0}{2\pi RT}\frac{N_i}{\rho_i}} P_0 \left( \varphi_{sat,i} \left( \frac{\varphi_{i,simu}^{vap}}{\varphi_{sat,i}} \right)^{N_i} - \varphi_i^{\infty} \right) \tag{16}$$

Here, $\varphi_i^{\infty}$ is defined as $\varphi_i^{\infty} = P_i^{\infty}/P_0$, with $P_i^{\infty}$ being the partial pressure in the environment and $\alpha$ being the evaporation-condensation coefficient. $\varphi_{i,simu}^{vap}$ is the simulated volume fraction of the material *i* in the vapor phase. This equation is the classical Hertz-Knudsen formula, [99] where the term $\left( \frac{\varphi_{i,simu}^{vap}}{\varphi_{sat,i}} \right)^{N_i}$ compensates the assumption of constant densities. However, in order to recover exactly the Hertz-Knudsen behavior, the outflux needs to take into account the (not always negligible) amount of solvent in the gas phase. The associated mass balance leads to the final expression of the outflux implemented at the upper boundary of the simulation box:

$$j_i^{z=z_{max}} = j_{i,HK} - \left(\varphi_{i,simu}^{vap} + \Delta\varphi_{i,simu}^{vap}\right) \left( \sum_{k \in \{solv\}} j_{k,HK} \right) + \frac{z_{max}\Gamma_{vap}}{dt} \Delta\varphi_{i,simu}^{vap} \tag{17}$$

Here, $\Delta\varphi_{i,simu}^{vap}$ is the volume fraction variation in the vapor phase during a single time step, $z_{max}$ the height of the simulation box and $\Gamma_{vap}$ the proportion of vapor phase in the whole box.

## 2.4. Kinetic equations: crystallization

The evolution of the order parameters of the $k = 1 \dots n_{cryst}$ crystalline materials obeys the advective stochastic Allen-Cahn equation:

$$\frac{\partial \phi_k}{\partial t} + \boldsymbol{v}\boldsymbol{\nabla}\phi_k = -\frac{N_k v_0}{RT} M_k \left( \frac{\partial \Delta G_V}{\partial \phi_k} - \nabla \left( \frac{\partial \Delta G_V}{\partial (\boldsymbol{\nabla}\phi_k)} \right) \right) + \sigma_{AC} f\big(\phi_k, d_\zeta, c_\zeta, w_\zeta\big) {\zeta_{AC}}^k \tag{18}$$

The mobility coefficient $M_k$ for the solid-liquid interface for crystals of material $k$ can be chosen to be either constant, $M_k = M_{k,0}$ or related to the self-diffusion coefficient of that material in the amorphous phase $M_k = M_{k,0}\, D_{s,k}^{liq}(\{\varphi\})/D_{s,kk}^{liq}$. Such a dependence is expected, [71] because crystal growth is not only driven by the thermodynamic properties, but also by the local mobility of the crystallizing atoms or molecules that have to spatially arrange in order to attach to the crystal. In this second case, the mobility is substantially increased when a solute is dispersed in a very mobile phase, for instance in a solvent. ${\zeta_{AC}}^k$ is again a Gaussian white noise preserving the fluctuation-dissipation theorem, [100] i.e. for each material $k$, $\langle {\zeta_{AC}}^k(\boldsymbol{r},t)\rangle = 0$ and $\langle {\zeta_{AC}}^k(\boldsymbol{r},t){\zeta_{AC}}^k(\boldsymbol{r}',t')\rangle = \frac{2N_k v_0}{N_a} M_k \delta_D(t-t')\delta_D(\boldsymbol{r}-\boldsymbol{r}')$. This noise term is responsible for nucleation and grain coarsening in the simulations. Here again, $\sigma_{AC}$ is a prefactor used to adjust the intensity of the noise. $f\big(\phi_k, d_\zeta, c_\zeta, w_\zeta\big)$ is again defined by Equation 13 with the parameters $d_\zeta$, $c_\zeta$, $w_\zeta$ defining the amplitude, center and width of the interpolation function. This can be used to damp the fluctuations in the already crystalline domains, mainly in order to improve numerical stability without impacting the physical behavior of the simulation. Other methods have been proposed to simulate nucleation without using these numerically expensive fluctuations, typically by manually introducing new nuclei with a given (composition-dependent) nucleation rate, [101] [102] [103] [104] including calculation of the proper nucleation rate, as well as radius [105] [52] [53] and shape [106] [107]. While being quite simple for binary blends with relatively homogeneous volume fractions in the liquid phase, these methods can become complex in the situation we are after in this work, with three or more materials and a possibly very strong composition inhomogeneity. Another classical argument against the use of the stochastic Allen-Cahn equation is that nucleation events are expected to be rare, so that a considerable amount of time steps and thus a prohibitive computation time is required until nuclei arise spontaneously from the thermal fluctuations. Fortunately, for applications like solution-processed photovoltaics, nucleation occurs within time scales comparable to the drying time (and hence the simulated physical time), so that the nucleation events are indeed frequent. These reasons led us to consider the stochastic Allen-Cahn equation in order to generate the nucleation events.

In addition to the order parameter fields $\phi_k$ , the marker field $\theta_k$ has to be updated in order to evolve in line with the order parameter field. In contrast to the usual OFPF models where a kinetic equation is prescribed, [51] [95] [54] we use a simple heuristic procedure for the generation and the evolution of the marker field for each material $k$, basing on previous work. [88] First, the detection of the nucleation events is done in the following way: initially, out of the crystalline domains the marker field is undefined. A given area in the simulation domain is assumed to correspond to a new nucleus if it does not have a defined marker value yet, and if the order parameter $\phi_k$ and the volume fraction $\varphi_k$ exceed at the same time given threshold values $t_{\phi,k}$ and $t_{\varphi,k}$, respectively. All the nodes belonging to the same new nucleus are detected using a 'connected component labelling' procedure, and the same, new marker value is attributed to all of them. Note that such nucleation events are forbidden in the direct neighborhood of already existing crystals (typically 2-4 mesh points around the crystals depending on the interface thickness) in order to avoid false detections due to fluctuations in the diffuse interface of the crystal. Second, once crystals are formed, the associated order parameter at a given mesh point can decrease (unstable nuclei smaller than the critical nucleus, grain coarsening, advection from an old position), so that the mesh point cannot be considered as crystalline anymore. Thus, at each time step, the marker value of all nodes where the order parameter $\phi_k$ is less than the threshold values $t_{\phi,k}$ or the volume fraction $\varphi_k$ less than $t_{\varphi,k}$ is set to 'undefined' again. Third, around already existing crystals, the associated order parameter at a given mesh point can also increase (crystal growth for nuclei bigger than the critical nucleus, advection of the crystal to a new position) so that they can be considered as crystal nodes and should be attributed a marker value. This is done in the following way: when the value of the order parameter and the volume fraction on these surrounding nodes exceeds the threshold values $t_{\phi,k}$ and $t_{\varphi,k}$, they are attributed the marker value of the (already crystallized) neighboring node with the highest crystallinity $\phi_k\varphi_k$ . The marker value in a single crystallite is thus uniform, so that the orientation mismatch energy term in Equation 7 is only non-zero at the boundaries between two crystals with different marker values, as desired.

Whenever a gas phase is present in the system, crystals can reach the film surface so that a complex area with liquid-solid, liquid-vapor and solid-vapor interfaces may arise. We define the pure vapor as the area where $\phi_{vap} > 1 - t_{\phi_{vap}}$ and the condensed-gas phase interface as the area where $1 - t_{\phi_{vap}} > \phi_{vap} > t_{\phi_{vap}}$. Here, $t_{\phi_{vap}}$ is a threshold value that should be small. All crystalline order parameters and marker values are set to zero in the pure vapor. In order to avoid problems in the numerically critical solid-vapor interfaces, both noise contributions in the Allen-Cahn and Cahn-Hilliard equations are switched off, and no new crystal can appear in the condensed-gas phase interface as well as in the pure vapor. Despite of this, crystals formed in the film can reach the condensed-gas phase interface. However, a high volume fraction of crystalline solute in the vapor phase and in the condensed-gas phase interface is energetically very unlikely, so that crystals tend to be unstable when they reach the condensed-gas phase interface. Since this is an area often featuring a high volume fraction of solvent, [90] and therefore the Allen-Cahn mobility of the crystals are very high, this leads to a fast disappearance of any crystal in the condensed-gas phase interface. To avoid this unphysical effect due to the diffuse interface approach, an interaction energy has been defined in the solid-vapor interfaces (see Equation 5). This strongly inhibits the overlap of the order parameters of the crystals $\phi_k$ and of the vapor $\phi_{vap}$ . The interaction is active only inside the crystals (where a marker value is defined) and the directly surrounding areas representing the remaining diffuse crystal interface (typically 2-4 mesh points). The interaction energy is given by

$$E_k(\varphi_k, \phi_k) = E_{k,0} \frac{d_{sv}}{f(\varphi_k \phi_k, d_{sv}, c_{sv}, w_{sv})} \quad (19)$$

where $E_{k,0}$ is the interaction energy for a perfectly crystalline region. Here, we use again the interpolation function $f$ (Equation 13), and the parameters $d_{sv}$, $c_{sv}$, $w_{sv}$ define the amplitude, center and width of the interpolation function. This means that the interaction energy typically increases progressively over orders of magnitude from zero to $E_{k,0}$ when the product $\varphi_k \phi_k$ exceeds the value $c_{sv}$, preventing the vapor order parameter $\phi_{vap}$ to enter well-formed crystals. Unfortunately, this is still not sufficient to ensure the stability of emerging crystals at the film surface in a dilute solution because both order parameters $\phi_k$ and $\phi_{vap}$ overlap in the diffuse interface, promoting the reduction of the order parameter $\phi_k$. Since the solute volume fraction in the diffuse interface is far from 1, the diffusion coefficients and the Allen-Cahn mobility might be very high and the interface progressively disappears, which leads to a shrinking crystal. To avoid this, the Allen-Cahn mobility $M_k$ is also strongly damped in the regions where the solid-vapor interaction is active, using $M_k = f(\varphi_k \phi_k, d_{sv}, c_{sv}, w_{sv}) M_{k,0} \, D_{s,k}^{liq}(\{\varphi_k\}) / D_{s,kk}^{liq}$. As soon as the product $\varphi_k \phi_k$ becomes higher than $c_{sv}$, a crystal at the film surface becomes stable. This will be illustrated and discussed in Section 4.3.

## 2.5. Kinetic equations: fluid dynamics

The Cahn-Hilliard and the Allen-Cahn equations together ensure that the system progressively relaxes towards its thermodynamic equilibrium, by minimizing its free energy relative to the volume fraction and the order parameter variables. In addition to this, an advection term is introduced in the phase field equations (Equations 9, 15, 18) in order to take into account the impact of fluid motion on the system evolution. At the micro- or nanometer scale considered in this work, fluid motion is dominantly induced by capillary forces. These capillary forces are due to the numerous interfaces present in the system. In this section, we present the approach used to calculate the capillary forces from the phase fields and thereafter the velocity field to be used in the advective Cahn-Hilliard and Allen-Cahn equations.

The starting point are the continuity and momentum conservation equations, the energy conservation not being taken into account in the current model:

$$\begin{cases} \dfrac{\partial \rho}{\partial t} + \nabla(\rho \boldsymbol{v}) = 0 \\ \rho \left( \dfrac{\partial \boldsymbol{v}}{\partial t} + \boldsymbol{v} \boldsymbol{\nabla} \boldsymbol{v} \right) = \boldsymbol{F} - \boldsymbol{\nabla} P + \boldsymbol{\nabla \Sigma} \end{cases} \quad (20)$$

In these equations, $\boldsymbol{v}$ is the velocity, $\boldsymbol{F}$ are the applied forces, $P$ is the pressure and $\sum$ the viscous stress tensor. We assume the following in the current work:

- Even if we deal with multicomponent mixture, we make use of only one single velocity field that will be used for the advection of all materials present in the system.
- We assume perfect incompressibility, even in the vapor phase, as we did for the phase-field equations. Moreover, the density is assumed to be independent of the composition, which is a reasonable assumption for organic materials we wish to investigate.
- Since we are targeting at thin film applications, all length scales are sub-micrometer, and the Reynolds number is expected to be orders of magnitude smaller than 1. Thus, we are dealing with Stokes flows, whereby the inertial terms can be neglected.

- For these length scales, the dominant forces are the capillary forces stemming from the volume fraction and order parameter gradients $\boldsymbol{F_\varphi}$ and $\boldsymbol{F_\phi}$, respectively. In particular, gravity forces are neglected.
- The single fluid considered for the calculation of the velocity field is assumed to be a Newtonian fluid with inhomogeneous shear viscosity $\eta_{mix}$. Since we might be dealing with solute-solvent systems, including polymers for example, the viscosity is supposed to be strongly dependent on the composition. Moreover, the simulated mixture can contain crystals, which will be represented as highly viscous domains, and a vapor phase of very low viscosity. Hence, the viscosity depends on the order parameters $\phi_k$ and $\phi_{vap}$. Except when dealing with soft crystals, crystallites should be so viscous that they do not deform because of the flow. On the other hand, even when dealing with polymer mixtures, non-Newtonian viscous properties as well as visco-elastic or thixotropic behavior are ignored for the sake of simplicity.

As a consequence, we use the following simplified continuity and momentum conservation equations:

$$\begin{cases} \nabla \boldsymbol{v} = 0 \\ -\boldsymbol{\nabla} P + \boldsymbol{\nabla}\left(2\eta_{mix}\left(\{\varphi_i\},\{\phi_k\},\phi_{vap}\right)\boldsymbol{S}\right) + \boldsymbol{F_\varphi} + \boldsymbol{F_\phi} = \boldsymbol{0} \end{cases} \tag{21}$$

where $\boldsymbol{S}$ is the strain rate tensor. Regarding the viscosity, our aim is neither to propose nor develop a physical model or a constitutive law for multicomponent solute-solvent mixtures, nor to use any already-existing sophisticated model. We rather use a very simple functional form that renders qualitatively the basic phenomena, namely a high viscosity increase upon liquid-solid phase transition, a high viscosity decrease upon liquid-gas phase transition, and the viscosity evolution upon mixture composition, notably the viscosity increase of drying films upon evaporation. We propose the following equation for the viscosity in the condensed phase (where $\phi_{vap} \leq t_{\phi_{vap}}$, condensed-gas phase interface not included):

$$\frac{1}{\eta_{mix}} = \frac{1}{\eta_{cond}} = f\left(\sum_{k=1}^{n_{cryst}} \delta_D(\theta_k)\varphi_k\phi_k\,, d_\eta, c_\eta, w_\eta\right)\sum_{i=1}^{n}\frac{\varphi_i}{\eta_i} \tag{22}$$

As can be seen from the equation above, the viscosity in the amorphous domains is a weighted harmonic mean of the pure material viscosities $\eta_i$. In the crystalline areas (where a marker value is defined, as indicated by $\delta_D(\theta_k)$), the viscosity is increased according to the crystallinity $\varphi_k\phi_k$, using the interpolation function Equation 13 with parameters $d_\eta$, $c_\eta$, $w_\eta$ for the amplitude, center and width of the penalty. Concerning the vapor phase, the viscosity is expected to be orders of magnitude lower than in the condensed phase. Taking this into account would lead to intractable simulation times. Therefore, we use unrealistically high viscosity values for the vapor phase. Thus, the calculated flow in this area is not expected to be correct, which is fully acceptable since our concern is the morphology formation within the condensed phase. Nevertheless, the vapor phase viscosity is chosen much smaller than the condensed phase viscosity. The main advantage of this, in the context of a diffuse interface model, is to avoid the implementation of a free surface boundary condition at the condensed-gas phase interface, because the vapor phase is a low-viscosity layer which allows mimicking the free surface boundary condition. [108] [109] [110] The viscosity of the vapor phase (interface and pure vapor phase) is defined as

$$\eta_{mix} = \eta_{vap} = max\left({\eta_{mix}}^{1-\phi_{vap}}{\eta_g}^{\phi_{vap}}, min(\eta_{cond})/k_\eta\right) \tag{23}$$

where $\eta_g$ is a (small) user-defined viscosity value for the gas, $min(\eta_{cond})$ the smallest viscosity value found in the condensed phase and $k_\eta$ the acceptable ratio between both. This allows for a sufficient but limited viscosity contrast between the condensed and the vapor phase and avoids numerical problems that could occur during simulation of film drying: in order to mimic the free surface condition, the gas phase viscosity has to be significantly smaller than the film viscosity right from the beginning. Upon drying, the viscosity of the film might increase tremendously (for polymer solutions for instance) and our implementation allows for a corresponding increase of the vapor viscosity. This prevents the formation of a huge viscosity contrast at the film surface.

The coupling between the fluid dynamics and the phase-field equations is ensured, on the one hand, by the advection term in the Cahn-Hilliard and Allen-Cahn equations and, on the other hand, by the calculation of the capillary forces from the phase fields for the momentum conservation equation. Different ways of calculating the capillary forces have been proposed in the literature, which we call, following Jaensson, [87] the 'stress form' $\boldsymbol{F}(u) \sim a|\boldsymbol{\nabla} u|^2\mathbf{I} - \boldsymbol{\nabla} u \times \boldsymbol{\nabla} u$ with $a=0$, [77] $a=1/2$ [78] [79] [80] [81] or $a=1$ [87], the 'first potential form' $\boldsymbol{F}(u) \sim \frac{\delta\Delta G_V}{\delta u}\boldsymbol{\nabla} u$ [82] [87] and the 'second potential form' $\boldsymbol{F}(u) \sim u\boldsymbol{\nabla}\frac{\delta\Delta G_V}{\delta u}$ [83] [87] . In fact, these expressions have exactly the same deviatoric part and differ only by their isotropic part (see [87] [83] [84] and the derivation in the Supporting information S1). In other words, they differ only by the definition of the pressure. Therefore, when solving Equation 21 for $P$ and $\boldsymbol{v}$ with the different possible expressions of the capillary forces, the solution for the velocity field will exactly be the same, and only the calculated pressure field will be different. All three forms have been evaluated and it turns out from these tests that the stress form allows for the best numerical convergence properties. Generalizing the expression above for multicomponent volume fraction and order parameter fields, the expression of the capillary forces read (the expression for the potential forms can be found in the Supporting Information S1):

$$\begin{cases} \boldsymbol{F_\varphi} = \boldsymbol{\nabla}\left[\sum_{i=1}^{n} \kappa_i\left(|\boldsymbol{\nabla}\varphi_i|^2\boldsymbol{I} - \boldsymbol{\nabla}\varphi_i \times \boldsymbol{\nabla}\varphi_i\right)\right] \\ \boldsymbol{F_\phi} = \boldsymbol{\nabla}\left[\sum_{k=1}^{n_{cryst}} \varepsilon_k{}^2\left(|\boldsymbol{\nabla}\phi_k|^2\boldsymbol{I} - \boldsymbol{\nabla}\phi_k \times \boldsymbol{\nabla}\phi_k\right) + \varepsilon_{vap}{}^2\left(\left|\boldsymbol{\nabla}\phi_{vap}\right|^2\boldsymbol{I} - \boldsymbol{\nabla}\phi_{vap} \times \boldsymbol{\nabla}\phi_{vap}\right)\right] \end{cases} \quad (24)$$

# 3. Implementation

## 3.1. Dimensionless equations

The dimensionless kinetic equation can be derived using the scaling coefficients for energy density $g_{sc}$, for diffusion coefficients $D_{sc}$ and for length $l_{sc}$. They are chosen as $g_{sc} = RT/v_0$, $D_{sc} = max\left(N_i D_{s,i}\right)$, and $l_{sc} = \sqrt{max\left(\kappa_{1\dots n}, \varepsilon_{1\dots n_{cryst}}{}^2, \varepsilon_{vap}{}^2\right) / g_{sc}}$ to be consistent with the size of the thinnest interface of the system. The scaling coefficient for time is given by $t_{sc} = l_{sc}^2/D_{sc}$.

Defining the reduced variables $\tilde{t} = t/t_{sc}$, $\hat{\boldsymbol{v}} = \boldsymbol{v}t_{sc}/l_{sc}$, $\widetilde{\boldsymbol{\nabla}} = l_{sc}\boldsymbol{\nabla}$, $\widetilde{\Lambda}_{ij} = \Lambda_{ij}/D_{sc}$, $\widetilde{\Delta G_V^{loc}} = \Delta G_V^{loc}/g_{sc}$, $\hat{\kappa}_i = \kappa_i/(g_{sc}l_{sc}^2)$ and $\widehat{\zeta_{CH}}^i = t_{sc}\zeta_{CH}{}^i$, and making use of the non-local free energy density (Equation 7) the *n-1* dimensionless advective Cahn-Hilliard-Cook equations read

$$\frac{\partial\varphi_i}{\partial\hat{t}} + \hat{\boldsymbol{v}}\widehat{\boldsymbol{\nabla}}\varphi_i = \widehat{\nabla}\left[\sum_{j=1}^{n-1}\hat{\Lambda}_{ij}\widehat{\boldsymbol{\nabla}}\left(\frac{\partial\widehat{\Delta G_V}}{\partial\varphi_i} - \frac{\partial\widehat{\Delta G_V}}{\partial\varphi_n} - \hat{\kappa}_i\widehat{\boldsymbol{\nabla}}^2\varphi_i + \hat{\kappa}_n\widehat{\boldsymbol{\nabla}}^2\varphi_n\right)\right] + \sigma_{CH}\widehat{\zeta_{CH}}^i \quad (25)$$

Defining the additional reduced variables $\widehat{M}_{vap} = t_{sc}M_{vap}$, $\hat{\varepsilon}_{vap} = \varepsilon_{vap}/\sqrt{g_{sc}l_{sc}^2}$, $\widehat{M}_k = t_{sc}M_k$, , $\hat{\varepsilon}_k = \varepsilon_k/\sqrt{g_{sc}l_{sc}^2}$, $\widehat{\varepsilon_{g,k}} = \varepsilon_{g,k}/(g_{sc}l_{sc})$, $\widehat{\zeta_{AC}}^k = t_{sc}\zeta_{AC}{}^k$, the dimensionless advective Allen-Cahn equation for the condensed-gas phase transition reads:

$$\frac{\partial\phi_{vap}}{\partial\hat{t}} + \hat{\boldsymbol{v}}\widehat{\boldsymbol{\nabla}}\phi_{vap} = -\widehat{M}_{vap}\left(\frac{\partial\widehat{\Delta G_V}}{\partial\phi_{vap}} - \hat{\varepsilon}_{vap}{}^2\widehat{\boldsymbol{\nabla}}^2\phi_{vap}\right) \quad (26)$$

with the boundary condition $\hat{j}_i^{z=z_{max}} = j_i^{z=z_{max}}t_{sc}/l_{sc}$. The $k$ dimensionless advective stochastic Allen-Cahn equations for liquid-solid phase transition read:

$$\frac{\partial\phi_k}{\partial\hat{t}} + \hat{\boldsymbol{v}}\widehat{\boldsymbol{\nabla}}\phi_k = \begin{array}{l} -N_k\widehat{M}_k\left(\frac{\partial\widehat{\Delta G_V}}{\partial\phi_k} - \hat{\varepsilon}_k{}^2\widehat{\boldsymbol{\nabla}}^2\phi_k + p'\left(\Phi_k, \xi_{0,k}\right)\frac{\pi\widehat{\varepsilon_{g,k}}}{2}\left|\widehat{\nabla}\right|\delta\left(\widehat{\boldsymbol{\nabla}}\theta_k\right)\right) \\ +\sigma_{AC}f\left(\phi_k, d_\zeta, c_\zeta, w_\zeta\right)\widehat{\zeta_{AC}}^k \end{array} \quad (27)$$

Finally, defining $\hat{P} = P/g_{sc}$, $\hat{\eta}_{mix} = \eta_{mix}/(g_{sc}t_{sc})$, $\widehat{\boldsymbol{S}} = t_{sc}\boldsymbol{S}$, $\widehat{\boldsymbol{F}}_{\boldsymbol{\varphi}} = \boldsymbol{F_\varphi}l_{sc}/g_{sc}$ and $\widehat{\boldsymbol{F}}_{\boldsymbol{\phi}} = \boldsymbol{F_\phi}l_{sc}/g_{sc}$, the dimensionless continuity and momentum conservation equations read:

$$\begin{cases} \widehat{\nabla}\hat{\boldsymbol{v}} = 0 \\ -\widehat{\boldsymbol{\nabla}}\hat{P} + \widehat{\boldsymbol{\nabla}}\left(2\hat{\eta}_{mix}\widehat{\boldsymbol{S}}\right) + \widehat{\boldsymbol{F}}_{\boldsymbol{\varphi}} + \widehat{\boldsymbol{F}}_{\boldsymbol{\phi}} = \boldsymbol{0} \end{cases} \quad (28)$$

## 3.2. Discretization

The dimensionless equations are discretized using a second order finite volume scheme on a 3D regular Cartesian staggered grid (see Figure 1) with centered differences for the phase field part and upwind/downwind differences for the advection and fluid dynamics part. Neumann, Dirichlet or periodic boundary conditions can be applied in each of the three directions. The use of staggered grids simplifies the discretization of the momentum (written at the respective velocity nodes) and continuity (written at the pressure nodes) equations. [110] Moreover, choosing the phase-field nodes to lie together with the pressure nodes simplifies the discretization of the capillary forces (written at the velocity nodes) and of the advection term (written at the phase-field nodes). The advection is calculated using a second order MUSCL scheme [111] with a Kurganov and Tabor scheme [112] and a Superbee flux limiter. [113] Indeed, other flux limiters (MinMod, van Leer, van Alaba, Sweby…) have been implemented and evaluated on various test cases of pure advection of crystalline or droplet structures. But the results turns out to be more satisfactory with the Superbee, in the sense that the structure shape are best conserved with this flux limiter.

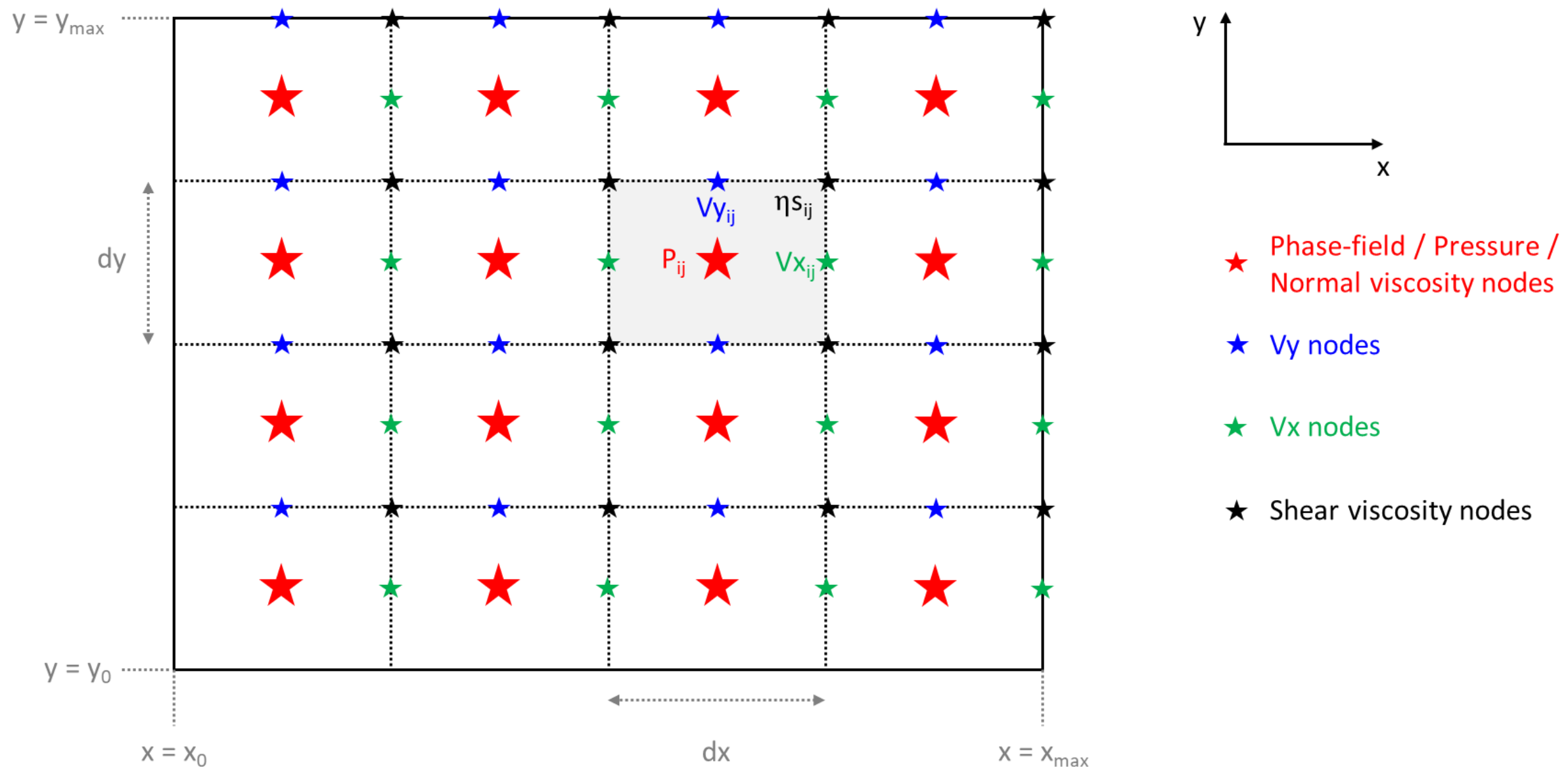

*Figure 1: representation of the staggered grids used for the discretization in the 2D case*

The discretization of the noise terms in the Cahn-Hilliard-Cook equations deserves some details. Basing of the work of Petschek, [114] Schaefer proposed an implementation of this noise using uncorrelated random numbers for a three component system. [64] In this work, we generalize the method of Schaefer to a multicomponent mixture. The final expression of the discretized noise for the material $i$ at the mesh point $x_a$ in the $x$-direction is (see Supporting Information S2):

$$\zeta_{CH,a}{}^{i} = \begin{aligned} &\sum_{k=1}^{i-1}\left( sign\left(\Lambda_{ik}{}^{a+\frac{1}{2}}\right)\sqrt{\left|\Lambda_{ik}{}^{a+\frac{1}{2}}\right|}\, B_{ik}{}^{a+1} - sign\left(\Lambda_{ik}{}^{a-\frac{1}{2}}\right)\sqrt{\left|\Lambda_{ik}{}^{a-\frac{1}{2}}\right|}\, B_{ik}{}^{a}\right) \\ &+\sqrt{2\Lambda_{ii}{}^{a+\frac{1}{2}}}B_{ii}{}^{a+1} - \sqrt{2\Lambda_{ii}{}^{a-\frac{1}{2}}}B_{ii}{}^{a} \\ &+\sum_{k=i+1}^{n}\left(\sqrt{\left|\Lambda_{ik}{}^{a+\frac{1}{2}}\right|}\, B_{ik}{}^{a+1} - \sqrt{\left|\Lambda_{ik}{}^{a-\frac{1}{2}}\right|}\, B_{ik}{}^{a}\right) \end{aligned} \tag{29}$$

where the $B_{ik}$ are $n$ fields of Gaussian random numbers with variance $\frac{2v_0}{N_a \Delta x^3 \Delta y \Delta z \Delta t}$ coupling the fluctuations of the materials $i$ and $k$. Similar expressions can be found in the $y$- and $z$-direction so that in 3D, the thermal fluctuations are calculated from $3(n(n+1)/2-1)$ independent fields of Gaussian random numbers.

## 3.3. Time stepping

The time stepping scheme proceeds in the following way: first, in order to obtain the velocity field, Equation 28 is solved for $\boldsymbol{v}$ and $P$ using Equation 22 and 23 for the viscosity and Equation 24 for the capillary forces. Second, making use of the calculated velocity field, the coupled advective phase-field Equations 25-27 are solved for $\{\varphi_i\}, \{\phi_k\}, \phi_{vap}$ using the local free energy defined by Equations 2-6 and the boundary condition Equations 16-17. Third, the marker fields $\theta_k$ are updated with the procedure described above (see *Figure 2*).

In realistic simulations, one of the challenges is that the different time scales of the problem (for diffusion, crystallization, advection, evaporation…) might differ by orders of magnitude. This makes implicit time stepping methods necessary to solve the phase field equations. Indeed, using explicit time stepping would require very small time steps for the sake of numerical stability, for instance because of the fourth order spatial derivative in the Cahn-Hilliard equation or the very high mobility values used for the Allen-Cahn mobility for the condensed-gas phase transition. This would lead to intractable simulation times even for 1D simulations. As a consequence, a set of unconditionally A- and L-stable diagonally implicit Runge-Kutta (DIRK) time schemes [115] [116] have been implemented, from the one-stage, first-order Euler backward method to the five-stage, fourth- order SDIRK4(3)5L[1]SA_C(2) method. In practice, we find that in almost all of the performed simulations, the two-stage, second-order Pareschi-Russo (see Supporting Information S3) method leads to the best compromise between time convergence properties and simulation time. Note that the advection part of the phase-field equations could also be solved separately using high-order explicit Runge-Kutta methods, but we obtain better results by solving the whole, coupled advective phase-field equations at a time.

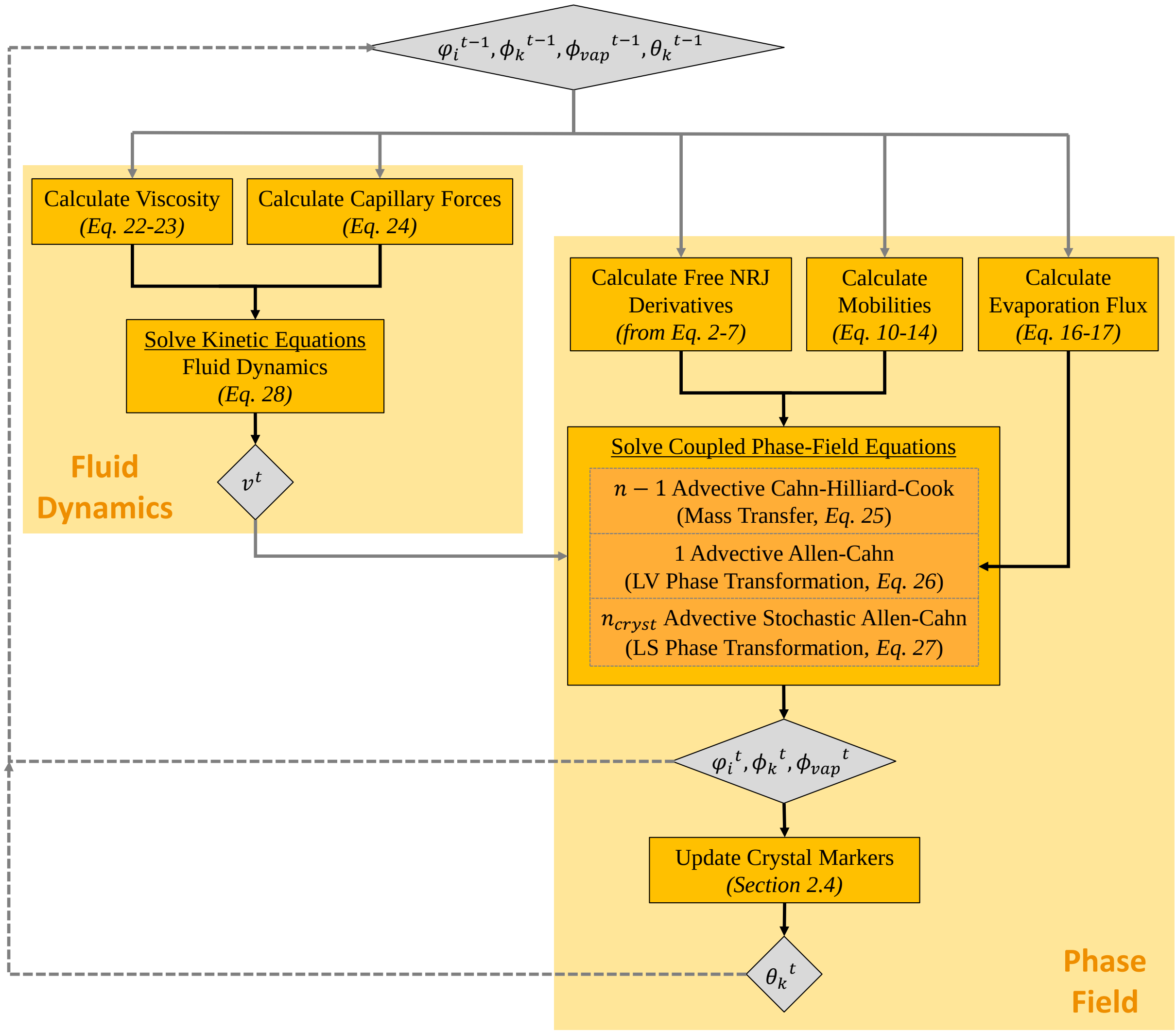

*Figure 2: Overview of the solution procedure for one time step in a single stage time stepping method. In the case of multistage time stepping, this fluid mechanics + phase field resolution procedure is repeated at each stage before calculating the phase fields at the next time from the results at all stages according to the Butcher matrix.*

For a computationally efficient simulation, it is desired that the time steps become as large as possible. In practice, the time steps can vary over orders of magnitude during one single simulation. We use adaptive time steps with a heuristic strategy basing on a simple principle: it is required that all volume fractions everywhere in the simulation box lie in the ]0,1[ interval. If this condition is not fulfilled with the calculated solution, the time increment is rejected and recalculated with a time step reduced by 50%. Otherwise, the time step is increased by 20% for the next time increment. In order to reach longer time steps, the local part of the free energy is linearized relative to the field variables $\{\varphi_i\}, \{\phi_k\}, \phi_{vap}$. On top of this, several upper limits are set to the time step and constantly updated during the simulation.

- The first upper limit is given by the Courant-Friederichs-Lewy (CFL) criterion calculated using the expected interface velocity $v_{int}(t)$ that can be derived from the outflux boundary condition (Equation 17) $\Delta t \frac{v_{int}(t)}{\Delta x} < C_{CFL1}$, where $\Delta x$ is the grid spacing and $C_{CFL1}$ a number chosen smaller than 1.
- The second upper limit is given by the Courant-Friederichs-Lewy criterion calculated using the velocity field $\boldsymbol{v}$ obtained from the fluid mechanics equations, $\Delta t \cdot max_{\boldsymbol{r}} \left| \sum_{i=1}^{n_{dim}} \frac{v_i(\boldsymbol{r})}{\Delta x} \right| < C_{CFL2}$, where $n_{dim}$ is the dimensionality of the simulation, $\boldsymbol{r}$ the position of a grid point and $v_i(\boldsymbol{r})$ the projection of the velocity in the $i^{th}$-direction.

- The third upper limit is given by the stability criterion for explicit time stepping relative to the Allen-Cahn equations for a liquid-solid transition, $\Delta t < \frac{RT}{v_0} \frac{2^{-n_{dim}} dx^2}{max_{r,k}(N_k M_k(r)) max_k(\varepsilon_k{}^2)}$. Even if this criterion might in principle be overcome with implicit time stepping, it turns out that the time convergence is often not guaranteed beyond this value, and that the updates of the marker fields might become problematic, especially at grain boundaries.

Using large time steps might in principle prevent the simulation to converge in time. In practice, it turns that the rules for time step management listed above together with the second order Pareschi-Russo time stepping method ensure proper time convergence. For each simulation, the volume conservation for all non-evaporating materials and the decrease of the total energy $G_{tot}$ with increasing time are checked for. If the volume conservation is not fulfilled with sufficient precision, the problem can be simply solved by limiting the time steps to smaller values.

When strong fluctuations of the volume fraction and/or order parameter fields are present, the capillary forces and the viscosity calculated from the phase fields can be very noisy, which may lead to a very irregular velocity field and a severe associated Courant-Friederichs-Lewy condition. In order to regularize the velocity field, we use modified, blurred phase fields (filtered with a Gaussian filter) for the calculation of the capillary forces in Equation 24. The viscosity field calculated from Equations 22, 23 is also filtered with the same Gaussian filter before solving the equation for the fluid dynamics. The impact of this filtering procedure on the overall physical behavior of the simulation turns out to be negligible, while significantly enhancing the numerical efficiency.

It should be pointed out that the presence of many different time scales results in various numerical and modelling challenges.

- First, if the physical processes to be simulated occur at different time scales, this might strongly impact the computational time. As stated above, three upper boundaries related to evaporation, crystallization and advection limit the time step size. On the one hand, the balance between these processes determines the computational cost. For instance, in the simulation of a drying film, if evaporation is limiting the size of the time step, the computational cost for the simulation is minimal. If advection is limiting the time step instead, e.g. a factor 10 smaller, then the computational cost is simply 10 times higher. At some point, prohibitive computational cost restricts the accessible parameter space. In this example, for realistic drying film simulations, this leads us to use unrealistically high viscosity values in the vapor and in the dilute solution, as discussed in Section 2.5 above and Section 4.3 below. On the other hand, coarsening processes (after LLPS and/or for grain coarsening) are very slow processes as compared to the phase build up process itself (LLPS or crystal nucleation and growth process). Therefore, investigation of coarsening requires very long simulations.
- Second, the strong mobility gradients that may arise (due to composition and order parameter dependent diffusion coefficients, Allen-Cahn mobilities, viscosities) in the simulation domain, especially at interfaces, can in principle lead to numerical difficulties, where the LV (or even the SV) interface appears to be the most problematic. It has to be sufficiently thick and the viscosity gradient between gas and condensed phase needs to be limited to enable proper solver convergence. Otherwise, no problematic restrictions regarding the resolution of both systems of equations (phase field on the one hand and fluid mechanics on the other hand) have been observed so far. Beyond this, having a fine mesh is desirable in the regions of high mobility gradients for accurate numerical solutions. There is off course a trade-off between computational cost and accuracy of the solution, but we have not been confronted so far to problematic situations that would jeopardize the results regarding the morphology development.
- Third, the strong mobility gradients at interfaces might unfortunately lead to unphysical phenomena in the simulation. This is especially the case for crystals, whereby the (in the real world very sharp) interface is represented by a diffuse interface in the phase-field framework. In order to avoid unphysical effects regarding crystal nucleation and stability, specific adjustments of the model are necessary. This is a major concern in this work, and this will be discussed in detail in Sections 4.2 and 4.3.

The code is implemented in parallel. The advective phase-field equations on the one side, and the fluid mechanics equations on the other side, are solved using the MUMPS direct solver [117] [118] through the PETSc library. [119] [120] [121] The computational time strongly depends on the number of time steps required to simulate the desired physical time, on the number of used cores, and of course on the number of degrees of freedom, which is the number of grid points times the number of coupled equations solved. For the simulations presented in this papers, this ranges from one hour on 4 cores (simulations of pure crystal advection, Section 4.3) to 5 days on 32 cores (full simulations, Section 5).

As described above, the equations are formulated and implemented for any number $n$ of materials. Thus, the extension of the simulations presented in this work to more components is straightforward and does not require any code modification. However, the computational cost increases significantly, especially because the number of degrees of freedom in the Cahn-Hilliard equation system increases with $(n-1)^2$. Most importantly, however, the analysis of the interactions between all the physical processes at play, and thus the understanding and the interpretation of the simulation results, might become quickly very challenging.

## 4. Benchmarks

In this section, we summarize some of the benchmark and test cases used to check the behavior of the model. To begin with, let us mention shortly some results on the individual, uncoupled building blocks. The behavior of our implementation of the Cahn-Hilliard equation has already been described in details in previous works. [88] [122] Additionally, we check that the implementation of the noise in the Cahn-Hilliard-Cook equation (Equation 9) gives consistent and meaningful results even with more than three materials, although we are unfortunately not aware of any benchmark test for such mixtures. The behavior of the stochastic Allen-Cahn equation is detailed in Section 4.2 below. Concerning the fluid dynamics part, we verify the code by comparing the simulation data to analytical solutions of simple flow problems (Poiseuille and Couette flow). We also check the implementation in situations with non-constant viscosity by simulating a Couette flow with a constant viscosity gradient, as well as a test case proposed by Gerya regarding advection at the boundary between two areas with different viscosities and densities. [123] Next, we investigate the advective part alone in the advective Cahn-Hilliard and Allen-Cahn equations. For this, we simulate the pure advection at constant velocity of a structure typically encountered in phase-field simulations, namely a domain of 5-10 grid points radius with a diffuse interface of 4-8 grid points thickness. It turns out that the second order MUSCL scheme is necessary to avoid numerical diffusion and thus to preserve the shape of the structure. As mentioned before, it is best preserved with the Superbee limiter. Moreover, with such a coarse mesh in the interface region and for large time steps (typically for a Courant-Friederichs-Lewy criterion fixed by $C_{CFL2} > 0.1$), first-order time stepping schemes generate a significant numerical diffusion so that a second or higher-order time stepping scheme is required for proper calculations.

The coupling of the Allen-Cahn and Cahn-Hilliard equations has also already been described in our previous work, for the simulation of crystallization and liquid-liquid phase separation [88] as well as for the simulation of evaporation. [90] In the following, we thus mainly focus on the new building blocks of the framework. First, we benchmark the behavior of the coupled Navier-Stokes-Cahn-Hilliard system in the case of spinodal decomposition in a binary mixture. Second, we illustrate the coupling of the stochastic Allen-Cahn equation with the Cahn-Hilliard equation by simulating nucleation, growth and coarsening of crystallites in a polymer solution. Third, we discuss the coupling of the whole set of equations by simulating drying films containing one or several crystallites.

### 4.1. Liquid-liquid phase separation with the coupled Navier-Stokes-Cahn-Hilliard equations

To check the implementation of the coupling of the Cahn-Hilliard equation with the fluid mechanics equations, and especially of the capillary forces, we first perform a static benchmark test regarding the Laplace pressure: starting from a 2D droplet in an immiscible binary mixture, we let the droplet grow until the equilibrium is reached. We then measure the droplet radius $R$, the pressure drop $\Delta P$ between the inside and the outside of the droplet, and the surface tension $\sigma$. The procedure is repeated for different initial concentrations (and hence different final radii) and $\kappa$ values (and hence different surface tensions) and we verify that the Laplace law $\Delta P = \sigma / R$ is well recovered.

We then perform 2D simulations of spinodal decomposition with the coupled Cahn-Hilliard-Navier-Stokes equations. The investigated system is an incompatible binary symmetric critical mixture (50:50 blend, $N_1 = N_2 = 1$, $\chi_{ll} = 4$, molar masses $1\ kg \cdot mol^{-1}$) with initially homogeneous composition. The mobility $\Lambda_{11}$ is assumed to be constant and equal to the diffusion coefficient, $\Lambda_{11} = D = 10^{-11}\ m^2 \cdot s^{-1}$. The homogeneous, constant viscosity $\eta_{mix}$ is varied between $10^{-2}\ Pa \cdot s$ to $10^4\ Pa \cdot s$ (see Supporting Information S4.1 for more details). The chosen parameters can be thought of as representative of an oil mixture. Demixing takes place spontaneously due to the concentration fluctuations $\zeta_{CH}$ in Equation 9. With typical structure sizes in the range $L = 50 - 500\ nm$, the Reynolds numbers remain significantly smaller than 1 over the whole viscosity range, which is consistent with the assumption that inertial effects can be neglected. The Peclet number $Pe = Lv/D \approx L\sigma/(\eta D)$ ranges approximately from $10^{-2}$ to $10^3$ which means that diffusion fluxes are dominant at high viscosities whereas advection fluxes are dominant at low viscosities. We investigate the coarsening behavior of the phase-separated system depending on the viscosity.

As long as the morphology remains self-similar over time, the coarsening behavior can classically be described by the equation $L(t)^{1/\delta} - {L_0}^{1/\delta} \propto (t - t_0)$ where $L_0$ is the initial size of the separated phases, $t_0$ the time for the onset of demixing and $\delta$ the so-called coarsening exponent. Theoretical works have shown that the coarsening exponent $\delta$ is expected to vary with decreasing viscosity from 1/3 for a purely diffusive behavior to 1 (in 3D) or 1/2 (in 2D) in the viscous regime and 2/3 in the inertial regime, which is not considered here. [124] [125] However,

Lattice Boltzmann [126] [127] as well as coupled Navier-Stokes-Cahn-Hilliard simulations [128] [129] [130] [131] have shown that the coarsening behavior is very complex when both diffusive and advective fluxes are active, with the emergence of distinct length scales, break-down of self-similarity, and no simple coarsening kinetics. In the following, we illustrate this sophisticated behavior with our results, which are fully in-line with the well-established findings of the literature. Note that noise can contribute significantly to the growth mechanism and impact the simulated morphologies,[126] [131] but the level of noise used in the calculations presented below have been kept sufficiently low so that this effect can be neglected.

In order to characterize the coarsening behavior, we calculate the 2D structure factor of the volume fraction field, the probability distribution $p(q,t)$ of wave numbers $q$ by angular integration and define a characteristic length scale as the inverse of the mean $q$ value, $L(t) = 2\pi/\int qp(q,t)dq$. Note that other length scale indicators can be chosen, which behave quite differently when self-similarity is not respected, as has been highlighted by Wagner. [126] The result of this procedure is shown in Figure 3. For the highest viscosities, the advection fluxes are negligible and the coarsening exponent is equal to 1/3 as expected. With decreasing viscosities, the coarsening kinetics becomes faster and the apparent coarsening exponent increases. However, for the lowest viscosities (below approximately 1 $Pa \cdot s$), the coarsening exponent exceeds significantly the theoretically expected ½-value at short times and decreases strongly with time. [130] This is due to the break-down of self-similarity, and in such cases the system cannot be described by a single length scale. As highlighted by Fan and Camley, at intermediate viscosity values ($3 - 30\ Pa \cdot s$ in our case, blue and green curves in Figure 3), an 'apparent' coarsening exponent $\delta = 0.5$ can be observed, even if the morphology evolution is already not self-similar anymore. [130] [131]

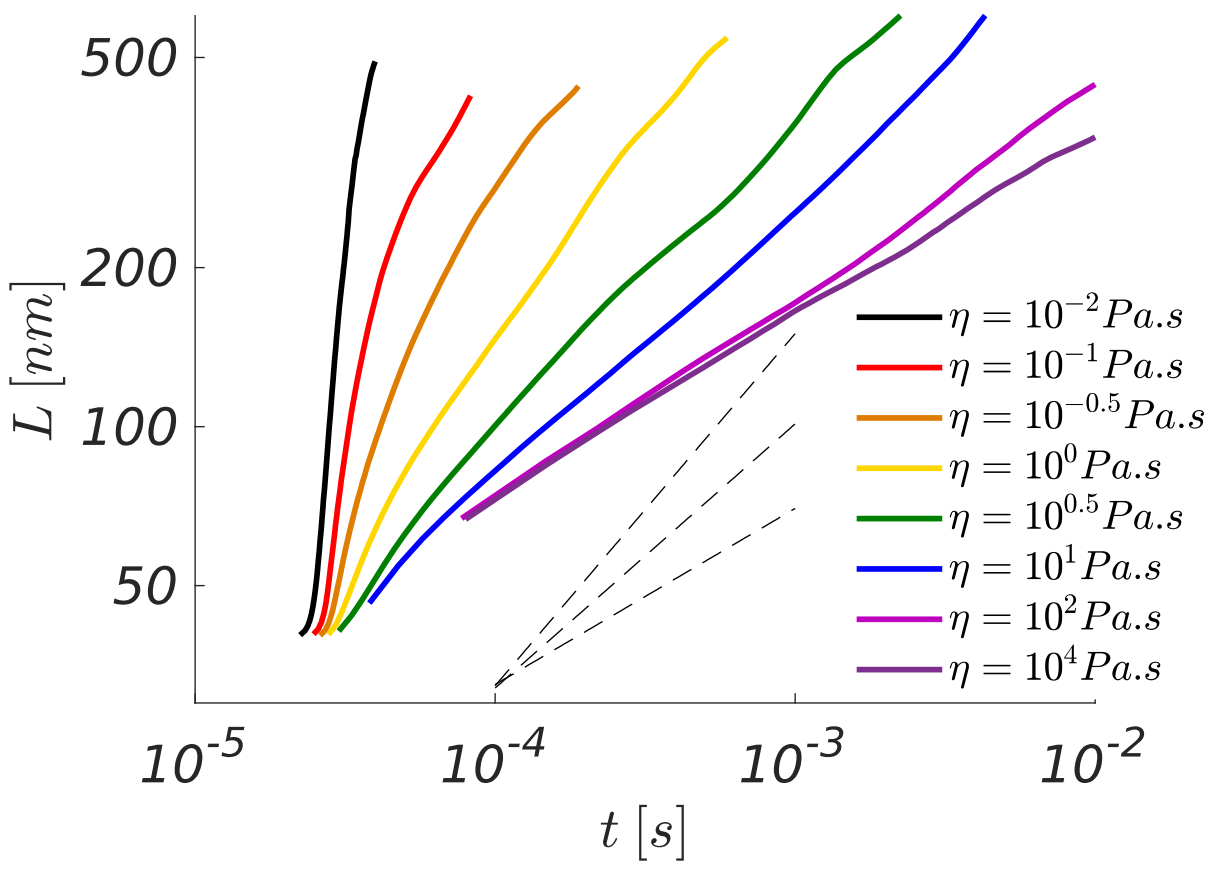

*Figure 3: characteristic length scale L(t) depending on time for various viscosities. The dashed lines show 1/3, 1/2 and 2/3 asymptotic behaviours. The box size is 2048 nm x 2048 nm .*

The evolution of the morphology for different viscosities is illustrated in Figure 4. The structure is perfectly self-similar at high viscosities ($\eta_{mix} = 10^4\ Pa \cdot s$, Figure 4a-c), with a co-continuous structure of both phases with few droplet-like inclusions and well-defined thin interfaces. For this condition, advection is negligible. This corresponds to the classical behavior obtained with the Cahn-Hilliard(-Cook) equation for a symmetric blend. However, with decreasing viscosity, the advection plays an increasing role which leads to deviations from the purely diffusive behavior. For $\eta_{mix} = 1\ Pa \cdot s$, the break-down of self-similarity can be clearly observed from the presence of smaller droplets in the larger structures, leading to a variety of length scales (Figure 4d-f). There is no clear scaling behavior when this effect occurs.[127] [130] There, in addition to the diffusion process, the hydrodynamic flow promotes the reduction of interface length and therefore the coarsening, at least until the domains are nearly circular. The circular shape is obtained faster for smaller domains. Then, the coarsening of these small domains is not assisted by the hydrodynamic flow anymore and they coarsen slower than the large domains, which leads to a morphology with small spherical inclusions in the large domains. [126] Still, these droplets occasionally merge through the diffusion-enhanced collision mechanism (Figure 4f). [126] [131] At even lower viscosity, the diffusive fluxes, which are responsible for phase separation, are not fast enough to promote and/or maintain the compositional equilibrium (Figure 4g-i). The concentration gradients at the interfaces are smoother at the beginning of the LLPS, and the volume fractions in the separated phases hardly reach the equilibrium values with time. This leads to a secondary phase separation process, initially identified by Tanaka, that takes place inside the phase separated domain, because the concentrations there are still in the unstable domain of the phase diagram. [128] [129]

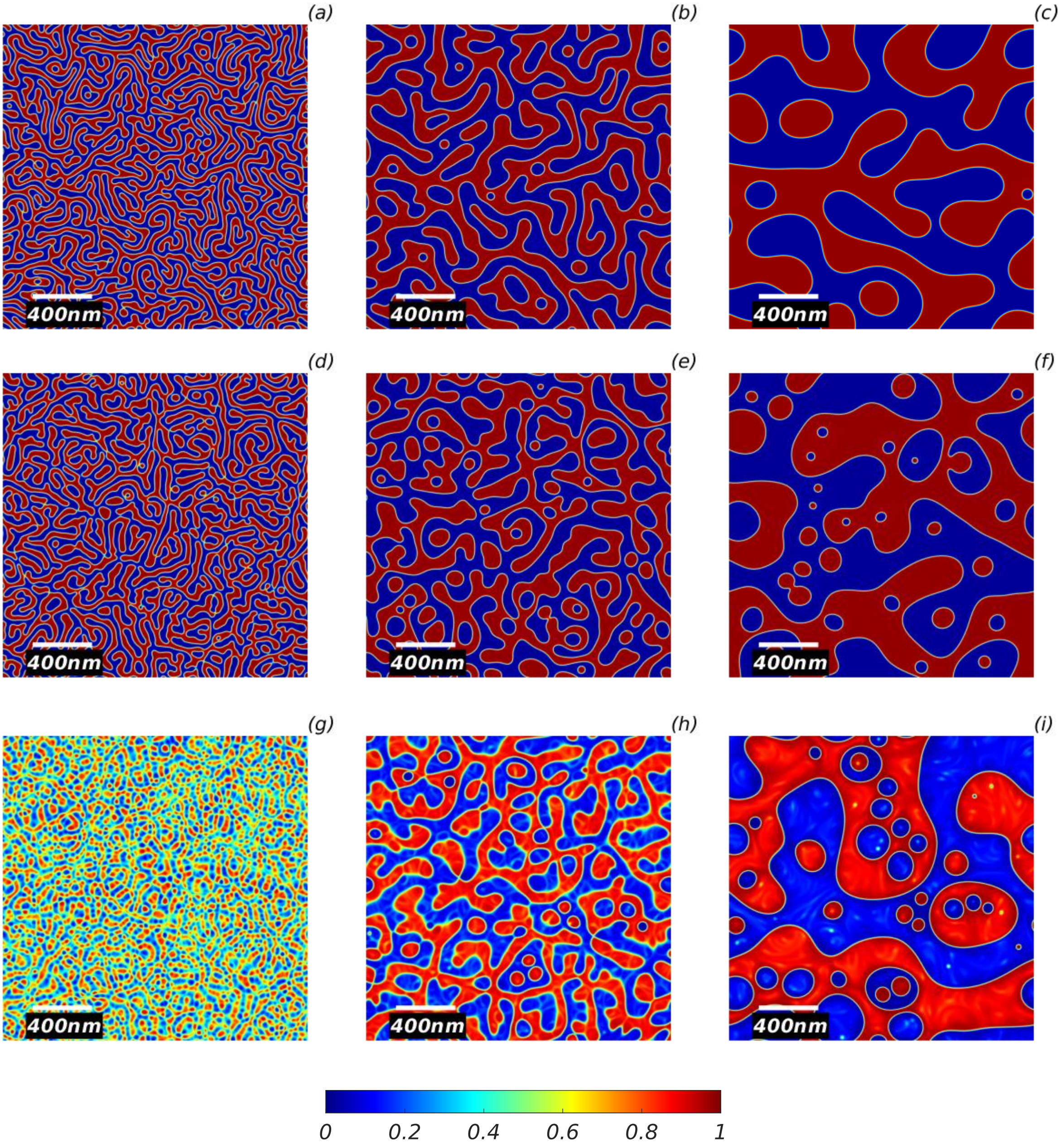

*Figure 4: snapshots showing the volume fraction of the first fluid, at different stages of the coarsening, for $\eta_{mix} = 10^4, 1, 10^{-2}\ Pa \cdot s$ (from top to bottom). The snapshots correspond to a measured characteristic length L(t) of 70, 150, 350 nm (from left to right). The smallest droplets (R < 15 nm) observed in the last snapshot (i) for $\eta_{mix} = 10^{-2}\ Pa \cdot s$ are formed by secondary phase separation.*

## 4.2. Coupling of the stochastic Allen-Cahn and Cahn-Hilliard equation: crystallization in a blend

***Nucleation, growth and coarsening of a pure material***

First, we provide some insights in the general behavior of the stochastic Allen-Cahn equation for a pure material (Figure 5). The parameters of the simulations can be found in Supporting Information S4.2. Thanks to the fluctuations, the order parameter can overcome locally the energy barrier for nucleation (see first term of the RHS in Equation 3), which leads to the formation of a crystal, defined as an area with an order parameter above $t_{\phi,k}$, and which is given a random, uniform marker value $\theta$ . The order parameter quickly increases to $\xi_{0,k}$.With increasing time, several nuclei appear, grow and impinge, which results in an increasing overall crystallinity as shown in Figure 5.

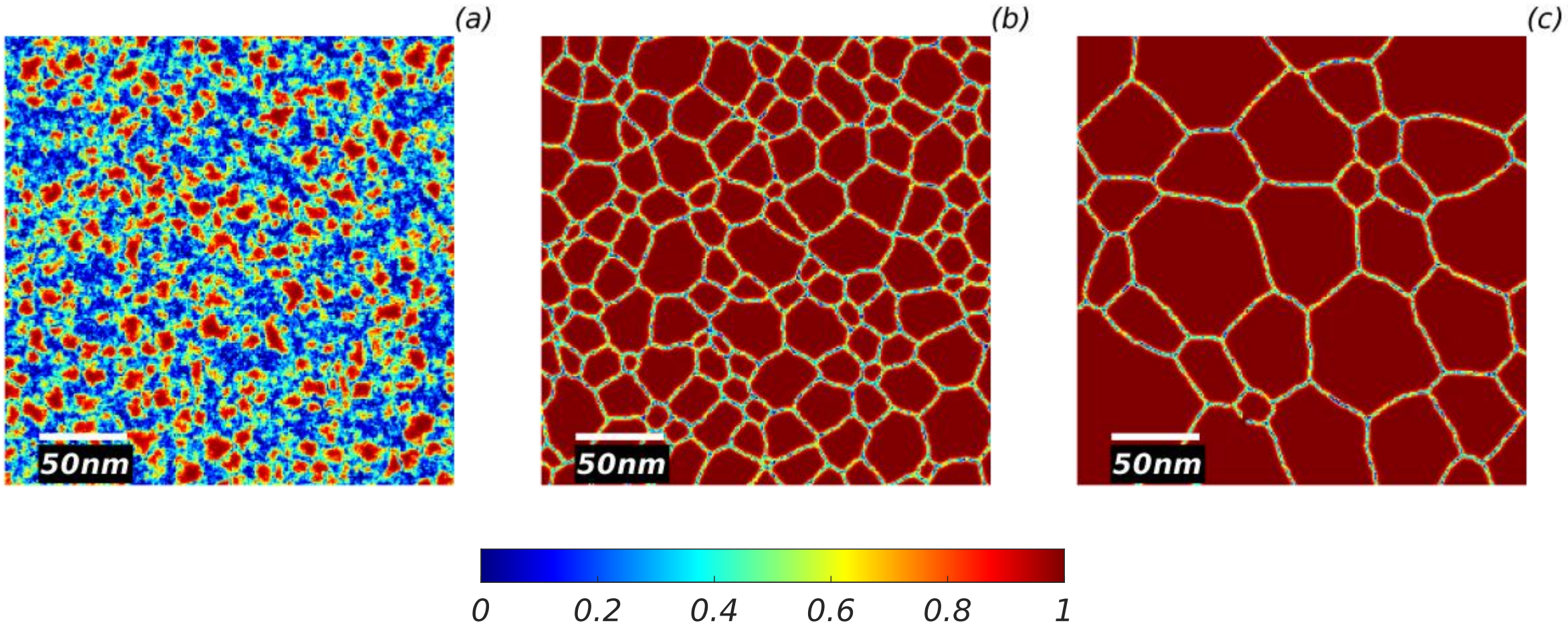

*Figure 5: order parameter field at different stages of the crystallization process for $W_{fus} = 7.5 \cdot 10^4\ J \cdot kg^{-1}$ and $\varepsilon = 10^{-5}\ (J \cdot m^{-1})^{0.5}$. (a) During nucleation for a crystallinity of 50%, (b) fully crystalline system at the beginning of the coarsening and (c) later stage during coarsening. Shown is a 256 nm x 256 nm area of the 512 nm x 512 nm simulation box.*

The time-dependent crystallinity $\chi(t)$ can be described by the Johnson-Mehl-Amravi equation $\chi(t) = \chi_{max}\left(1 - e^{-K(t-t_i)^n}\right)$, where $\chi_{max}$ is the maximum crystallinity and $t_i$ the incubation time for the onset of nucleation. In 2D, the exponent is expected to be $n = 2$ for pure growth, while $n = 3$ is expected for if homogeneous nucleation and growth occur at the same time. This behavior is recovered in the stochastic Allen-Cahn model. [132] Thereby, $\chi_{max}$ is assumed to be the crystallinity reached as soon as the system is fully covered by crystallites and grain boundaries: at this point, there is no amorphous material anymore. This is illustrated for some sets of parameters in Figure 6a. Once the system is fully crystallized at $t = t_0$, the crystallites coarsen with a power law $R(t)^2 - {R_0}^2 \propto (t - t_0)$, as expected for purely surface directed growth (Figure 6b).[133] [134] [135]

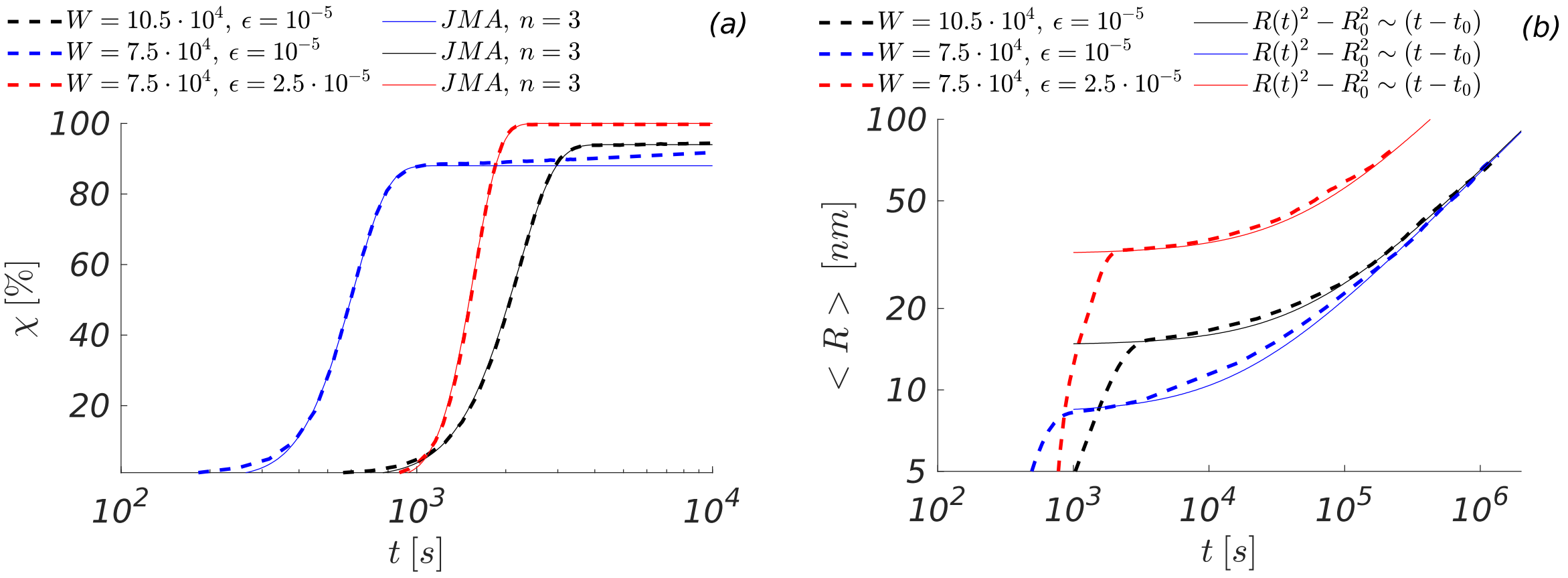

*Figure 6: Evolution of (a) the proportion of crystalline material and (b) average radius of the crystallites for various model parameters, as well as fits with the Johnson-Mehl-Avrami equation and quadratic grain growth, respectively. $W_{fus}$ is in $J \cdot kg^{-1}$ and $\varepsilon$ in $(J \cdot m^{-1})^{0.5}$. The full parameter sets can be found in Supporting Information S4.2*

***Nucleation, growth and coarsening in a mixture***

The situation in a blend is more complicated, because the nucleation and growth rates are strongly composition-dependent. It is not in the scope of this paper to extensively describe how the nucleation and growth rates depend on all the thermodynamic and kinetic properties of the system. For a general understanding, it is here sufficient to keep in mind that the nucleation rate of a material $k$ can be written as the product of three terms:

$$\frac{1}{t_{nucl}} \propto M_k(\{\varphi_k\}) A(H_k) e^{-\frac{\Delta G^*}{RT}} \qquad (30)$$

In the equation above, $t_{nucl}$ is the mean formation time of a nucleus, $H_k$ is the height of the energy barrier for the liquid-solid transformation (see first term of the RHS in Equation 3) and $\Delta G^*$ is the energy of a critical nucleus. The product of both first terms gives the frequency at which a local fluctuation of the order parameter may overcome the energy barrier upon crystallization. The first one is a purely kinetic factor related to the mobility of the atoms/molecules in the mixture and is assumed here to be proportional to the Allen-Cahn mobility (and therefore to the self-diffusion coefficient). The second one is related to the probability of a fluctuation overcoming the energy barrier and depends only on the height of the barrier and thus on the thermodynamic properties of the blend. The last term is a purely thermodynamic factor. This is the energy barrier to be overcome for the formation of a stable nucleus, $\Delta G^*$ being the energy of the critical germ for which the energy gain upon crystallization balances the surface energy. In a mixture containing one crystalline material (which we will call 'solute' here because we focus on crystallization in a solution), the last two terms contribute to a considerable decrease (orders of magnitudes) of the nucleation rate with decreasing solute concentration. In a solution, since the mobility gets significantly higher upon dilution, this is balanced by the fact that the first term strongly increases with decreasing solute concentration. As a consequence, and depending on the relative weight of these factors, the nucleation rate in a solution might typically have a maximum at intermediate concentrations. The same holds for the crystal growth rate, the effect of increasing mobilities upon dilution being balanced by the decreasing thermodynamic driving force for phase change. This results in the typical behavior shown in Figure 7 depicting the crystallization kinetics of a polymer in solution (the parameters of the simulation can be found in Supporting Information S4.3) for different polymer volume fractions $\varphi_0$. The crystallization properties of the polymer are the same as for the blue curves in Figure 6. The diffusion coefficient of the polymer and therefore the Allen-Cahn mobility is assumed to vary over five orders of magnitude from the pure polymer to infinite dilution. This is the dominant effect for concentrations ranging from $\varphi_0 = 0.9$ to roughly $\varphi_0 = 0.2$, so that the nucleation rate increases with decreasing polymer volume fractions. For $\varphi_0 \lesssim 0.15$, the two thermodynamic contributions become dominant, and the nucleation rate abruptly drops. No nucleation can be observed any more during the simulated time.

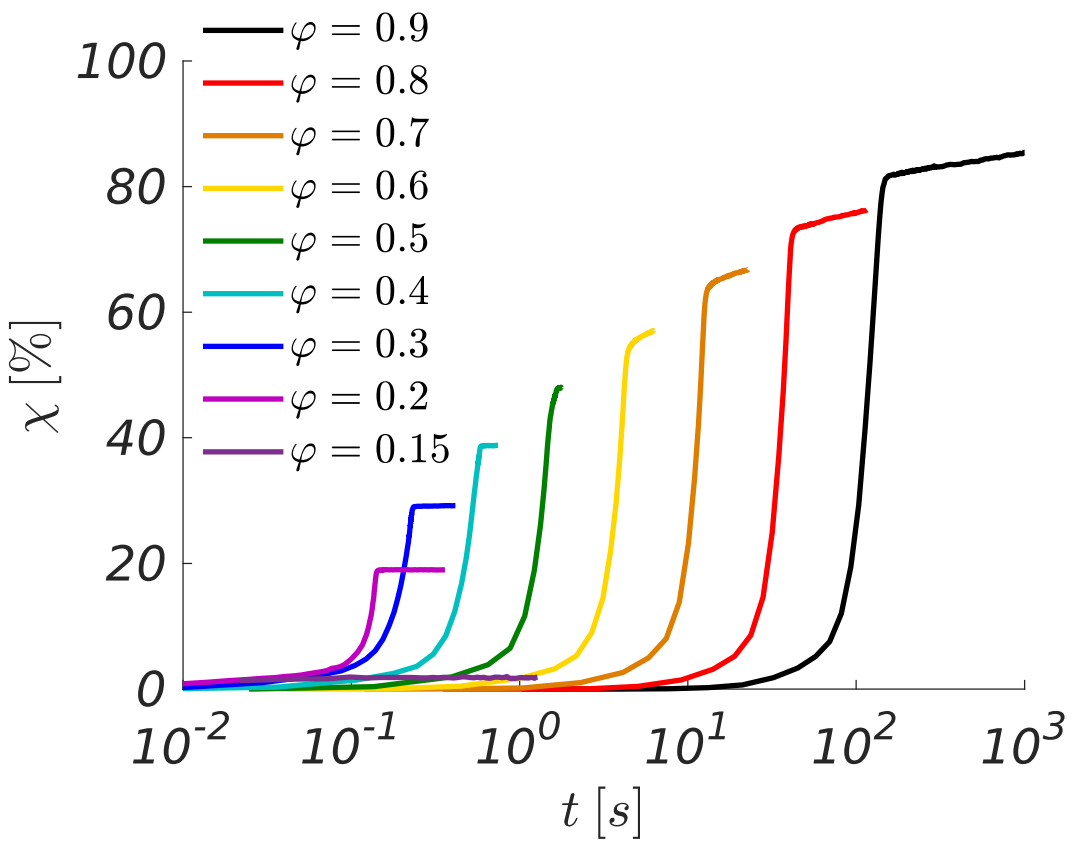

*Figure 7: evolution of the proportion of crystalline material in a solute-solvent blend with time, for various initial solute volume fractions $\varphi$. The crystallinity is defined as the total volume of crystalline solute over the whole system volume.*

***Nucleus build-up***

From a physical point of view, the nucleus can be seen as a solid body with a sharp interface, namely an infinitely steep composition jump between the mixture outside and the solid inside the nucleus. Therefore, in the simulation, it is desired that once the stable nucleus starts to form, the order parameter and volume fraction field inside the crystal quickly reach their equilibrium values. In our phase field framework, however, the formation of a nucleus is not instantaneous. This is in most cases no problem since the crystal build-up is in general very fast. Nevertheless, in some 'pathological' situations where the concentration dependence of the Allen-Cahn mobility is strong, together with a weak driving force for crystallization, the increase of the order parameter and of the volume fraction field in the emerging nucleus might get slow as compared to the lateral growth of the crystal. This is typically encountered when polymer crystallites nucleate in dilute solutions. Here, the Allen-Cahn mobilities strongly decrease in the diffuse interface of the forming nucleus, from the outside to the inside. This leads to the unacceptable situation that the crystal composition can be significantly different from the expected equilibrium value. To overcome this drawback of our diffuse interface model, we first evaluate the composition and therefore the Allen-Cahn mobility in the environment directly around each emerging nucleus. Then, this Allen-Cahn mobility calculated from the environment is used inside the nucleus, whatever the composition, at every grid point where the crystal forms, namely $\frac{\delta \Delta G_V}{\delta \phi_k} < 0$. The nucleus formation following this procedure is shown in Figure 8, corresponding to the volume fraction $\varphi_0 = 0.3$ of Figure 7. The order parameter in the emerging nuclei quickly reaches its maximum value and the volume fraction reaches a value (0.8) which is higher than the solid concentration ($\varphi_s = 0.78$ in this case) because the liquid part is still very concentrated (Figure 8a). The order parameter and volume fraction fields inside the crystals remain homogeneous with further growth (Figure 8b). Upon further crystallization, the polymer volume fraction in the liquid phase decreases, the smallest nuclei that might thus not be stable any more disappear. This, together with a classical coarsening process, leads to the growth of the largest crystals. The volume fraction in the crystals at the end of the simulation (0.75) is a little lower than the solid volume fraction due to diffusional limitations in the largest crystals, but it is still very close to it. For comparison, the mean volume fraction in the crystals without application of the correction described above is strongly inhomogeneous and still remains below 0.65 for identical simulation times. Note that the order parameter field is much noisier than the volume fraction field. This is because of the location $\phi_b$ of the energy barrier for solid-liquid phase transformation on the order parameter axis (see first term on the RHS of Equation 3). Order parameter fluctuations in the range of $\phi_b$ (typically $\phi_b = 0.1 \dots 0.3$.) are necessary in order to reach order parameter values with $\frac{\partial \Delta G_V}{\partial \phi_k} < 0$ and thus a driving force for crystal nucleation. Turning to the volume fraction field, for the onset of LLPS, fluctuations that are order of magnitudes smaller are actually sufficient. Moreover, in the miscible blend considered here they tend to be smeared out anyway.

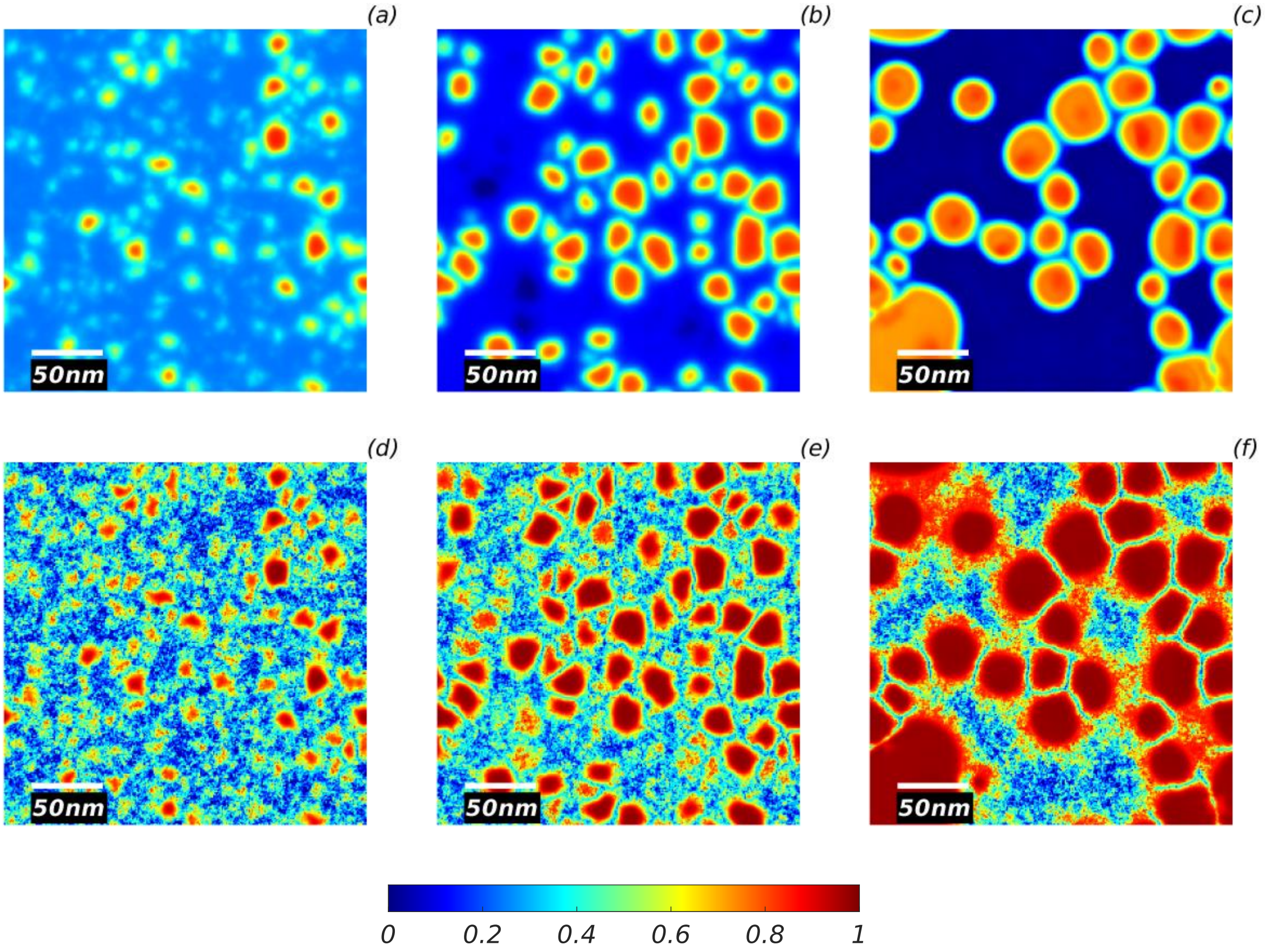

*Figure 8: solute volume fraction field (top row) and order parameter field (bottom row) at different stages of the crystallization process for the initial solute volume fraction $\varphi = 0.3$. (a,d) During nucleation for a crystallinity of 15% (half of the solute material is crystalline), (b,e) system with nearly fully crystallized solute at the beginning of the coarsening and (c) later stage during coarsening. The size of the simulation box is 256 nm x 256 nm.*

## 4.3. Advection and stability of crystals in a drying film

***Advection in a drying film***

We now investigate the coupling of the full phase-field model, including crystallization and evaporation, to the solver for the dynamics of the fluids. First, we focus on a test case regarding the advection of a single crystal in a drying film. The simulation setup is as follows (see Figure 9a-b): we consider a polymer solution with 30% polymer concentration and a small crystal sitting initially close to the top of the film. The crystal has a diameter of 15 grid points and the liquid-solid interface thickness is 8 grid points. The solvent evaporates and the film is therefore drying. The parameters for the polymer solution and the crystallization properties are the same as in the previous section, except that the Allen-Cahn mobility of the crystalline polymer is set to zero so that the crystallization process is inactive during the evaporation (the full set of simulation parameters can be found in Supporting Information S4.4). In such a situation, we expect the film surface to come in contact with the crystal and push it downwards, provided the capillary forces between the surface and the crystal are strong enough to compensate for the viscous forces arising from the crystal's displacement. The final result of the simulation is shown in Figure 9c-d. As desired, the crystal is advected vertically, without surface area modification and with very limited deformation of the order parameter field. This shows that the advection works properly, the MUSCL scheme together with the Superbee flux limiter ensuring that the crystal's shape is almost conserved, in particular preventing numerical diffusion.

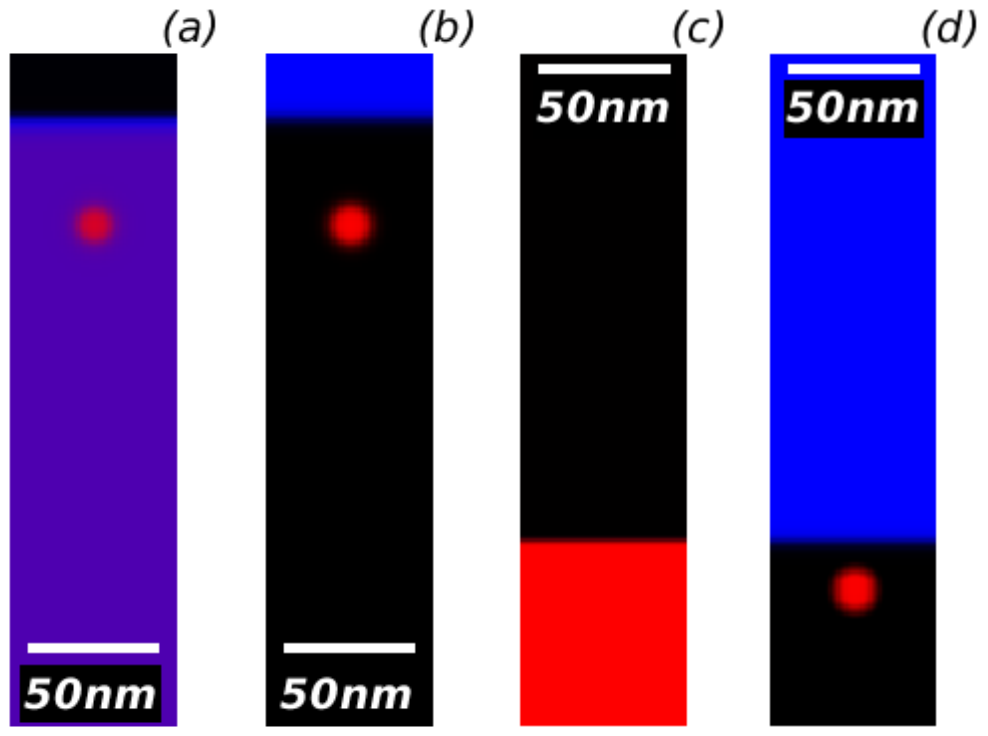

*Figure 9: simulation of a drying polymer solution with a single initial crystal. Volume fractions and order parameters at the beginning (respectively (a) and (b)) and at the end (respectively (c) and (d)) of the simulation. For the volume fraction fields, the polymer is represented in red and the solvent in blue. For the order parameter fields, the polymer crystal is represented in red and the vapor in blue. The system size is 256 nm x 64 nm.*

Second, we perform the same simulation but with three crystals initially present in the wet film (Figure 10). The upper crystals is pushed downwards as soon as it touches the film surface. The second crystal also reaches the film surface and is then pushed downwards. Due to hydrodynamic interactions, the three crystals finally stick to one another (see the final state in Figure 10c-d). Here again, we could check that, despite these complex displacements and the agglomeration process, the surface area of the crystals is conserved and the evolution of the interface profiles is very limited. To conclude these advection test cases, we point out a limitation of our framework concerning the viscosity. The viscosity values used in our simulations are unrealistically high, at least for the regime of very dilute solutions at the beginning of the drying. Indeed, using realistic values would lead to high velocities in the simulation box, and thus very small admissible time steps due to the Courant-Friederich-Levy condition for advection. Simulating realistic evaporation time would require a huge amount of time steps and thereafter inaccessible calculation times. The associated underestimation of the role of advection in the simulations of dilute solutions has to be kept in mind.

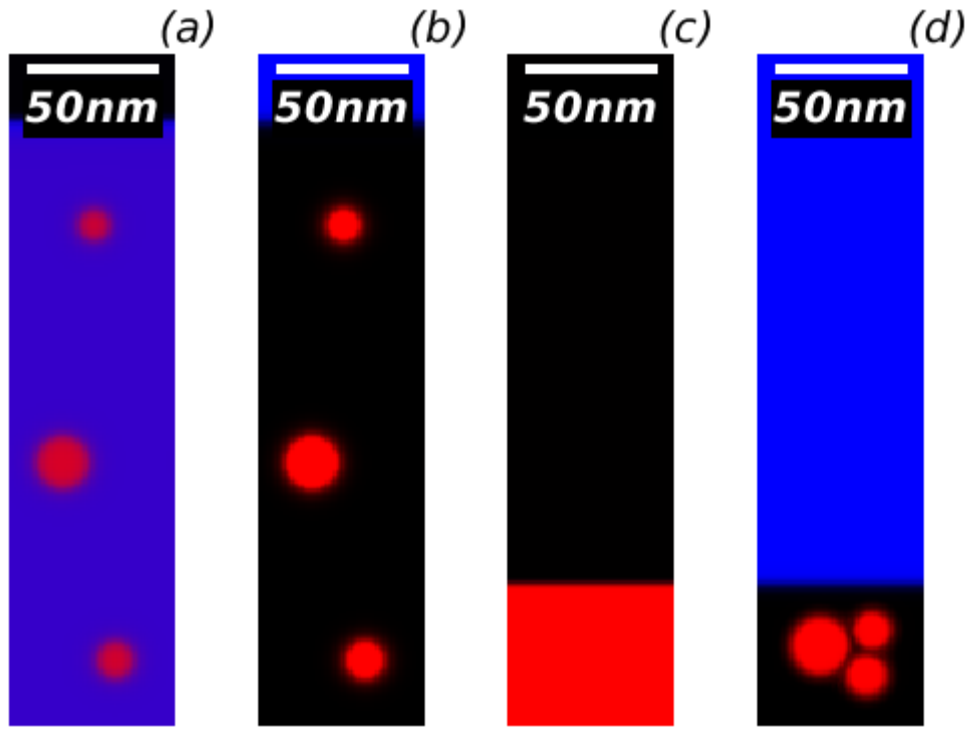

*Figure 10: simulation of a drying polymer solution with three initial crystals. Volume fractions and order parameters at the beginning (respectively (a) and (b)) and at the end (respectively (c) and (d)) of the simulation. For the volume fraction fields, the polymer is represented in red and the solvent in blue. For the order parameter fields, the polymer crystal is represented in red and the vapor in blue. The system size is 256 nm x 64 nm.*

***Stability of the crystals at the film surface***

As already explained in Section 2.4, the stability of a crystalline structure touching the film surface is a critical issue. This is a consequence of the diffuse nature of the interfaces in the phase-field framework, which leads, at the condensed-gas phase interface, to the overlap of the air, solvent and solute fraction fields on the one side, and to the overlap of the crystalline and vapor order parameters on the other side. Since a substantial overlap of the phase fields is in fact energetically very unfavorable, this leads to an evolution of the phase fields in this region. If the kinetic properties are such that the diffusion processes and the crystallization/dissolution processes are fast in the solid-gas interface, the order parameter and volume fraction fields of the crystalline material are reduced in the interface, allowing the vapor phase of the drying film to progress, so that the crystal finally disappear. Such a fast kinetic is typically encountered when investigating crystallization processes upon drying in dilute solution, and crystals reaching the surface may dissolve during the drying. Moreover, even in the case of well-formed crystals in highly concentrated films, the kinetics within the diffuse solid-vapor interface is generally fast in the outer region of the crystal, because the solute volume fraction is relatively low. This means that even well-formed crystals or dry structures are in principle hardly stable. As a consequence, handling crystal nucleation together with ensuring crystal stability in simulations of drying films is a complicated issue. To overcome this unphysical effect, we have introduced in Section 2.4 an interaction energy between crystals and vapor phases, as well as a penalty function for the crystallization kinetics. In the following, we illustrate how this allows for stability of crystalline structures at the interface. The simulation setup is shown in Figure 11. We calculate the evolution of a mixture initially composed of 30% crystalline polymer, 50% solvent and 20% additional amorphous small molecule solute. The parameters for the polymer and the solvent are the same as in the previous section. The full parameter sets can be found in Supporting Information S4.5. Before starting the evaporation, we let a columnar like crystalline structure grow in the film. Then, we let the film dry and allow for a fast Allen-Cahn mobility, which means that the crystal has time to evolve within the time required for evaporation, and we then investigate whether the columnar structure is stable.

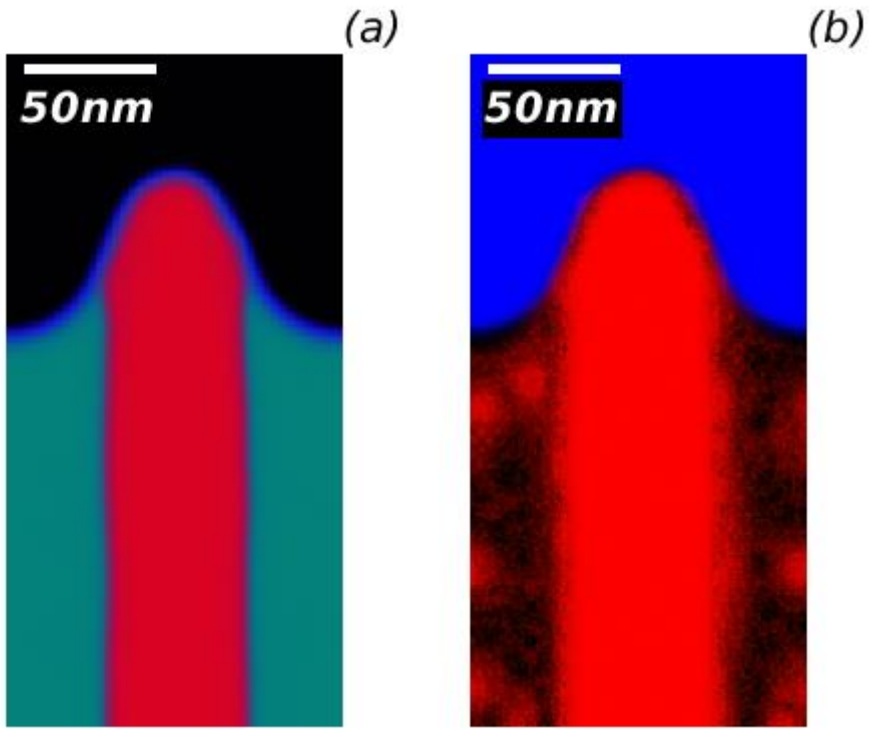

*Figure 11: drying of a ternary polymer-small molecule-solvent blend, with an initial columnar polymer crystal. The volume fractions (a) and order parameters fields (b) during the simulation, after impingement of the vapour and the solid, are shown. For the volume fraction fields, the polymer is represented in red, the small molecule in green and the solvent in blue. For the order parameter fields, the polymer crystal is represented in red and the vapor in blue. The system size is 256 nm x 128 nm.*

Figure 12 shows the position of the top of the columnar crystal depending on time for various simulation parameters and illustrates how the instability can be prevented. In a first step, we perform simulations only with the phase field model (dashed lines). As detailed above, without solid-vapor interaction, as soon as the crystal is in contact with the film surface ($t \approx 0.2\ s$), it is not stable, even during the evaporation time (roughly 10s) and its maximal height finally drops down to 120 nm, which is actually the height of the final flat film (red curve). Introducing a solid-vapor interaction energy greatly increases the stability (orange curve), but the columnar structure still evolves quickly as compared to the drying time. Only by introducing also a kinetic penalty to the evolution of the crystallinity at the solid-vapor interface, the crystal can be made perfectly stable (yellow curve).

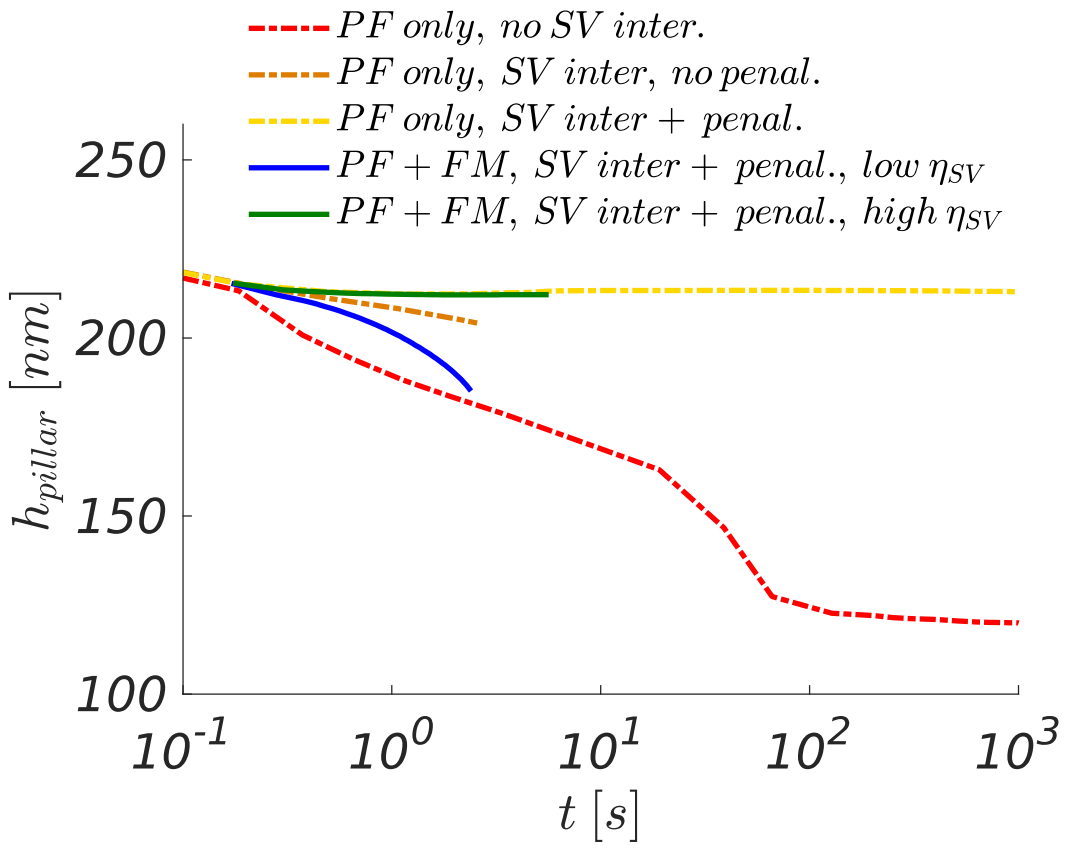

*Figure 12: time-dependent position of the top of the columnar crystalline structure shown in Figure 11 for different simulation setups. (red) phase field model only, no SV interaction energy (orange) phase field model only with SV interaction energy, (yellow) phase field model only with SV interaction energy and penalty to the Allen-Cahn mobility in the SV interface, (blue) phase field coupled to fluid mechanics, low viscosity in the SV interface, (green) phase field coupled to fluid mechanics, high viscosity in the SV interface*

In a second step, we couple the phase-field model (including solid-vapor interaction and kinetic penalty in the SV interface) to the mass and momentum conservation equations. The columnar structure turns out to be instable (blue curve), which highlights a second potential instability mechanism: again due to the diffuse nature of the SV interface, the viscosity values at the very top of the crystal are low, so that the velocity field is non zero. Advection takes place at the crystal border with a significant negative vertical component, which leads to the disappearance of the crystalline region at the surface. This can happen even if the crystal is set on the substrate and even if the viscosity inside the crystal is very high and therefore the velocities in the bulk negligible. To prevent this, the grid points in the SV interface corresponding to the whole crystal diffuse interface need to have a high viscosity. In practice, we set the centering $c_\eta$ of the viscosity penalty function inside the crystals (see Equation 22) in such a way that the viscosity drops over orders of magnitude as soon as the crystal detection limit is passed. In addition, the viscosity of the grid points in the SV interface around the crystal (2-4 points beyond the area where the marker value is defined) is also set to the bulk value. Using this procedure, the stability of the crystal with the full model can be ensured (green curve), which is crucial for applications where the roughness of the dry film has to be investigated. This is for example the case for solution-processed perovskite photoactive layers for solar cells or other optoelectronic applications. Finally, note that with the noise term in our phase-field equations, the phase fields, and especially the crystalline order parameter fields, can be very noisy. The same holds for the viscosity field, whereby abrupt viscosity changes can be found in the crystal interfaces. In order to be able to handle these fields numerically, they are smoothed with two successive Gaussian filters before calculating the capillary forces and the velocity fields. The calculated velocity fields are therefore only approximate solutions of the problem. Fortunately, this does not affect the global physics of the structuring film: in fact, the simulations of crystal advection presented above have been performed with this filtering, and it has been shown that the expected behavior can be recovered.

To conclude this section, two important points related to the stability need to be pointed out. First, remember that the phase field equations lead to the minimization of the energy of the system. This means that the interface area has to be reduced, and that the thermodynamically stable final structure in the example above is a flat dry film anyway. Nevertheless, the kinetics of this flattening is virtually infinitely slow due to the vanishing transport properties of the crystal. In our simulations, we make sure that the crystals are stable over time scales at least comparable with the drying time. Second, the instability of floating crystals at the surface of the drying film can be prevented with the same procedure. Note that the instability mechanism discussed here has nothing to do with the thermodynamic instability of a small germ ($r<r^*$) or with the suppression of the smallest crystals due to coarsening. Indeed, this is an instability mechanism due to the diffuse nature of the interfaces. The stability of emerging nuclei floating at the surface strongly depends on the SV interaction energy and of the penalty to the Allen-Cahn mobility, notably whether the phase fields inside the crystals is such that $\varphi_k \phi_k > c_{sv}$. In other words, nuclei with $\varphi_k \phi_k < c_{sv}$ are unstable and disappear at the film surface while nuclei with $\varphi_k \phi_k > c_{sv}$ are stable and tend to gather at the film surface. This leads to the fact that the overall vertical location of the emerging crystals in the drying film not only depend on the physics, but also on purely numerical, somewhat arbitrary parameters. Therefore, conclusions on the vertical position of crystalline structures have to be handled very cautiously.

## 5. Simulations of film structuration upon drying

In this section, we present simulations performed with the full model for the case of a drying thin film. The mixture is a ternary polymer-small molecule-solvent blend, initially perfectly mixed with either 13:20:67 or 20:13:67 volume fraction ratio (polymer-small molecule blend ratio 40:60 or 60:40). Both polymer and small molecule materials are crystalline and are immiscible. We investigate the evaporation-induced morphology formation until the film is dry. In such a sophisticated mixture, many morphology formation pathways are in principle possible, depending on the material properties and on the process parameters. It is not the topic of this paper to systematically investigate and understand which material properties and processing conditions lead to which morphologies. Instead, we simply illustrate a couple of different morphology formation processes and dry structures, but still in the 'most complex situation' where LLPS as well as crystallization of both materials can occur. The objective is to demonstrate that our numerical method can handle such cases. Again, the parameters are the same as compared to the previous section, except that the blend ratio, the Allen-Cahn mobility (and thus the crystallization kinetics) of both polymer and small molecule, and the polymer-small-molecule Flory-Huggins interaction parameter are varied. The full parameter sets can be found in Supporting Information S4.6. The simulation box is 512 nm x 256 nm and the initial film height is 450 nm.

Figure 13 shows the morphology formation upon drying for a highly incompatible 40:60 polymer-small molecule blend. In this example, a liquid-liquid phase separation occurs first as the ternary mixture reaches the unstable region of the phase diagram (Figure 13a,e). Then, polymer crystals form in the polymer majority phase (Figure 13b,f). Upon further drying, both polymer and small molecule phases become more pure and the crystallization process progresses, with purification of the crystals and coarsening (Figure 13c-d,f-h). The crystals at the film surface are pushed downwards (Figure 13b-d,e-h). After 3s of drying, the film surface hits a hardly deformable crystal structure connected to the substrate and starts to bend (Figure 13c,g). With further drying, the whole solvent finally evaporates, some re-organization of the crystalline morphology occurs due to coarsening and residual advection, but the dry structure remains rough (Figure 13d,h). The crystallization of the small molecule material is very limited here during drying because the critical concentration for nucleation is high (for 40% volume fraction) and at such a concentration the crystallization kinetics is slow. The dry structure can be seen as a quasi 'two phase system' with a purely crystalline polymer phase and a purely amorphous small molecule phase. Nevertheless, it is important to keep in mind that the morphology at the end of the drying is not at thermodynamic equilibrium. Therefore, it still evolves with further crystallization of both materials. However, this happens at much longer time scales, because in the solvent-free system all kinetic properties are slower. In this case, very limited changes have been observed between 10s and 30s simulation time.

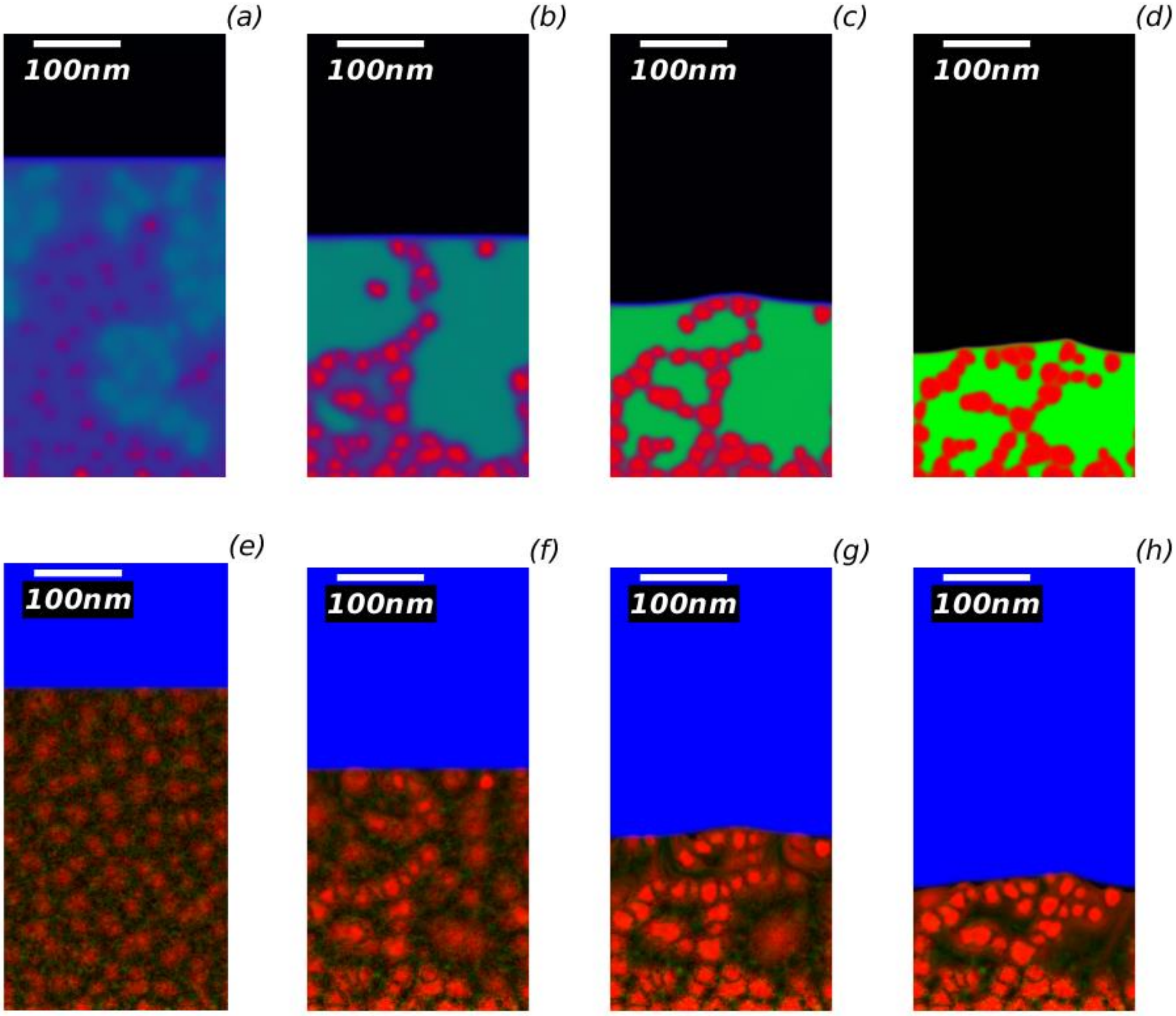

*Figure 13: drying of a ternary polymer-small molecule-solvent blend. The initial blend ratio is 13:20:67, the Allen-Cahn mobilities for the polymer and the small molecule are respectively $M_{p,0} = 1.5 \cdot 10^{-5}\ s^{-1}$ and $M_{sm,0} = 8 \cdot 10^{-5}\ s^{-1}$, and the polymer-small molecule Flory-Huggins interaction parameter $\chi_{ll,psm} = 2$. The volume fractions (top row, (a-d)) and order parameter fields (bottom row, (e-h)) are shown after 1s, 2s, 3s, 10s of drying (from left to right). The film is completely dry after 7s (Figure 14). For the volume fraction fields, the polymer is represented in red, the small molecule in green and the solvent in blue. For the order parameter fields, the polymer crystals are represented in red, the small molecule crystals in green and the vapor in blue. The system size is 512 nm x 256 nm.*

By only varying slightly the blend ratio, the kinetic or thermodynamic properties of the mixture, very different formation pathways and morphologies can be obtained. The time-dependent film height for all the simulations in this section are shown in Figure 14 and the final morphologies in Figure 15.

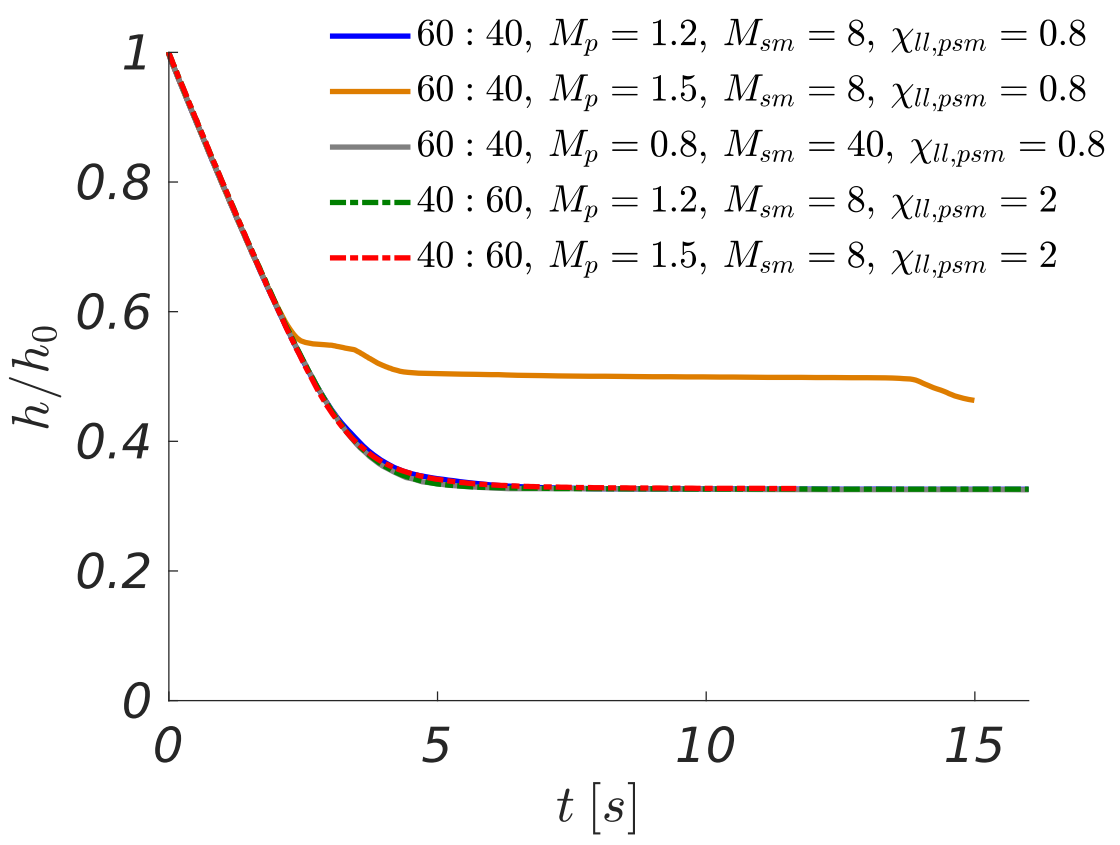

*Figure 14: normalized time-dependent film height the simulations presented in Figure 13 and Figure 15. h is the mean film height and $h_0$ the initial film height.*

In the case of an increased miscibility of the polymer and the small molecule materials, the crystallization process starts before the spinodal decomposition and is responsible for the phase separation (Figure 15a-c and e-g). A slow crystallization process for both polymer and small molecule materials (Figure 15a,e) leads to a quasi '3 phase structure' with an amorphous, impure small molecule phase, an impure amorphous polymer phase and pure polymer crystals. A slightly faster crystallization process for the polymer (Figure 15b,d) leads to a quasi '2 phase structure' with a purely amorphous small molecule phase, and pure polymer crystals, like in the example above. However, polymer crystals gather at the surface and solvent is trapped below this crystalline layer, so that the evaporation process nearly stops, even if it is not completely blocked (Figure 14). The significant difference between both simulations can be explained by a tipping point in the competition between the crystallization and the drying process. The solvent removal contributes to the increase of solute volume fraction in the amorphous phase, whereas the crystals take over materials from the amorphous phase. If the crystallization is too slow (Figure 15a,e), the mean polymer concentration in the amorphous phase increases, which in turn slows down the crystallization (dominant kinetic factor in Equation 30). If the crystallization is faster (Figure 15b,f), the mean polymer concentration in the amorphous phase decreases, which in turn dramatically accelerates the crystallization. The polymer volume fraction then quickly reaches the liquidus value. A slow crystallization process for the polymer and a fast crystallization process for the small molecule materials (Figure 15c,g) leads to a quasi '4 phase structure' with impure amorphous polymer and small molecule phases, and pure polymer and small molecule crystals. Finally, going back to a highly immiscible 40:60 system, but with slower polymer crystallization as compared to the case of Figure 13, we can obtain again a quasi '3 phase structure' with an amorphous small molecule phase, an impure amorphous polymer phase and pure polymer crystals (Figure 15d,h). In comparison to Figure 15a,e, however, the small molecule phase is significantly more pure. This is because the phase separation process is triggered by spinodal decomposition before the onset of crystallization.

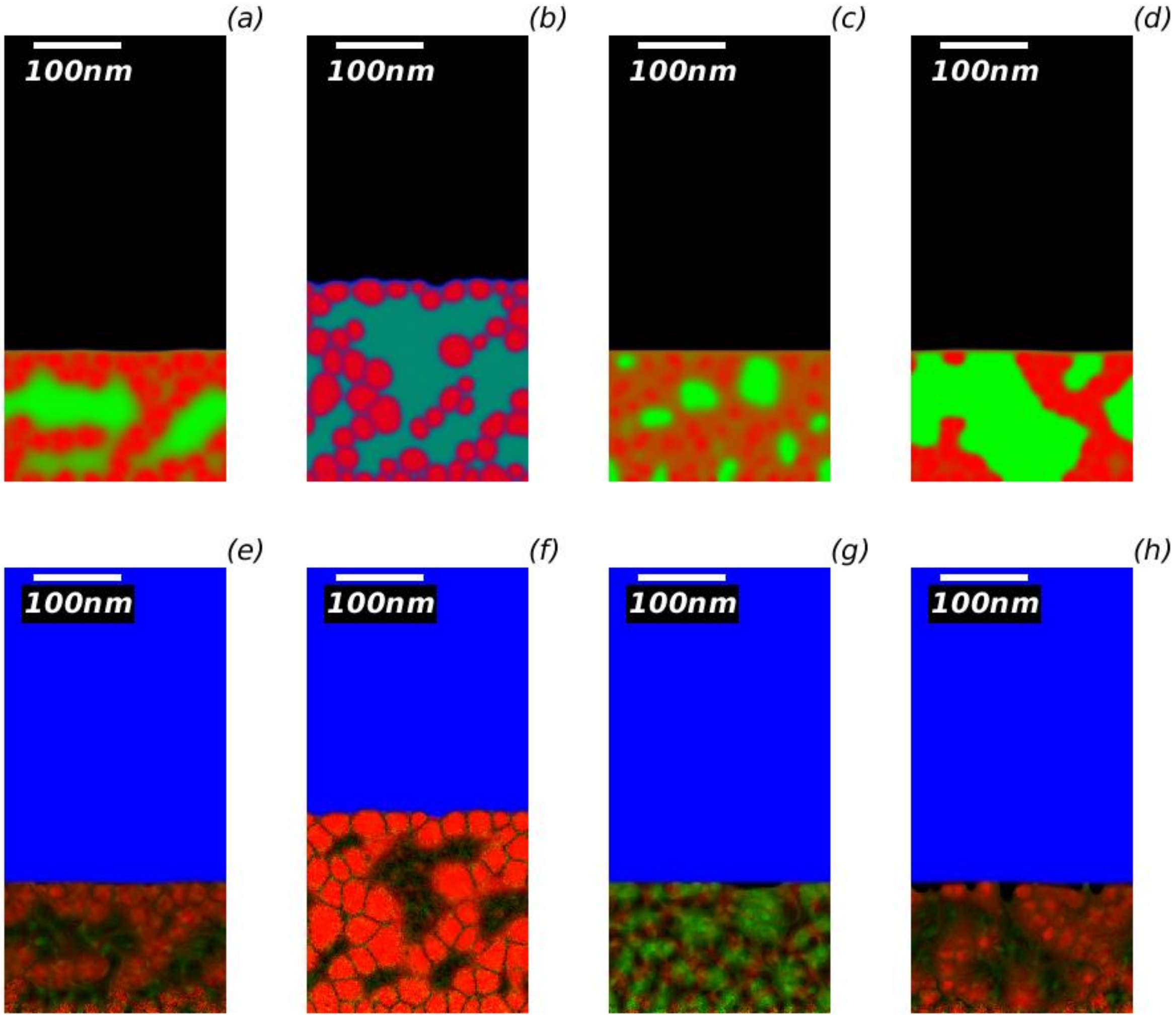

*Figure 15: morphology of a ternary polymer-small molecule-solvent blend after 10s drying time. The volume fraction (top row, (a-d)) and order parameter fields (bottom row, (e-h)) are shown for different parameter sets corresponding to Figure 14. (a,e) blend ratio 60:40,* $M_{p,0} = 1.2 \cdot 10^{-5}\ s^{-1}$ *and* $M_{sm,0} = 8 \cdot 10^{-5}\ s^{-1}$, $\chi_{ll,psm} = 0.8$ *(b,f) blend ratio 60:40,* $M_{p,0} = 1.5 \cdot 10^{-5}\ s^{-1}$ *and* $M_{sm,0} = 8 \cdot 10^{-5}\ s^{-1}$, $\chi_{ll,psm} = 0.8$ *(c,g) blend ratio 60:40,* $M_{p,0} = 0.8 \cdot 10^{-5}\ s^{-1}$ *and* $M_{sm,0} = 40 \cdot 10^{-5}\ s^{-1}$, $\chi_{ll,psm} = 0.8$ *(d,h) blend ratio 40:60,* $M_{p,0} = 1.2 \cdot 10^{-5}\ s^{-1}$ *and* $M_{sm,0} = 8 \cdot 10^{-5}\ s^{-1}$, $\chi_{ll,psm} = 2$. *The volume fraction field of the polymer is represented in red, the one of the small molecule in green and the one of the solvent in blue. For the order parameter field of the the polymer crystals is represented in red, the one of the small molecule crystals in green and the one of the vapor in blue. The system size is 512 nm x 256 nm.*

These examples show that even slight variations of the material parameters can lead to completely different morphology formation pathways and final structures. They also demonstrate the ability of the code to handle this miscellaneous physics with morphologically complex structures. In particular, in these sophisticated simulations, nucleation, growth, impingement, advection and stability at the film surface of the crystals and liquid-liquid phase separation work correctly as expected from the benchmark tests presented in the previous section. Finally, only an approximate and qualitative description of the morphologies has been given here. Even if it is not in the scope of the present paper, a precise qualitative analysis of the morphologies (in terms of composition, crystallinity, domain sizes, spatial organization…) is straightforward and shall be systematically performed in the near future.

## 6. Conclusion and perspectives

In this paper, we developed a coupled phase field-fluid mechanics simulation framework for the investigation of evaporation-induced morphology formation in multicomponent drying films. Using the Cahn-Hilliard-Cook, the stochastic Allen-Cahn, as well as the mass conservation and momentum conservation equations, physical processes like evaporation, liquid-liquid phase separation and coarsening, crystal nucleation, growth and coarsening in polycrystalline materials can be taken into account. Mass transport occurs by diffusion and advection. The simulation tool can handle any number of materials in the mixture. The implementation is based on a finite volume / finite difference discretization and uses advanced implicit time stepping methods and parallel computing based on the message passing protocol (MPI). It enables the handling of the different physical processes even if they occur on very different time scales. Even though only 2D simulations are presented in this paper, the code is implemented in 3D and can thus handle fully three dimensional systems as well. The basic working principles of the theoretical framework, results on benchmark tests, and the basic behavior in various cases (spinodal decomposition, crystallization in pure materials and mixtures, drying films with existing crystals) were presented. Finally, we presented some examples of the evaporation-induced structure formation in a sophisticated ternary polymer-small molecule-solvent mixture, whereby both polymers are crystalline and immiscible. Thereby, we showed that the simulation tool can handle various morphology formation pathways, and very different structures have been obtained.

The simulation tool is designed in a flexible way, so that the whole physics, or only part of it, can be taken into account. This can be recognized from the variety of simulations presented in the present work and renders our theoretical framework applicable not only for various material systems or applications, but also for various processing and solicitation conditions (solutions at fixed composition, solutions under flow, drying films, solvent vapor annealing, thermal annealing or ageing in dry systems…). We emphasize that the proposed framework should not be considered just as a toy model working in a parameter space that would not be accessible for real experiments. On the contrary, the model can also be used in a parameter window that is in line with realistic experimental values, as demonstrated in the present paper: in fact, the values that have been given to the various physical parameters can be thought of as representative of a real polymer-small molecule-solvent mixture. As pointed out above, the only major exception to this is the viscosity.

Thus, the model can be used not only to perform systematic studies on the material-process-structure relationship, but also to compare simulation results with measurements on real systems. Of course, the values of the parameters as well as their dependency on the composition should be discussed in much more detail and refined, and this will be the topic of future work. Therefore, we hope that this simulation tool can produce results that will be at least qualitatively comparable with experimental results. However, a precise quantitative match is probably out of reach, due to several limitations. First, as usual in phase-field or other diffuse interface simulation frameworks, the thickness of the interfaces is significantly larger than in the real world. Second, the viscosities accessible for reasonable calculation times, are, at least for realistic simulations of drying dilute solutions, orders of magnitude higher than the real values. Thus, the impact of convective fluxes on the film morphology formation might be underestimated in such situations. Third, it has been already highlighted above that some purely numerical parameters may quantitatively influence the crystallization process, especially in drying films.

Although one should therefore handle quantitative results with caution, we believe that this framework opens the way to a vast horizon of investigations on the morphology of sophisticated multicomponent, crystalline systems, thus fulfilling the objectives detailed in the introduction of this paper. The interaction between evaporation, miscibility in the amorphous state and crystallization of each material can be studied, depending on the thermodynamic and kinetic properties, and on the time and length scale of each physical process. This allows in principle to systematically sort out, analyze and characterize the different possible structure formation pathways and the various associated final morphologies, and hopefully to make predictions for given material systems processed in specific conditions. At the end, the overarching goal is to help gaining control on the process-structure relationship.

The physics taken into account in the phase-field framework can be extended. On a short term perspective, specific interactions between the substrate and the various materials will be implemented following ideas already described in the literature. [62] [67] [85], as well as strongly anisotropic crystal growth. [56] [136] [137] On a longer term perspective, we plan to take stochastic fluctuations in the momentum conservation equation into account, so as to handle crystal diffusion, which might be an important physical process for nanometer sized particles in dilute solutions.

Furthermore, a major research topic in the near future will be the investigation of real systems. Our approach will be applied in the field of solution processed solar cells, in particular for understanding the formation and morphological stability of bulk heterojunctions in organic photoactive layers, and of polycrystalline perovskite layers in perovskite solar cells. Beyond the progress on the simulation side, additional challenges will be the proper measurement of the material parameters, in particular their composition dependency and their mapping onto simulation parameters. The framework will be validated by comparing the simulation results with the accurate experimental characterizations of the morphology. This requires not only advanced, multi-technique in-situ characterizations, but also the development of quantitative morphology analysis tools for the simulated structures. Then, the simulation framework will be used thoroughly to unravel the possible structure formation pathways in these systems and propose improved processing conditions.

Finally, we believe that our framework could also be advantageously applied to many other material systems, notably drying thin or thick films, as soon as the morphology formation process involves liquid-liquid or liquid-solid phase transformations.

## Acknowledgements

The authors acknowledge financial support by the German Research Foundation (DFG, project HA 4382/14-1), the European commission (project 101008701), and the Impulse and Networking Fund of the Helmholtz Society. They gratefully thank Dr. Olga Wodo and Dr. Vincent Auvray for fruitful discussions.

# Supporting Information

# "Phase-field simulations of the morphology formation in evaporating crystalline multicomponent films"

*Olivier J.J. Ronsin [a,b]* and Jens Harting [a,b,c]**

[a] *Helmholtz Institute Erlangen-Nürnberg for Renewable Energy, Forschungszentrum Jülich, Fürther Straße 248, 90429 Nürnberg, Germany*

[b] *Department of Chemical and Biological Engineering, Friedrich-Alexander-Universität Erlangen-Nürnberg, Fürther Straße 248, 90429 Nürnberg, Germany*

[c] *Department of Chemical and Biological Engineering and Department of Physics, Friedrich-Alexander-Universität Erlangen-Nürnberg, Fürther Straße 248, 90429 Nürnberg, Germany*

**Corresponding Authors**

*E-mail: o.ronsin@fz-juelich.de (O.R.)*
*E-mail: j.harting@fz-juelich.de (J.H.)*

# S1. Relationship between the 'stress form' and the 'potential forms' of the capillary forces

The general expression of the 'stress form' of the capillary forces is given by

$$\begin{cases} \boldsymbol{F_{\varphi}} = \boldsymbol{\nabla}\left[\sum_{i=1}^{n} \kappa_i\left(a|\boldsymbol{\nabla}\varphi_i|^2\boldsymbol{I} - \boldsymbol{\nabla}\varphi_i \times \boldsymbol{\nabla}\varphi_i\right)\right] \\ \boldsymbol{F_{\phi}} = \boldsymbol{\nabla}\left[\sum_{k=1}^{n_{cryst}} {\varepsilon_k}^2\left(a|\boldsymbol{\nabla}\phi_k|^2\boldsymbol{I} - \boldsymbol{\nabla}\phi_k \times \boldsymbol{\nabla}\phi_k\right) + {\varepsilon_{vap}}^2\left(a\left|\boldsymbol{\nabla}\phi_{vap}\right|^2\boldsymbol{I} - \boldsymbol{\nabla}\phi_{vap} \times \boldsymbol{\nabla}\phi_{vap}\right)\right] \end{cases} \tag{S31}$$

whereby in the main text of the paper (Equation 24), the choice $a=1$ has been made. Starting from this expression, we derive the 'first potential form' and the 'second potential form' as available in the literature.

Reminding the formula $\boldsymbol{\nabla}(\boldsymbol{u} \times \boldsymbol{v}) = (\nabla\boldsymbol{u})\boldsymbol{v} + (\boldsymbol{u}\boldsymbol{\nabla})\boldsymbol{v}$ for two vectors $\boldsymbol{u}$ and $\boldsymbol{v}$, we get the general formula for any scalar field $\psi$:

$$\boldsymbol{\nabla}(\boldsymbol{\nabla}\psi \times \boldsymbol{\nabla}\psi) = (\boldsymbol{\nabla}^2\psi)\boldsymbol{\nabla}\psi + \frac{1}{2}\boldsymbol{\nabla}(|\boldsymbol{\nabla}\psi|^2) \tag{S32}$$

which allows to perform the transformation:

$$\boldsymbol{\nabla}(a|\boldsymbol{\nabla}\psi|^2\boldsymbol{I} - \boldsymbol{\nabla}\psi \times \boldsymbol{\nabla}\psi) = \left(a - \frac{1}{2}\right)\boldsymbol{\nabla}(|\boldsymbol{\nabla}\psi|^2) - (\boldsymbol{\nabla}^2\psi)\boldsymbol{\nabla}\psi \tag{S33}$$

Applying this to all $\{\varphi_i\}$, $\{\phi_k\}$ and $\phi_{vap}$ found in Equation S31, the total capillary force $\boldsymbol{F_{cap}} = \boldsymbol{F_{\varphi}} + \boldsymbol{F_{\phi}}$ can be written as

$$\boldsymbol{F_{cap}} = \begin{array}{l} \boldsymbol{\nabla}\left[\left(a - \frac{1}{2}\right)\left(\sum_{i=1}^{n} \kappa_i|\boldsymbol{\nabla}\varphi_i|^2 + \sum_{k=1}^{n_{cryst}} {\varepsilon_k}^2|\boldsymbol{\nabla}\phi_k|^2 + {\varepsilon_{vap}}^2\left|\boldsymbol{\nabla}\phi_{vap}\right|^2\right)\right] \\ -\left(\sum_{i=1}^{n} \kappa_i(\boldsymbol{\nabla}^2\varphi_i)\boldsymbol{\nabla}\varphi_i + \sum_{k=1}^{n_{cryst}} {\varepsilon_k}^2(\boldsymbol{\nabla}^2\phi_k)\boldsymbol{\nabla}\phi_k + {\varepsilon_{vap}}^2\left(\boldsymbol{\nabla}^2\phi_{vap}\right)\boldsymbol{\nabla}\phi_{vap}\right) \end{array} \tag{S34}$$

The first term is the pressure term for the stress form which is written as $-\boldsymbol{\nabla}P_{cap,SF}$ and the stress form reads

$$\boldsymbol{F_{cap}} = \begin{array}{l} -\boldsymbol{\nabla}P_{cap,SF} \\ -\left(\sum_{i=1}^{n} \kappa_i(\boldsymbol{\nabla}^2\varphi_i)\boldsymbol{\nabla}\varphi_i + \sum_{k=1}^{n_{cryst}} {\varepsilon_k}^2(\boldsymbol{\nabla}^2\phi_k)\boldsymbol{\nabla}\phi_k + {\varepsilon_{vap}}^2\left(\boldsymbol{\nabla}^2\phi_{vap}\right)\boldsymbol{\nabla}\phi_{vap}\right) \end{array} \tag{S35}$$

Now, using the relationship $\frac{\delta\Delta G_V}{\delta\phi} = \frac{\partial\Delta G_V}{\partial\phi} - \nabla\left(\frac{\partial\Delta G_V}{\partial(\boldsymbol{\nabla}\phi)}\right)$, we can write on the one hand

$$\varepsilon^2(\boldsymbol{\nabla}^2\phi) = \frac{\partial\Delta G_V}{\partial\phi} - \frac{\delta\Delta G_V}{\delta\phi} \tag{S36}$$

for all the order parameters. On the other hand, for the volume fractions, we are using Equations 7 and 8 of the main text:

$$\Delta\mu_{V,i}^{gen} = \mu_{V,i}^{gen} - \mu_{V,i}^{gen} = \frac{\partial\Delta G_V}{\partial\varphi_i} - \frac{\partial\Delta G_V}{\partial\varphi_n} - \kappa_i\boldsymbol{\nabla}^2\varphi_i + \kappa_n\boldsymbol{\nabla}^2\varphi_n \tag{S37}$$

We can rewrite the second line of Equation S35 with the help of Equations S36 and S37 in order to obtain

$$\boldsymbol{F_{cap}} = \begin{array}{l} -\boldsymbol{\nabla} P_{cap,SF} \\ -\sum_{i=1}^{n} \left( \frac{\partial \Delta G_V}{\partial \varphi_i} - \frac{\partial \Delta G_V}{\partial \varphi_n} - \Delta\mu_{V,i}^{gen} + \kappa_n \boldsymbol{\nabla}^2 \varphi_n \right) \boldsymbol{\nabla} \varphi_i \\ -\sum_{k=1}^{n_{cryst}} \left( \frac{\partial \Delta G_V}{\partial \phi_k} - \frac{\delta \Delta G_V}{\delta \phi_k} \right) \boldsymbol{\nabla} \phi_k - \left( \frac{\partial \Delta G_V}{\partial \phi_{vap}} - \frac{\delta \Delta G_V}{\delta \phi_{vap}} \right) \boldsymbol{\nabla} \phi_{vap} \end{array} \quad \text{(S38)}$$

Using the volume conservation, $\sum_{i=1}^{n} \varphi_i = 1$, we remark that $\sum_{i=1}^{n} \boldsymbol{\nabla} \varphi_i = 0$ so that the $\frac{\partial \Delta G_V}{\partial \varphi_n}$ and $\kappa_n \boldsymbol{\nabla}^2 \varphi_n$ terms vanish from the second line of the equation above. Rearranging the remaining of the second and third line leads to

$$\boldsymbol{F_{cap}} = \begin{array}{l} -\boldsymbol{\nabla} P_{cap,SF} \\ -\left( \sum_{i=1}^{n} \frac{\partial \Delta G_V}{\partial \varphi_i} \boldsymbol{\nabla} \varphi_i + \sum_{k=1}^{n_{cryst}} \frac{\partial \Delta G_V}{\partial \phi_k} \boldsymbol{\nabla} \phi_k + \frac{\partial \Delta G_V}{\partial \phi_{vap}} \boldsymbol{\nabla} \phi_{vap} \right) \\ +\sum_{i=1}^{n-1} \Delta\mu_{V,i}^{gen} \boldsymbol{\nabla} \varphi_i + \sum_{k=1}^{n_{cryst}} \frac{\delta \Delta G_V}{\delta \phi_k} \boldsymbol{\nabla} \phi_k + \frac{\delta \Delta G_V}{\delta \phi_{vap}} \boldsymbol{\nabla} \phi_{vap} \end{array} \quad \text{(S39)}$$

At the second line of the equation above, we recognize the total derivative of $\Delta G_V$ with respect to the spatial variable, $\nabla(\Delta G_V)$, with leads to the result:

$$\boldsymbol{F_{cap}} = \begin{array}{l} -\boldsymbol{\nabla} \left[ \Delta G_V + P_{cap,SF} \right] \\ +\sum_{i=1}^{n-1} \Delta\mu_{V,i}^{gen} \boldsymbol{\nabla} \varphi_i + \sum_{k=1}^{n_{cryst}} \frac{\delta \Delta G_V}{\delta \phi_k} \boldsymbol{\nabla} \phi_k + \frac{\delta \Delta G_V}{\delta \phi_{vap}} \boldsymbol{\nabla} \phi_{vap} \end{array} \quad \text{(S40)}$$

Equation S40 is the ‘first potential form’ of the capillary forces. The pressure term for this form is related to the one for the stress form by $P_{cap,PF1} = \Delta G_V + P_{cap,SF}$ and the deviatoric part is written in such a way that the potentials relative to each phase field variable appear clearly in the equation.

In order to obtain the ‘second potential form’, we simply apply the general formula $\boldsymbol{\nabla}(fg) = f\boldsymbol{\nabla} g + g\boldsymbol{\nabla} f$ to the $\Delta\mu_{V,i}^{gen} \boldsymbol{\nabla} \varphi_i$ terms and the $\frac{\delta \Delta G_V}{\delta \phi} \boldsymbol{\nabla} \phi$ terms in Equation S40 :

$$\boldsymbol{F_{cap}} = \begin{array}{l} -\boldsymbol{\nabla} \left[ \Delta G_V + P_{cap,SF} - \left( \sum_{i=1}^{n-1} \varphi_i \Delta\mu_{V,i}^{gen} + \sum_{k=1}^{n_{cryst}} \phi_k \frac{\delta \Delta G_V}{\delta \phi_k} + \phi_{vap} \frac{\delta \Delta G_V}{\delta \phi_{vap}} \right) \right] \\ -\left( \sum_{i=1}^{n-1} \varphi_i \boldsymbol{\nabla} \left( \Delta\mu_{V,i}^{gen} \right) + \sum_{k=1}^{n_{cryst}} \phi_k \boldsymbol{\nabla} \left( \frac{\delta \Delta G_V}{\delta \phi_k} \right) + \phi_{vap} \boldsymbol{\nabla} \left( \frac{\delta \Delta G_V}{\delta \phi_{vap}} \right) \right) \end{array} \quad \text{(S41)}$$

In practice, different pressure terms (which means different definitions of the pressure) do not have an impact on the calculation of the velocity field. The pressure terms due to the capillary forces may be dropped. In this case, when the fluid dynamics equations are solved for the unknown fields $P$ and $\boldsymbol{v}$, we obtain identical velocity fields but different pressure fields that can be matched onto each other using the relationships detailed above. [87] The momentum equation can be written as

$$\rho\left(\frac{\partial \boldsymbol{v}}{\partial t}+\boldsymbol{v}\nabla\boldsymbol{v}\right)=\boldsymbol{F}_{\boldsymbol{cap}}-\ \nabla P+\nabla\boldsymbol{\Sigma} \tag{S42}$$

using indiscriminately for $\boldsymbol{F}_{\boldsymbol{cap}}$ either the 'stress form'

$$\boldsymbol{F}_{\boldsymbol{cap,SF}}=-\left(\sum_{i=1}^{n}\kappa_i(\boldsymbol{\nabla}^2\varphi_i)\boldsymbol{\nabla}\varphi_i+\sum_{k=1}^{n_{cryst}}{\varepsilon_k}^2(\boldsymbol{\nabla}^2\phi_k)\boldsymbol{\nabla}\phi_k+{\varepsilon_{vap}}^2\left(\boldsymbol{\nabla}^2\phi_{vap}\right)\boldsymbol{\nabla}\phi_{vap}\right) \tag{S43}$$

or the 'first potential form'

$$\boldsymbol{F}_{\boldsymbol{cap,PF1}}=\sum_{i=1}^{n-1}\Delta\mu_{V,i}^{gen}\boldsymbol{\nabla}\varphi_i+\sum_{k=1}^{n_{cryst}}\frac{\delta\Delta G_V}{\delta\phi_k}\boldsymbol{\nabla}\phi_k+\frac{\delta\Delta G_V}{\delta\phi_{vap}}\boldsymbol{\nabla}\phi_{vap} \tag{S44}$$

or the 'second potential form'

$$\boldsymbol{F}_{\boldsymbol{cap,PF2}}=-\left(\sum_{i=1}^{n-1}\varphi_i\boldsymbol{\nabla}\left(\Delta\mu_{V,i}^{gen}\right)+\sum_{k=1}^{n_{cryst}}\phi_k\boldsymbol{\nabla}\left(\frac{\delta\Delta G_V}{\delta\phi_k}\right)+\phi_{vap}\boldsymbol{\nabla}\left(\frac{\delta\Delta G_V}{\delta\phi_{vap}}\right)\right) \tag{S45}$$

# S2. Derivation of the noise for the multicomponent Cahn-Hilliard-Cook equation

The noise term in the Cahn-Hilliard-Cook equation needs to be computed from uncorrelated random numbers, for any number of components. The objective is to generate a noise term which respects following rules for all materials *i* and *j* ranging from *1* to *n-1*:

$$\begin{cases} \langle \zeta^i(\boldsymbol{r},t) \rangle = 0 \\ \langle \zeta^i(\boldsymbol{r},t)\zeta^j(\boldsymbol{r}',t') \rangle = -\dfrac{2v_0}{N_a} \boldsymbol{\nabla}\left[\Lambda_{ij}\delta_D(t-t')\boldsymbol{\nabla}\big(\delta_D(\boldsymbol{r}-\boldsymbol{r}')\big)\right] \end{cases} \tag{S46}$$

To find a general rule for the expression of the random numbers we are after, we first review briefly the discretization of this noise as detailed by Schaefer. [64] We write $\delta_D(\boldsymbol{r}-\boldsymbol{r}')$ for any couple of points $\{\boldsymbol{r},\boldsymbol{r}'\} = \{(\boldsymbol{x},\boldsymbol{y},\boldsymbol{z}),(\boldsymbol{x}',\boldsymbol{y}',\boldsymbol{z}')\} = \{(\boldsymbol{x_a},\boldsymbol{y_c},\boldsymbol{z_e}),(\boldsymbol{x_b},\boldsymbol{y_d},\boldsymbol{z_f})\}$ using the factorization $\delta_D(\boldsymbol{r}-\boldsymbol{r}') = \delta_D(\boldsymbol{x_a}-\boldsymbol{x_b})\delta_D(\boldsymbol{y_c}-\boldsymbol{y_d})\delta_D(\boldsymbol{z_e}-\boldsymbol{z_f})$. We can discretize the gradient in the *x*-direction as $\frac{d}{dx}\delta_D(\boldsymbol{x_a}-\boldsymbol{x_b}) \approx \frac{1}{\Delta x^2}\left(\delta_{a+\frac{1}{2},b} - \delta_{a-\frac{1}{2},b}\right)$, which leads, using similar expressions in the *y*- and *z*-direction, to

$$\Lambda_{ij}\boldsymbol{\nabla}\big(\delta_D(\boldsymbol{r}-\boldsymbol{r}')\big) \approx \frac{\Lambda_{ij}}{\Delta x\Delta y\Delta z}\begin{cases} \dfrac{1}{\Delta x}\left(\delta_{a+\frac{1}{2},b} - \delta_{a-\frac{1}{2},b}\right)\delta_{cd}\delta_{ef} \\ \dfrac{1}{\Delta y}\left(\delta_{c+\frac{1}{2},d} - \delta_{c-\frac{1}{2},d}\right)\delta_{ab}\delta_{ef} \\ \dfrac{1}{\Delta z}\left(\delta_{e+\frac{1}{2},f} - \delta_{e-\frac{1}{2},f}\right)\delta_{ab}\delta_{cd} \end{cases} \tag{S47}$$

Taking the gradient again and remembering that the mobilities $\Lambda_{ij}$ are variable in space, we obtain the 3D final expression for the discrete noise:

$$\begin{aligned} &\langle \zeta^i(\boldsymbol{r},t)\zeta^j(\boldsymbol{r}',t') \rangle \\ &\approx \frac{2v_0\delta_{tt'}}{N_a\Delta x\Delta y\Delta z\Delta t}\begin{cases} \dfrac{\delta_{cd}\delta_{ef}}{\Delta x^2}\left(-\delta_{a+1,b}{\Lambda_{ij}}^{a+\frac{1}{2},ce} + \delta_{ab}\left({\Lambda_{ij}}^{a+\frac{1}{2},ce} + {\Lambda_{ij}}^{a-\frac{1}{2},pe}\right) - \delta_{a-1,b}{\Lambda_{ij}}^{a-\frac{1}{2},ce}\right) \\ \dfrac{\delta_{ab}\delta_{ef}}{\Delta y^2}\left(-\delta_{c+1,d}{\Lambda_{ij}}^{a,c+\frac{1}{2},e} + \delta_{cd}\left({\Lambda_{ij}}^{a,c+\frac{1}{2},e} + {\Lambda_{ij}}^{a,c-\frac{1}{2},e}\right) - \delta_{c-1,d}{\Lambda_{ij}}^{a,c-\frac{1}{2},e}\right) \\ \dfrac{\delta_{ab}\delta_{cd}}{\Delta z^2}\left(-\delta_{e+1,f}{\Lambda_{ij}}^{ac,e+\frac{1}{2}} + \delta_{ef}\left({\Lambda_{ij}}^{ac,e+\frac{1}{2}} + {\Lambda_{ij}}^{ac,e-\frac{1}{2}}\right) - \delta_{e-1,f}{\Lambda_{ij}}^{ac,e-\frac{1}{2}}\right) \end{cases} \end{aligned} \tag{S48}$$

Now, the next step is to write each $\zeta_i$ as the sum of *n* Gaussian random numbers in each direction (remember that *n* is the number of materials). Schaefer et al. [64] proposed an implementation for three materials. Building on their ideas, we propose below a general approach for any number of materials. Considering from now on only the *x*-direction for simplicity, we write

$$\zeta^i(\boldsymbol{r},t) \approx {\zeta_a}^i = \sum_{k=1}^{n}\left({\epsilon_{ik}}^{a+1}{B_{ik}}^{a+1} - {\epsilon_{ik}}^{a}{B_{ik}}^{a}\right) \tag{S49}$$

with $B_{ik} = B_{ki}$ being a (Gaussian) random number with variance $\sigma^2 = \frac{2v_0\delta_{tt'}}{N_a\Delta x^3 \Delta y \Delta z \Delta t}$, coupling fluctuations of material *i* and *k*:

$$\begin{cases} \langle B_{ik} \rangle = 0 \\ \langle {B_{ik}}^a {B_{i'k'}}^b \rangle = \begin{cases} \sigma^2 \delta_{ab} \; if \; \{i,k\} = \{i',k'\} \\ 0 \; otherwise \end{cases} \end{cases} \quad \text{(S50)}$$

Let us develop $\langle \zeta^i(x,t)\zeta^j(x',t') \rangle \approx \langle {\zeta_a}^i {\zeta_b}^j \rangle$ using Equation S49. First, we investigate the case $i \neq j$. Using the fact that only two-material correlations are non-zero, we get:

$$\frac{\langle {\zeta_a}^i {\zeta_b}^j \rangle}{\sigma^2} = \begin{matrix} {\epsilon_{ij}}^{a+1}{\epsilon_{ji}}^{b+1} \langle {B_{ij}}^{a+1} {B_{ij}}^{b+1} \rangle - {\epsilon_{ij}}^{a+1}{\epsilon_{ji}}^{b} \langle {B_{ij}}^{a+1} {B_{ij}}^{b} \rangle \\ + \quad {\epsilon_{ij}}^{a}{\epsilon_{ji}}^{b} \langle {B_{ij}}^{a} {B_{ij}}^{b} \rangle - {\epsilon_{ij}}^{a}{\epsilon_{ji}}^{b+1} \langle {B_{ij}}^{a} {B_{ij}}^{b+1} \rangle \end{matrix} \quad \text{(S51)}$$

By using the definition of $B_{ik}$ (Equation S50) this leads to

$$\frac{\langle {\zeta_a}^i {\zeta_b}^j \rangle}{\sigma^2} = {\epsilon_{ij}}^{a+1}{\epsilon_{ji}}^{b+1}\delta_{a+1,b+1} + {\epsilon_{ij}}^{a}{\epsilon_{ji}}^{b}\delta_{a,b} - {\epsilon_{ij}}^{a+1}{\epsilon_{ji}}^{b}\delta_{a+1,b} - {\epsilon_{ij}}^{a}{\epsilon_{ji}}^{b+1}\delta_{a,b+1} \quad \text{(S52)}$$

and finally we obtain

$$\frac{\langle {\zeta_a}^i {\zeta_b}^j \rangle}{\sigma^2} = -\delta_{a+1,b}{\epsilon_{ij}}^{a+1}{\epsilon_{ji}}^{b} + \delta_{a,b}\left({\epsilon_{ij}}^{a+1}{\epsilon_{ji}}^{b+1} + {\epsilon_{ij}}^{a}{\epsilon_{ji}}^{b}\right) - \delta_{a,b+1}{\epsilon_{ij}}^{a}{\epsilon_{ji}}^{b+1} \quad \text{(S53)}$$

By comparing Equation S53 with the first line of Equation S48 and dropping the indices for the *y*- and *z*-directions, we identify the following terms:

$$\begin{cases} {\epsilon_{ij}}^{a+1}{\epsilon_{ji}}^{a+1} = {\Lambda_{ij}}^{a+\frac{1}{2}} \\ {\epsilon_{ij}}^{a+1}{\epsilon_{ji}}^{a+1} + {\epsilon_{ij}}^{a}{\epsilon_{ji}}^{a} = {\Lambda_{ij}}^{a+\frac{1}{2}} + {\Lambda_{ij}}^{a-\frac{1}{2}} \\ {\epsilon_{ij}}^{a}{\epsilon_{ji}}^{a} = {\Lambda_{ij}}^{a-\frac{1}{2}} \end{cases} \quad \text{(S54)}$$

If the mobilities were all positive, we could simply use ${\epsilon_{ij}}^a = {\epsilon_{ji}}^a = \sqrt{{\Lambda_{ij}}^{a-\frac{1}{2}}}$, but the cross mobilities might be negative. As a consequence, we have to consider ${\epsilon_{ij}}^a \neq {\epsilon_{ji}}^a$ and choose arbitrarily:

$$\begin{cases} {\epsilon_{ij}}^a = \sqrt{\left|{\Lambda_{ij}}^{a-\frac{1}{2}}\right|} & for \; i < j \\ {\epsilon_{ij}}^a = sign\left({\Lambda_{ij}}^{a-\frac{1}{2}}\right)\sqrt{\left|{\Lambda_{ij}}^{a-\frac{1}{2}}\right|} & for \; i > j \end{cases} \quad \text{(S55)}$$

Now the case $i = j$ has to be investigated. Using the same approach, we get

$$\frac{\langle \zeta_a{}^i \zeta_b{}^i \rangle}{\sigma^2} = \begin{aligned} &-\delta_{a+1,b} \sum_{k=1}^{n} \epsilon_{ik}{}^{a+1} \epsilon_{ki}{}^{b} \\ &+ \quad \delta_{a,b} \sum_{k=1}^{n} (\epsilon_{ik}{}^{a+1} \epsilon_{ki}{}^{b+1} + \epsilon_{ik}{}^{a} \epsilon_{ki}{}^{b}) \\ &-\delta_{a,b+1} \sum_{k=1}^{n} \epsilon_{ik}{}^{a} \epsilon_{ki}{}^{b+1} \end{aligned} \tag{S56}$$

Isolating the contributions of the $i = k$ term and using the results of Equation S55 for the $i \neq k$ terms, we obtain by identifying again with the first line of Equation S48:

$$(\epsilon_{ii}{}^a)^2 = \Lambda_{ii}{}^{a-\frac{1}{2}} - \sum_{\substack{k=1 \\ k\neq i}}^{n} \Lambda_{ik}{}^{a-\frac{1}{2}} \tag{S57}$$

Using the fact that, for the slow-mode and fast-mode theory, $\sum_{k\neq i} \Lambda_{ik} = -\Lambda_{ii}$, this simplifies to

$$\epsilon_{ii}{}^a = \sqrt{2\Lambda_{ii}{}^{a-\frac{1}{2}}} \tag{S58}$$

Inserting Equation S55 and Equation S58 into Equation S49, we get the final expression of the discretized noise for the material *i* at the mesh point $x_a$ in the *x*-direction:

$$\zeta_a{}^i = \begin{aligned} &\sum_{k=1}^{i-1} \left( sign\left(\Lambda_{ik}{}^{a+\frac{1}{2}}\right) \sqrt{\left|\Lambda_{ik}{}^{a+\frac{1}{2}}\right|}\, B_{ik}{}^{a+1} - sign\left(\Lambda_{ik}{}^{a-\frac{1}{2}}\right) \sqrt{\left|\Lambda_{ik}{}^{a-\frac{1}{2}}\right|}\, B_{ik}{}^{a} \right) \\ &+ \sqrt{2\Lambda_{ii}{}^{a+\frac{1}{2}}} B_{ii}{}^{a+1} - \sqrt{2\Lambda_{ii}{}^{a-\frac{1}{2}}} B_{ii}{}^{a} \\ &+ \sum_{k=i+1}^{n} \left( \sqrt{\left|\Lambda_{ik}{}^{a+\frac{1}{2}}\right|}\, B_{ik}{}^{a+1} - \sqrt{\left|\Lambda_{ik}{}^{a-\frac{1}{2}}\right|}\, B_{ik}{}^{a} \right) \end{aligned} \tag{S59}$$

In the equation above, the $B_{ik}$ are *n* fields of Gaussian random numbers with variance $\frac{2v_0}{N_a \Delta x^3 \Delta y \Delta z \Delta t}$. The same approach can be applied in the *y*- an *z*-direction so that in 3D, the thermal fluctuations of each material are calculated from 3*n* independent fields of Gaussian random numbers.

## S3. Butcher matrices of various A- and L- stable DIRK methods

Runge-Kutta methods are used to solve the differential equation

$$\frac{dy}{dt} = f(t, y) \tag{S60}$$

and the solution for time step $t + 1 = t + \Delta t$ is calculated from the solution for time step $t$ by

$$y_{t+1} = y_t + \Delta t \sum_{s=1}^{n_s} b_s k_s \tag{S61}$$

with $k_s$ given by

$$k_s = f\left(t + c_s \Delta t, y_t + \Delta t \sum_{r=1}^{n_s} a_{sr} k_r\right) \tag{S62}$$

In the Butcher matrices below,[116] the left column is the vector of $\{c_s\}$ values, the bottom line the vector of $\{b_s\}$ values and the matrix corresponds to the $\{a_{sr}\}$ values.

<u>Euler backward (one stage, first order)</u>

| | |
|---|---|
| $1$ | $1$ |
| | $1$ |

<u>Pareschi-Russo (two stages, second order)</u>

$\lambda = 1 - \frac{\sqrt{2}}{2}$

| | | |
|---|---|---|
| $\lambda$ | $\lambda$ | $0$ |
| $1 - \lambda$ | $1 - 2\lambda$ | $\lambda$ |
| | $\frac{1}{2}$ | $\frac{1}{2}$ |

<u>DIRK3 (three stages, third order)</u>

$\lambda = \frac{1767732205903}{4055673282236}$

| | | | |
|---|---|---|---|
| $\lambda$ | $\lambda$ | $0$ | $0$ |
| $\frac{1+\lambda}{2}$ | $\frac{1-\lambda}{2}$ | $\lambda$ | $0$ |
| $1$ | $-\frac{3\lambda^2}{2} + 4\lambda - \frac{1}{4}$ | $\frac{3\lambda^2}{2} - 5\lambda + \frac{5}{4}$ | $\lambda$ |
| | $-\frac{3\lambda^2}{2} + 4\lambda - \frac{1}{4}$ | $\frac{3\lambda^2}{2} - 5\lambda + \frac{5}{4}$ | $\lambda$ |

<u>SDIRK4(3)5L[1]SA_C(2) (five stages, fourth order)</u>

$\lambda = \frac{1}{4}$

| | | | | | |
|---|---|---|---|---|---|
| $\frac{1}{4}$ | $\lambda$ | $0$ | $0$ | $0$ | $0$ |
| $\frac{2-\sqrt{2}}{4}$ | $\frac{1-\sqrt{2}}{4}$ | $\lambda$ | $0$ | $0$ | $0$ |
| $\frac{13+8\sqrt{2}}{41}$ | $\frac{-1676+145\sqrt{2}}{6724}$ | $3\frac{709+389\sqrt{2}}{6724}$ | $\lambda$ | $0$ | $0$ |
| $\frac{41+9\sqrt{2}}{49}$ | $\frac{-371435-351111\sqrt{2}}{470596}$ | $\frac{98054928+73894543\sqrt{2}}{112001848}$ | $\frac{56061972+30241643\sqrt{2}}{112001848}$ | $\lambda$ | $0$ |
| $1$ | $0$ | $4\frac{74+273\sqrt{2}}{5253}$ | $\frac{19187+5031\sqrt{2}}{55284}$ | $\frac{116092-100113\sqrt{2}}{334956}$ | $\lambda$ |
| | $0$ | $4\frac{74+273\sqrt{2}}{5253}$ | $\frac{19187+5031\sqrt{2}}{55284}$ | $\frac{116092-100113\sqrt{2}}{334956}$ | $\lambda$ |

# S4. Simulation parameters

## S4.1 Liquid-liquid phase separation

Subscripts '1', '2' stand for 'material 1' and 'material 2'.

| Parameters | Values | Units |
|---|---|---|
| GENERAL | | |
| *Grid spacing* | *2* | *nm* |
| *Grid size* | *1024 x 1024* | *nm* |
| *Initial film blend ratio* | *50:50* | - |
| $T$ | *300* | *K* |
| $\rho_1, \rho_2$ | *1000, 1000* | $kg{\cdot}m^{-3}$ |
| $m_1, m_2$ | *1, 1* | $kg{\cdot}mol^{-1}$ |
| THERMODYNAMICS | | |
| $\chi_{12,ll}$ | *4* | - |
| $\sigma_{CH}$ | $10^{-5}$ | - |
| $\beta$ | $10^{-5}$ | $J{\cdot}m^{-3}$ |
| $\gamma_b$ | *1* | - |
| $\kappa_p, \kappa_s$ | $10^{-10}, 10^{-10}$ | $J{\cdot}m^{-1}$ |
| KINETICS | | |
| $D_{s,1}^{\varphi_1\to1}, D_{s,1}^{\varphi_2\to1}, D_{s,2}^{\varphi_1\to1}, D_{s,2}^{\varphi_2\to1}$ | $10^{-11}, 10^{-11}, 10^{-11}, 10^{-11}$ | $m^2{\cdot}s^{-1}$ |
| DETECTION THRESHOLDS | | |
| | | |
| FLUID MECHANICS | | |
| $\eta_1 = \eta_2$ | ***See main text*** | $Pa{\cdot}s^{-1}$ |

## S4.2 Crystallization in a pure material

Subscript 'p', stands for 'polymer'.

| Parameters | Values | Units |
|---|---|---|
| GENERAL | | |
| *Grid spacing* | *1* | *nm* |
| *Grid size* | *512 x 512* | *nm* |
| $T$ | *330* | *K* |
| $\rho_p$ | *1100* | $kg{\cdot}m^{-3}$ |
| $m_p$ | *30* | $kg{\cdot}mol^{-1}$ |
| THERMODYNAMICS | | |
| $T_{m,p}$ | *510* | *K* |
| $L_p$ | $5{\cdot}10^4$ | $J{\cdot}kg^{-1}$ |
| $W_p$ | ***See main text*** | $J{\cdot}kg^{-1}$ |
| $\xi_{0,p}$ | *1* | - |
| $\sigma_{AC}$ | *1* | - |
| $\varepsilon_p$ | ***See main text*** | $(J{\cdot}m^{-1})^{0.5}$ |
| $\varepsilon_{g,p}$ | $3{\cdot}10^{-2}$ | $J{\cdot}m^{-2}$ |
| KINETICS | | |
| $M_{p,0}$ | $10^{-5}$ | $s^{-1}$ |
| $d_\zeta, c_\zeta, w_\zeta$ | $10^{-2}$, *0.85, 15* | - |
| DETECTION THRESHOLDS | | |
| $t_{\phi,p}$ | *0.4* | - |
| FLUID MECHANICS | | |
| | | |

## S4.3 Crystallization in a binary mixture

In **bold** the parameters that have been added as compared to the previous section.
Underlined are the parameters that have been changed as compared to the previous section.
Subscripts ‘p’, ‘**s**’ stand for ‘polymer’, ‘**solvent**’, respectively.

| Parameters | Values | Units |
|---|---|---|
| GENERAL | | |
| *Grid spacing* | *1* | *nm* |
| *Grid size* | *256 x 256* | *nm* |
| ***Initial film blend ratio*** | ***See text*** | - |
| $T$ | *330* | *K* |
| $\rho_p$, $\boldsymbol{\rho_s}$ | *1100*, ***1300*** | $kg \cdot m^{-3}$ |
| $m_p$, $\boldsymbol{m_s}$ | *30*, ***0.147*** | $kg \cdot mol^{-1}$ |
| THERMODYNAMICS | | |
| $\boldsymbol{\chi_{ps,ll}}$ | ***0.05*** | - |
| $\boldsymbol{\chi_{ps,sl}}$ | $\boldsymbol{\chi_{ps,ll} + 0.35}$ | - |
| $T_{m,p}$ | *510* | *K* |
| $L_p$ | $5 \cdot 10^4$ | $J \cdot kg^{-1}$ |
| $W_p$ | $7.5 \cdot 10^4$ | $J \cdot kg^{-1}$ |
| $\xi_{0,p}$ | *1* | - |
| $\boldsymbol{\gamma_m}$ | **2** | - |
| $\boldsymbol{\sigma_{CH}}$, $\sigma_{AC}$ | $\boldsymbol{10^{-5}}$, *1* | - |
| $\boldsymbol{\beta}$ | $\boldsymbol{10^{-5}}$ | $\boldsymbol{J \cdot m^{-3}}$ |
| $\boldsymbol{\gamma_b}$ | **1** | - |
| $\boldsymbol{\kappa_p}$, $\boldsymbol{\kappa_s}$ | $\boldsymbol{10^{-10}, 10^{-10}}$ | $\boldsymbol{J \cdot m^{-1}}$ |
| $\varepsilon_p$ | $10^{-5}$ | $(J \cdot m^{-1})^{0.5}$ |
| $\varepsilon_{g,p}$ | $3 \cdot 10^{-2}$ | $J \cdot m^{-2}$ |
| KINETICS | | |
| $\boldsymbol{D_{s,p}^{\varphi_p \to 1}, D_{s,p}^{\varphi_s \to 1}}$<br>$\boldsymbol{D_{s,s}^{\varphi_p \to 1}, D_{s,s}^{\varphi_s \to 1}}$ | $\boldsymbol{10^{-16}, 5 \cdot 10^{-11}}$<br>$\boldsymbol{10^{-14}, 2 \cdot 10^{-9}}$ | $\boldsymbol{m^2 \cdot s^{-1}}$ |
| $M_{p,0}$ | $10^{-5}$ | $s^{-1}$ |
| $\boldsymbol{d_{sl}, c_{sl}, w_{sl}}$ | $\boldsymbol{10^{-6}}$, ***0.97, 35*** | - |
| $d_\zeta, c_\zeta, w_\zeta$ | $10^{-2}$, *0.85, 15* | - |
| DETECTION THRESHOLDS | | |
| $t_{\phi,p}$, $\boldsymbol{t_{\varphi,p}}$ | *0.4*, ***0.02*** | - |
| FLUID MECHANICS | | |
| | | |

## S4.4 Crystal advection

In **bold** the parameters that have been added as compared to the previous section.
Underlined are the parameters that have been changed as compared to the previous section.
Subscripts ‘p’, ‘s’, ‘**a**’ stand for ‘polymer’, ‘solvent’, ‘**air**’ respectively.

| Parameters | Values | Units |
|---|---|---|
| GENERAL | | |
| *Grid spacing* | *1* | *nm* |
| *Grid size* | *256 x 64* | *nm* |
| *Initial film blend ratio* | *30:70* | - |
| *T* | *330* | *K* |
| $\rho_p, \rho_s, \boldsymbol{\rho_a}$ | *1100, 1300,* ***1300*** | *kg·m$^{-3}$* |
| $m_p, m_s, \boldsymbol{m_a}$ | *30, 0.147,* ***0.03*** | *kg·mol$^{-1}$* |
| THERMODYNAMICS | | |
| $\chi_{ps,ll}, \boldsymbol{\chi_{pa,ll}}$<br>$\boldsymbol{\chi_{sa,ll}}$ | *0.05,* ***0***<br>***0*** | - |
| $\chi_{ps,sl}, \boldsymbol{\chi_{pa,sl}}$ | $\chi_{ps,ll} + 0.35$, ***0*** | - |
| $T_{m,p}$ | *510* | *K* |
| $L_p$ | *5·10$^{4}$* | *J·kg$^{-1}$* |
| $W_p$ | *7.5·10$^{4}$* | *J·kg$^{-1}$* |
| $\xi_{0,p}$ | *1* | - |
| $\gamma_m$ | *2* | - |
| $\sigma_{CH}, \sigma_{AC}$ | *10$^{-5}$, 1* | - |
| $\boldsymbol{P_0}$ | ***10$^{5}$*** | ***Pa*** |
| $\boldsymbol{P_{sat,p}, P_{sat,s}, P_{sat,a}}$ | ***1.7·10$^{2}$, 2·10$^{3}$, 10$^{8}$*** | ***Pa*** |
| $\boldsymbol{P_p^{\infty}, P_s^{\infty}, P_a^{\infty}}$ | ***0, 0, 0*** | ***Pa*** |
| $\boldsymbol{E_{p,0}}$ | ***5·10$^{9}$*** | ***J·m$^{-3}$*** |
| $\boldsymbol{\gamma_c, \gamma_v}$ | ***2, 2*** | **-** |
| $\beta$ | *10$^{-5}$* | *J·m$^{-3}$* |
| $\gamma_b$ | *1* | - |
| $\kappa_p, \kappa_s, \boldsymbol{\kappa_a}$ | *10$^{-10}$, 10$^{-10}$,* ***2·10$^{-9}$*** | *J·m$^{-1}$* |
| $\varepsilon_p$ | *10$^{-5}$* | *(J·m$^{-1}$)$^{0.5}$* |
| $\varepsilon_{g,p}$ | *3·10$^{-2}$* | *J·m$^{-2}$* |
| $\boldsymbol{\varepsilon_{vap}}$ | ***10$^{-4}$*** | ***(J·m$^{-1}$)$^{0.5}$*** |
| KINETICS | | |
| $D_{s,p}^{\varphi_p \to 1}, D_{s,p}^{\varphi_s \to 1}, \boldsymbol{D_{s,p}^{\varphi_a \to 1}}$<br>$D_{s,s}^{\varphi_p \to 1}, D_{s,s}^{\varphi_s \to 1}, \boldsymbol{D_{s,s}^{\varphi_a \to 1}}$<br>$\boldsymbol{D_{s,a}^{\varphi_p \to 1}, D_{s,a}^{\varphi_s \to 1}, D_{s,a}^{\varphi_a \to 1}}$ | *10$^{-16}$, 5·10$^{-11}$,* ***5·10$^{-11}$***<br>*10$^{-14}$, 2·10$^{-9}$,* ***2·10$^{-9}$***<br>***10$^{-14}$, 2·10$^{-9}$, 2·10$^{-9}$*** | *m$^{2}$·s$^{-1}$* |
| $\boldsymbol{D_p^{vap}, D_s^{vap}, D_a^{vap}}$ | ***10$^{-16}$, 2·10$^{-9}$, 2·10$^{-9}$*** | ***m$^{2}$·s$^{-1}$*** |
| $M_{p,0}$ | *0* | *s$^{-1}$* |
| $\boldsymbol{M_{vap}}$ | ***10$^{6}$*** | ***s$^{-1}$*** |
| $\boldsymbol{\alpha}$ | ***-2.3·10$^{-5}$*** | **-** |
| $d_{sl}, c_{sl}, w_{sl}$ | *10$^{-6}$, 0.97, 35* | - |
| $d_{\zeta}, c_{\zeta}, w_{\zeta}$ | *10$^{-2}$, 0.85, 15* | - |
| $\boldsymbol{d_{sv}, c_{sv}, w_{sv}}$ | ***10$^{-3}$, 0.3, 15*** | **-** |
| DETECTION THRESHOLDS | | |
| $t_{\phi,p}, t_{\varphi,p}, \boldsymbol{t_{\phi_{vap}}}$ | *0.4, 0.02,* ***0.02*** | - |
| FLUID MECHANICS | | |
| $\boldsymbol{\eta_p, \eta_s, \eta_a, \eta_g}$ | ***6·10$^{6}$, 10$^{3}$, 10$^{6}$, 30*** | ***Pa·s$^{-1}$*** |
| $\boldsymbol{d_{\eta}, c_{\eta}, w_{\eta}}$ | ***10$^{-7}$, 0.2, 20*** | **-** |
| $\boldsymbol{k_{\eta}}$ | ***10*** | **-** |

## S4.5 Columnar structure stability

In **bold** the parameters that have been added as compared to the previous section.
$\underline{\text{Underlined}}$ are the parameters that have been changed as compared to the previous section.
Subscripts 'p', '**sm**', 's', 'a' stand for 'polymer', '**small molecule**', 'solvent', 'air', respectively.

| Parameters | Values | Units |
|---|---|---|
| GENERAL | | |
| *Grid spacing* | *1* | *nm* |
| *Grid size* | $\underline{256 \text{ x } 128}$ | *nm* |
| *Initial film blend ratio* | $\underline{30{:}20{:}50}$ | - |
| $T$ | *330* | *K* |
| $\rho_p, \boldsymbol{\rho_{sm}}, \rho_s, \rho_a$ | 1100, **1600**, 1300, 1300 | $kg \cdot m^{-3}$ |
| $m_p, \boldsymbol{m_{sm}}, m_s, m_a$ | 30, **0.91**, 0.147, 0.03 | $kg \cdot mol^{-1}$ |
| THERMODYNAMICS | | |
| $\boldsymbol{\chi_{psm,ll}}, \chi_{ps,ll}, \chi_{pa,ll}$<br>$\boldsymbol{\chi_{sms,ll}}, \boldsymbol{\chi_{sma,ll}}, \chi_{sa,ll}$ | **0.8**, 0.05, 0<br>**0.6, 0**, 0 | - |
| $\boldsymbol{\chi_{psm,sl}}, \chi_{ps,sl}, \chi_{pa,sl}$<br>$\boldsymbol{\chi_{smp,sl}}, \boldsymbol{\chi_{sms,sl}}, \boldsymbol{\chi_{sma,sl}}$ | $\boldsymbol{\chi_{psm,ll} + 0.3}$, $\chi_{ps,ll} + 0.35$, 0<br>$\boldsymbol{\chi_{psm,ll} + 0.3}$, $\boldsymbol{\chi_{sms,ll} + 0.1}$, **0** | - |
| $\boldsymbol{\chi_{psm,ss}}$ | **0** | - |
| $T_{m,p}, \boldsymbol{T_{m,sm}}$ | 510, **558** | *K* |
| $L_p, \boldsymbol{L_{sm}}$ | $5 \cdot 10^{4}$, $\mathbf{2 \cdot 10^{4}}$ | $J \cdot kg^{-1}$ |
| $W_p, \boldsymbol{W_{sm}}$ | $7.5 \cdot 10^{4}$, $\mathbf{3 \cdot 10^{4}}$ | $J \cdot kg^{-1}$ |
| $\xi_{0,p}, \boldsymbol{\xi_{0,sm}}$ | 1, **1** | - |
| $\gamma_m$ | 2 | - |
| $\sigma_{CH}, \sigma_{AC}$ | $10^{-5}$, 1 | - |
| $P_0$ | $10^{5}$ | *Pa* |
| $P_{sat,p}, \boldsymbol{P_{sat,sm}}, P_{sat,s}, P_{sat,a}$ | $1.7 \cdot 10^{2}$, $\mathbf{10^{2}}$, $2 \cdot 10^{3}$, $10^{8}$ | *Pa* |
| $P_p^{\infty}, \boldsymbol{P_{sm}^{\infty}}, P_s^{\infty}, P_a^{\infty}$ | 0, **0**, 0, 0 | *Pa* |
| $E_{p,0}, \boldsymbol{E_{sm,0}}$ | $5 \cdot 10^{9}$, $\mathbf{5 \cdot 10^{9}}$ | $J \cdot m^{-3}$ |
| $\gamma_c, \gamma_v$ | 2, 2 | - |
| $\beta$ | $10^{-5}$ | $J \cdot m^{-3}$ |
| $\gamma_b$ | 1 | - |
| $\kappa_p, \boldsymbol{\kappa_{sm}}, \kappa_s, \kappa_a$ | $10^{-10}$, $\mathbf{10^{-10}}$, $10^{-10}$, $2 \cdot 10^{-9}$ | $J \cdot m^{-1}$ |
| $\varepsilon_p, \boldsymbol{\varepsilon_{sm}}$ | $10^{-5}$, $\mathbf{10^{-5}}$ | $(J \cdot m^{-1})^{0.5}$ |
| $\varepsilon_{g,p}, \boldsymbol{\varepsilon_{g,sm}}$ | $3 \cdot 10^{-2}$, $\mathbf{3 \cdot 10^{-2}}$ | $J \cdot m^{-2}$ |
| $\varepsilon_{vap}$ | $10^{-4}$ | $(J \cdot m^{-1})^{0.5}$ |
| KINETICS | | |
| $D_{s,p}^{\varphi_p \to 1}, \boldsymbol{D_{s,p}^{\varphi_{sm} \to 1}}, D_{s,p}^{\varphi_s \to 1}, D_{s,p}^{\varphi_a \to 1}$<br>$\boldsymbol{D_{s,sm}^{\varphi_p \to 1}}, \boldsymbol{D_{s,sm}^{\varphi_{sm} \to 1}}, \boldsymbol{D_{s,sm}^{\varphi_s \to 1}}, \boldsymbol{D_{s,sm}^{\varphi_a \to 1}}$<br>$D_{s,s}^{\varphi_p \to 1}, \boldsymbol{D_{s,s}^{\varphi_{sm} \to 1}}, D_{s,s}^{\varphi_s \to 1}, D_{s,s}^{\varphi_a \to 1}$<br>$D_{s,a}^{\varphi_p \to 1}, \boldsymbol{D_{s,a}^{\varphi_{sm} \to 1}}, D_{s,a}^{\varphi_s \to 1}, D_{s,a}^{\varphi_a \to 1}$ | $10^{-16}$, $\mathbf{5 \cdot 10^{-16}}$, $5 \cdot 10^{-11}$, $5 \cdot 10^{-11}$<br>$\mathbf{4 \cdot 10^{-15}}$, $\mathbf{10^{-14}}$, $\mathbf{5 \cdot 10^{-10}}$, $\mathbf{5 \cdot 10^{-10}}$<br>$10^{-14}$, $\mathbf{10^{-12}}$, $2 \cdot 10^{-9}$, $2 \cdot 10^{-9}$<br>$10^{-14}$, $\mathbf{10^{-12}}$, $2 \cdot 10^{-9}$, $2 \cdot 10^{-9}$ | $m^{2} \cdot s^{-1}$ |
| $D_p^{vap}, \boldsymbol{D_{sm}^{vap}}, D_s^{vap}, D_a^{vap}$ | $10^{-16}$, $\mathbf{10^{-14}}$, $2 \cdot 10^{-9}$, $2 \cdot 10^{-9}$ | $m^{2} \cdot s^{-1}$ |
| $M_{p,0}, \boldsymbol{M_{sm,0}}$ | $\underline{0.4 \cdot 10^{-5}}$, $\mathbf{2 \cdot 10^{-6}}$ | $s^{-1}$ |
| $M_{vap}$ | $10^{6}$ | $s^{-1}$ |
| $\alpha$ | $-2.3 \cdot 10^{-5}$ | - |
| $d_{sl}, c_{sl}, w_{sl}$ | $10^{-6}$, 0.97, 35 | - |
| $d_{\zeta}, c_{\zeta}, w_{\zeta}$ | $10^{-2}$, 0.85, 15 | - |
| $d_{sv}, c_{sv}, w_{sv}$ | $10^{-3}$, 0.3, 15 | - |
| DETECTION THRESHOLDS | | |
| $t_{\phi,p}, \boldsymbol{t_{\phi,sm}}, t_{\varphi,p}, \boldsymbol{t_{\varphi,sm}}, t_{\phi_{vap}}$ | 0.4, **0.4**, 0.02, **0.02**, 0.02 | - |
| FLUID MECHANICS | | |
| $\eta_p, \boldsymbol{\eta_{sm}}, \eta_s, \eta_a, \eta_g$ | $6 \cdot 10^{6}$, $\mathbf{6 \cdot 10^{3}}$, $10^{3}$, $10^{6}$, 30 | $Pa \cdot s^{-1}$ |
| $d_{\eta}, c_{\eta}, w_{\eta}$ | $10^{-7}$, 0.2, 20 | - |
| $k_{\eta}$ | 10 | - |

## S4.6 Full simulations

In **bold** the parameters that have been added as compared to the previous section.
Underlined are the parameters that have been changed as compared to the previous section.
Subscripts ‘p’, ‘sm’, ‘s’, ‘a’ stand for ‘polymer’, ‘small molecule’, ‘solvent’, ‘air’ respectively.

| Parameters | Values | Units |
|---|---|---|
| GENERAL | | |
| *Grid spacing* | *1* | *nm* |
| *Grid size* | *512 x 256* | *nm* |
| *Initial film blend ratio* | *See main text* | - |
| $T$ | *330* | *K* |
| $\rho_p, \rho_{sm}, \rho_s, \rho_a$ | *1100, 1600, 1300, 1300* | $kg \cdot m^{-3}$ |
| $m_p, m_{sm}, m_s, m_a$ | *30, 0.91, 0.147, 0.03* | $kg \cdot mol^{-1}$ |
| THERMODYNAMICS | | |
| $\chi_{psm,ll}, \chi_{ps,ll}, \chi_{pa,ll}$<br>$\chi_{sms,ll}, \chi_{sma,ll}, \chi_{sa,ll}$ | *See main text, 0.05, 0*<br>*0.6, 0, 0* | - |
| $\chi_{psm,sl}, \chi_{ps,sl}, \chi_{pa,sl}$<br>$\chi_{smp,sl}, \chi_{sms,sl}, \chi_{sma,sl}$ | $\chi_{psm,ll} + 0.3$, $\chi_{ps,ll} + 0.35$, *0*<br>$\chi_{psm,ll} + 0.3$, $\chi_{sms,ll} + 0.1$, *0* | - |
| $\chi_{psm,ss}$ | *0* | - |
| $T_{m,p}, T_{m,sm}$ | *510, 558* | *K* |
| $L_p, L_{sm}$ | $5 \cdot 10^4, 2 \cdot 10^4$ | $J \cdot kg^{-1}$ |
| $W_p, W_{sm}$ | $7.5 \cdot 10^4, 3 \cdot 10^4$ | $J \cdot kg^{-1}$ |
| $\xi_{0,p}, \xi_{0,sm}$ | *1, 1* | - |
| $\gamma_m$ | *2* | - |
| $\sigma_{CH}, \sigma_{AC}$ | $10^{-5}, 1$ | - |
| $P_0$ | $10^5$ | *Pa* |
| $P_{sat,p}, P_{sat,sm}, P_{sat,s}, P_{sat,a}$ | $1.7 \cdot 10^2, 10^2, 2 \cdot 10^3, 10^8$ | *Pa* |
| $P_p^\infty, P_{sm}^\infty, P_s^\infty, P_a^\infty$ | *0, 0, 0, 0* | *Pa* |
| $E_{p,0}, E_{sm,0}$ | $5 \cdot 10^9, 5 \cdot 10^9$ | $J \cdot m^{-3}$ |
| $\gamma_c, \gamma_v$ | *2, 2* | - |
| $\beta$ | $10^{-5}$ | $J \cdot m^{-3}$ |
| $\gamma_b$ | *1* | - |
| $\kappa_p, \kappa_{sm}, \kappa_s, \kappa_a$ | $10^{-10}, 10^{-10}, 10^{-10}, 2 \cdot 10^{-9}$ | $J \cdot m^{-1}$ |
| $\varepsilon_p, \varepsilon_{sm}$ | $10^{-5}, 10^{-5}$ | $(J \cdot m^{-1})^{0.5}$ |
| $\varepsilon_{g,p}, \varepsilon_{g,sm}$ | $3 \cdot 10^{-2}, 3 \cdot 10^{-2}$ | $J \cdot m^{-2}$ |
| $\varepsilon_{vap}$ | $10^{-4}$ | $(J \cdot m^{-1})^{0.5}$ |
| KINETICS | | |
| $D_{s,p}^{\varphi_p \to 1}, D_{s,p}^{\varphi_{sm} \to 1}, D_{s,p}^{\varphi_s \to 1}, D_{s,p}^{\varphi_a \to 1}$<br>$D_{s,sm}^{\varphi_p \to 1}, D_{s,sm}^{\varphi_{sm} \to 1}, D_{s,sm}^{\varphi_s \to 1}, D_{s,sm}^{\varphi_a \to 1}$<br>$D_{s,s}^{\varphi_p \to 1}, D_{s,s}^{\varphi_{sm} \to 1}, D_{s,s}^{\varphi_s \to 1}, D_{s,s}^{\varphi_a \to 1}$<br>$D_{s,a}^{\varphi_p \to 1}, D_{s,a}^{\varphi_{sm} \to 1}, D_{s,a}^{\varphi_s \to 1}, D_{s,a}^{\varphi_a \to 1}$ | $10^{-16}, 5 \cdot 10^{-16}, 5 \cdot 10^{-11}, 5 \cdot 10^{-11}$<br>$4 \cdot 10^{-15}, 10^{-14}, 5 \cdot 10^{-10}, 5 \cdot 10^{-10}$<br>$10^{-14}, 10^{-12}, 2 \cdot 10^{-9}, 2 \cdot 10^{-9}$<br>$10^{-14}, 10^{-12}, 2 \cdot 10^{-9}, 2 \cdot 10^{-9}$ | $m^2 \cdot s^{-1}$ |
| $D_p^{vap}, D_{sm}^{vap}, D_s^{vap}, D_a^{vap}$ | $10^{-16}, 10^{-14}, 2 \cdot 10^{-9}, 2 \cdot 10^{-9}$ | $m^2 \cdot s^{-1}$ |
| $M_{p,0}, M_{sm,0}$ | *See main text* | $s^{-1}$ |
| $M_{vap}$ | $10^6$ | $s^{-1}$ |
| $\alpha$ | $-2.3 \cdot 10^{-5}$ | - |
| $d_{sl}, c_{sl}, w_{sl}$ | $10^{-6}$, *0.97, 35* | - |
| $d_\zeta, c_\zeta, w_\zeta$ | $10^{-2}$, *0.85, 15* | - |
| $d_{sv}, c_{sv}, w_{sv}$ | $10^{-3}$, *0.3, 15* | - |
| DETECTION THRESHOLDS | | |
| $t_{\phi,p}, t_{\phi,sm}, t_{\varphi,p}, t_{\varphi,sm}, t_{\phi_{vap}}$ | *0.4, 0.4, 0.02, 0.02, 0.02* | - |
| FLUID MECHANICS | | |
| $\eta_p, \eta_{sm}, \eta_s, \eta_a, \eta_g$ | $6 \cdot 10^6, 6 \cdot 10^3, 10^3, 10^6, 30$ | $Pa \cdot s^{-1}$ |
| $d_\eta, c_\eta, w_\eta$ | $10^{-7}$, *0.2, 20* | - |
| $k_\eta$ | *10* | - |